\documentclass[fleqn,10pt,dvipsnames]{wlscirep}
\usepackage[utf8]{inputenc}
\usepackage[T1]{fontenc}
\usepackage[displaymath,mathlines]{lineno}
\usepackage{tikz,lipsum,lmodern}
\usepackage[most]{tcolorbox}
\usepackage{multicol}

\newcommand{\revision}[1]{{#1}}
\definecolor{mypink}{rgb}{0.968, 0.94, 0.968}
\definecolor{lightgreen_custom}{rgb}{0.93, 0.94, 0.93}

\usepackage{bm,amsmath,amssymb,xcolor,graphicx}
\usepackage{setspace}
\usepackage{accents}

\newcommand*{\bdot}[1]{\accentset{\mbox{\large\bfseries .}}{#1}}

\newcommand{\nocontentsline}[3]{}
\newcommand{\tocless}[2]{\bgroup\let\addcontentsline=\nocontentsline#1{#2}\egroup}

\title{
Controlling the speed and trajectory of evolution with counterdiabatic driving}

\author[1,*]{Shamreen Iram}
\author[2,*]{Emily Dolson}
\author[1,*]{Joshua Chiel}
\author[1]{Julia Pelesko}
\author[3]{Nikhil Krishnan}
\author[1]{\"Ozen\c{c} G\"ung\"or}
\author[1,6]{Benjamin Kuznets-Speck}
\author[4]{Sebastian Deffner}
\author[5]{Efe Ilker}
\author[1,2,3,$\dagger$]{Jacob G. Scott}
\author[1,$\dagger$]{Michael Hinczewski}

\affil[1]{Department of Physics, Case Western Reserve University, Cleveland, OH, 44106, USA}
\affil[2]{Translational Hematology Oncology Research, Cleveland Clinic, Cleveland OH, 44106, USA}
\affil[3]{Case Western Reserve University School of Medicine, Cleveland, OH, 44106, USA}
\affil[4]{Department of Physics, University of Maryland, Baltimore County, Baltimore, MD 21250, USA}
\affil[5]{Physico-Chimie Curie UMR 168, Institut Curie,
PSL Research University, 75248 Paris Cedex 05, France}
\affil[6]{Biophysics Graduate Group, University of California, Berkeley, CA 94720, USA}
\affil[*]{equal contribution}
\affil[$\dagger$]{to whom correspondence should be addressed: scottj10@ccf.org, michael.hinczewski@case.edu}


\let\oldaddcontentsline\addcontentsline
\renewcommand{\addcontentsline}[3]{}

\begin{abstract}
The pace and unpredictability of evolution are critically relevant in
a variety of modern challenges: combating drug resistance in pathogens
and cancer, understanding how species respond to environmental
perturbations like climate change, and developing artificial selection
approaches for agriculture.  Great progress has been made in
quantitative modeling of evolution using fitness landscapes, allowing
a degree of prediction for future evolutionary histories.  Yet
fine-grained control of the speed and the distributions of these
trajectories remains elusive.  We propose an approach to achieve this
using ideas originally developed in a completely different context --
counterdiabatic driving to control the behavior of quantum states for
applications like quantum computing and manipulating ultra-cold atoms.
Implementing these ideas for the first time in a biological context,
we show how a set of external control parameters (i.e. varying drug
concentrations / types, temperature, nutrients) can guide the
probability distribution of genotypes in a population along a
specified path and time interval.  This level of control, allowing
empirical optimization of evolutionary speed and trajectories, has
myriad potential applications, from enhancing adaptive therapies for
diseases, to the development of thermotolerant crops in preparation
for climate change, to accelerating bioengineering methods built on
evolutionary models, like directed evolution of biomolecules.
\end{abstract}

\begin{document}
\maketitle 


The quest to control evolutionary processes in areas like agriculture and medicine predates our understanding of evolution itself.  
Recent years have seen growing research efforts toward this goal, driven by rapid progress in quantifying genetic changes across a population~\cite{mira2015rational,ogbunugafor2016adaptive,brown_compensatory_2010} as well as a global rise in challenging problems like therapeutic drug resistance~\cite{WHO2014resistance,holohan2013resistance,WHO2019resistance}. New approaches that have arisen in response include prospective therapies that steer evolution of pathogens toward maximized drug sensitivity~\cite{nichol2015steering,maltas2018pervasive}, typically requiring multiple rounds of selective pressures and subsequent evolution under them. 
Since we cannot predict the exact progression of mutations that occur in the course of the treatment, the best we can hope for is to achieve control over probability distributions of evolutionary outcomes.  
However, our lack of precise control over the timing of these outcomes poses a major practical impediment to engineering the course of evolution.  
This naturally raises a question: \textit{Rather than being at the mercy of evolution's unpredictability and pace, what if we could simultaneously control the speed and the distribution of genotypes over time?}

Controlling an inherently stochastic process like evolution has close parallels to problems in other disciplines.  Quantum information protocols crucially depend on coherent control over the time evolution of quantum states under external driving~\cite{bason2012high,zhou2017accelerated}, in many cases requiring that a system remain in an instantaneous ground state of a time-varying Hamiltonian in applications like cold atom transport~\cite{walther2012controlling} and quantum adiabatic computation~\cite{farhi2001quantum}.  
The adiabatic theorem of quantum mechanics facilitates such control when the driving is infinitely slow, but over finite time intervals control becomes more challenging, because fast driving can induce random transitions to undesirable excited states.  
Overcoming this challenge---developing fast processes that mimic the perfect control of infinitely slow ones---has led to a whole subfield of techniques called ``shortcuts to adiabaticity''~\cite{torrontegui2013shortcuts,deffner2014classical,deffner2015njp,acconcia2015pre,campbell2017trade,guery2019sta}.  
One such method in particular, known as transitionless, or counterdiabatic (CD) driving, involves adding an auxiliary control field to the system to inhibit transitions to excited states~\cite{demirplak2003adiabatic,demirplak2005assisted,berry2009transitionless}.  
Intriguingly, the utility of CD driving is not limited to quantum contexts: requiring a quantum system to maintain an instantaneous ground state under driving is mathematically analogous to demanding a classical stochastic system remains in an instantaneous equilibrium state as external control parameters are changed~\cite{patra2017shortcuts,li2017shortcuts}.  
Extending CD driving ideas to the classical realm has already led to proof-of-concept demonstrations of accelerated equilibration in optical tweezer~\cite{martinez2016engineered} and atomic force microscope~\cite{le2016fast} experimental frameworks, and is closely related to optimal, finite-time control problems in stochastic thermodynamics~\cite{schmiedl2007optimal,aurell2012refined}.

Here we demonstrate the first biological application of CD driving, by using it to control the distribution of genotypes in a Wright-Fisher model~\cite{wright1932roles} describing evolution in a population of organisms.  
The auxiliary CD control field (implemented for example through varying drug concentrations or other external parameters that affect fitness) allows us to shepherd the system through a chosen sequence of genotype distributions, moving from one evolutionary equilibrium state to another in finite time.  We validate the CD theory through numerical simulations using an agent-based model of evolving unicellular populations, focusing on a system where sixteen possible genotypes compete via a drug dose-dependent fitness landscape derived from experimental measurements.

\section{Theory}

\subsection{Evolutionary model} 
We develop our CD driving theory in the framework of a Wright-Fisher diffusion model for the evolution of genotype frequencies in a population (see Methods for details).  Let us consider $M$ possible genotypes, where the $i$th genotype comprises a fraction $x_i$ of a population.
Since $\sum_{i=1}^{M} x_i = 1$, we can describe the state of the system through $M-1$ independent values of $x_i$, or equivalently through a frequency vector $\bm{x}=(x_1,\ldots,x_{M-1})$.  
Without loss of generality, we will take the $M$th genotype to be the reference (the ``wild type'') with respect to which the relative fitnesses of the others will be defined.  
Let $1+s_i$ be the relative fitness of genotype $i = 1,\ldots,M-1$ compared to the wild type, where $s_i$ is a selection coefficient, defining the $i$th component of a vector $\bm{s}$.  
We assume fitnesses are influenced by some time-dependent control parameter $\lambda(t)$, which we write as a scalar quantity, though it could in principle be a vector, reflecting a set of control parameters.  
These parameters could involve any environmental quantity amenable to external control: in the examples below we consider the concentration of a single drug applied to a population of unicellular organisms.
However we could have more complicated drug protocols (switching between multiple drugs)~\cite{nichol2015steering} or other perturbations in fitness secondary to microenvironmental change (e.g. nutrient or oxygenation levels).  
Our control protocol $\lambda(t)$ from initial time $t_0$ to final time $t_f$ defines a trajectory of the selection coefficient vector, $\bm{s}\big(\lambda(t)\big)$, shown schematically in Fig.~\ref{f1}A.  
Our population thus evolves under a time-dependent fitness landscape, or so-called ``seascape''~\cite{Mustonen2010}.  Note that all time variables, unless otherwise noted, are taken to be in units of Wright-Fisher generations.

For simplicity, the total population is assumed to be  fixed at a value $N$, corresponding to a scenario where the system stays at a time-independent carrying capacity over the time interval of interest.  
(Our approach is easily generalized to more complicated cases with time-dependent $N(t)$, as shown in the Supplementary Information [SI]). The final quantity characterizing the dynamics is an $M \times M$ dimensional mutation rate matrix $m$, where each off-diagonal entry $m_{\beta\alpha}$ represents the mutation probability (per generation) from the $\alpha$th to the $\beta$th genotype. 
For later convenience, the $\alpha$th diagonal entry of $m$ is defined as the opposite of the total mutation rate out of that genotype, $m_{\alpha\alpha} \equiv -\sum_{\beta \ne \alpha} m_{\beta\alpha}$. As in the case of $N$, we assume the matrix $m$ is time-independent, though this assumption can be relaxed.

\subsection{Driving the genotype frequency distribution} 
Given the system described above, we focus on $p(\bm{x},t)$, the probability to find genotype frequencies $\bm{x}$ at time $t$, calculated over an ensemble of possible evolutionary trajectories.  
The dynamics of this probability for the WF model can be described to an excellent approximation through a Fokker-Planck equation:

\begin{equation}\label{t1}
\partial_t p(\bm{x},t) = {\cal L}\big(\lambda(t)\big) p(\bm{x},t),
\end{equation}

\noindent where $\partial_t \equiv \partial /\partial t$ and ${\cal L}\big(\lambda(t)\big)$ is a differential operator, acting on functions of $\bm{x}$, described in the Methods. 
This operator involves $N$, $m$, and $\bm{s}\big(\lambda(t)\big)$, and we highlight the dependence on $\lambda(t)$.  
In setting up the analogy to driving in quantum mechanics, Eq.~\eqref{t1} corresponds to the Schr\"odinger equation, with $p(\bm{x},t)$ playing the role of the wavefunction and ${\cal L}\big(\lambda(t)\big)$ the time-dependent Hamiltonian operator.  \revision{The full analogy between quantum and evolutionary dynamics is described in more detail in Box 1 of the Methods.}  Though for our purposes we only employ this analogy qualitatively, in fact there exists in certain cases an explicit mapping from the Fokker-Planck to the Schr\"odinger equation (though not vice versa)~\cite{grabert1979quantum,van1992stochastic,risken1996fokker}. 
For a particular value of the control parameter $\lambda$, the analogue of the quantum ground state wavefunction is the eigenfunction $\rho(\bm{x};\lambda)$ with eigenvalue zero, the solution of the equation

\begin{equation}\label{t2}
{\cal L}(\lambda) \rho(\bm{x};\lambda) = 0.
\end{equation}

\noindent In the evolutionary context, $\rho(\bm{x};\lambda)$ has an additional meaning with no direct quantum correspondence: it is the \textit{equilibrium probability distribution of genotypes}.  
If one fixes the control parameter $\lambda(t) = \lambda$, the distribution $p(\bm{x},t)$ obeying Eq.~\eqref{t1} will approach $\rho(\bm{x};\lambda)$ in the limit $t \to \infty$.

\begin{figure}[t]
\includegraphics[width=\textwidth]{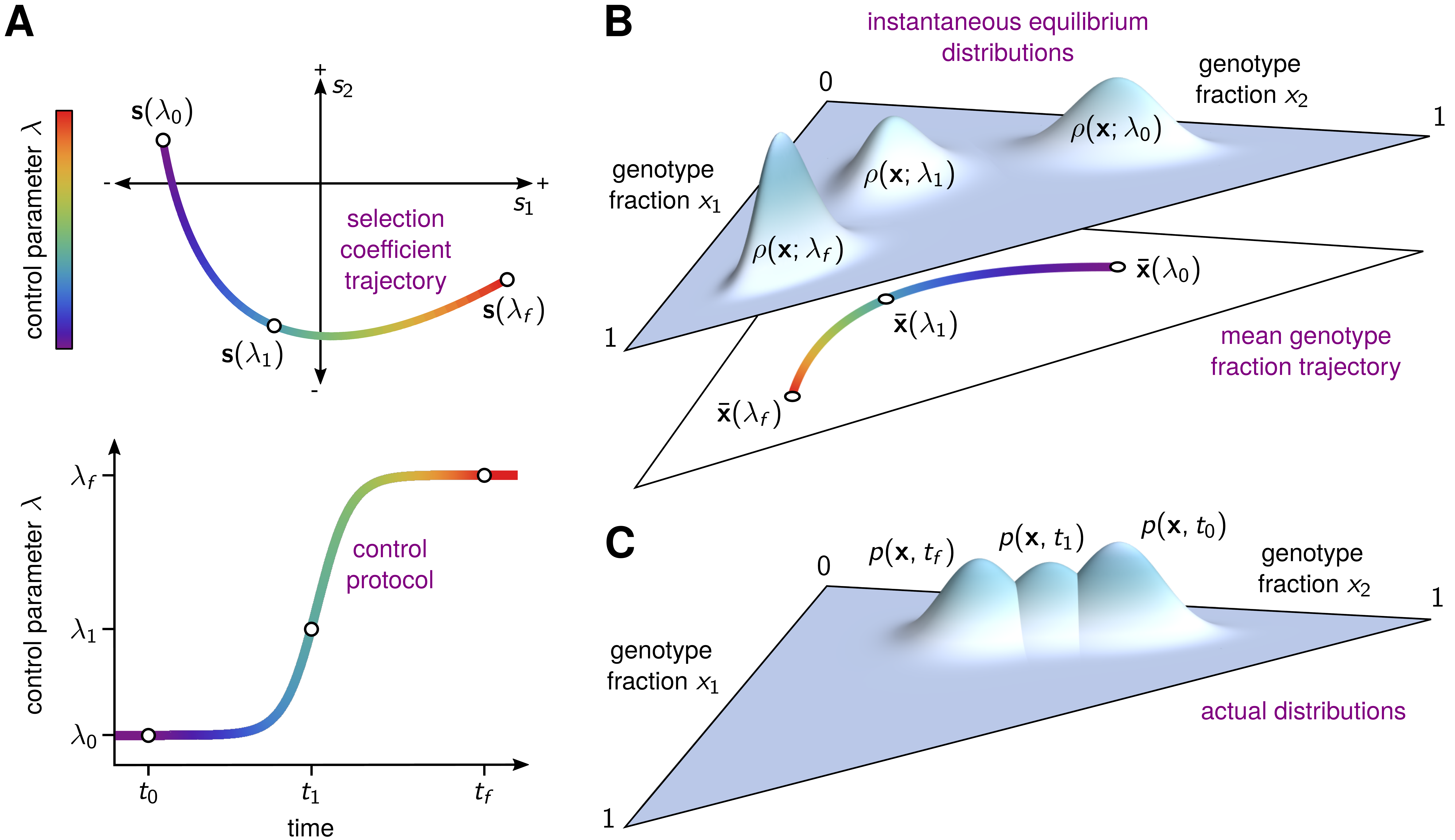}
\caption{\textbf{(A)} A schematic illustration for three genotypes ($M=3$) showing the trajectory of the selection coefficient vector $\bm{s}\big(\lambda(t)\big)$ as a function of a time-varying control parameter $\lambda(t)$ depicted in the bottom of the panel.  
This represents the fitness ``seascape'' under which the population evolves during driving.  
Three time points are highlighted: the initial time $t_0$, an intermediate time $t_1$, and the final time $t_f$, where the corresponding control parameter values are $\lambda_0$, $\lambda_1$, and $\lambda_f$.  
The amplitude of the control parameter along the trajectory is represented through a color gradient. 
\textbf{(B)} The instantaneous equilibrium (IE) distribution of genotypes $\rho\big(\bm{x};\lambda(t)\big)$ for the three highlighted values of the control parameter from panel \textbf{A}.  
These distributions are probability densities on the 2D simplex defined by $x_1+x_2 \le 1$ and $x_1$, $x_2 \ge 0$.  
In the lower part of the panel we show the curve of mean IE genotype frequencies $\overline{\bm{x}}\big(\lambda(t)\big)$.
\textbf{(C)} For driving over finite times, the actual distribution of genotypes $p(\bm{x},t)$ will generally lag behind the IE distribution while the control parameter is changing.  
Thus at $t_f$ the distribution $p(\bm{x},t_f)$ is still far from $\rho(\bm{x};\lambda_f)$, and will only catch up with it at times $t \gg t_f$ as the system re-equilibrates.}
\label{f1}
\end{figure}

Consider the following control protocol, where we start at one control parameter value, $\lambda(t) = \lambda_0$ for $t \le t_0$, and finish at another value, $\lambda(t) = \lambda_f$ for $t \ge t_f$, with some arbitrary driving function $\lambda(t)$ in the interval $t_0 < t < t_f$.  
We assume the system starts in one equilibrium distribution, $p(\bm{x},t_0) = \rho(\bm{x};\lambda_0)$, and we know that it will eventually end at a new equilibrium, $p(\bm{x},t) \to \rho(\bm{x};\lambda_f)$ for $t \gg t_f$.  
But what happens at intermediate times?  
If $\lambda(t)$ changes infinitesimally slowly during the driving (and hence $t_f \to \infty$) then the system would remain at each moment in the corresponding instantaneous equilibrium (IE) distribution, $p(\bm{x},t) = \rho\big(\bm{x};\lambda(t)\big)$ for all $t$.  
This result, derived in the SI, is the analogue of the quantum adiabatic theorem \cite{born1928adiabatic} applied to the ground state:  for a time-dependent Hamiltonian that changes extremely slowly, a quantum system that starts in the ground state of the Hamiltonian always remains in the same instantaneous ground state (assuming that at all times there is a gap between the ground state energy and the rest of the energy spectrum).  
Fig.~\ref{f1}B shows schematic snaphsots of $\rho\big(\bm{x};\lambda(t)\big)$ at three times, with the control parameter shifting them across the genotype frequency space.

When the driving occurs over finite times $(t_f < \infty)$, the above results break down: $p(\bm{x},t) \ne \rho\big(\bm{x};\lambda(t)\big)$ for $0 < t < t_f$, but is instead a linear combination of many instantaneous eigenfunctions of the Fokker-Planck operator, just as the corresponding quantum system under faster driving will generically evolve into a superposition of the instantaneous ground state and excited states.  
This will manifest itself as a lag, with $p(\bm{x},t)$ moving towards but not able to catch up with $\rho\big(\bm{x};\lambda(t)\big)$, as illustrated in Fig.~\ref{f1}C.  
For $t > t_f$, once $\lambda(t)$ stops changing, the system will eventually settle into equilibrium at $\rho(\bm{x};\lambda_f)$ in the long time limit.

\subsection{Control and counterdiabatic driving} 

This lag can be an obstacle if one wants to control the evolution of the system over finite time intervals.  
Since evolutionary trajectories are stochastic, we cannot necessarily guarantee that the system starts and ends at precise genotype frequencies, but we can attempt to specify initial and final target frequency distributions.  
At the end of the driving $t=t_f$, we would like our system to arrive at the target distribution, and then stay there as long as the control parameter is fixed.  
In this way we complete one stage of the control protocol, and have a known starting point for the next stage, since in practice we could imagine the interval $t_0 < t < t_f$ as just one step of a multi-stage protocol involving distinct interventions (i.e. a sequence of different drugs).  \revision{Completing each stage as quickly as possible, while accurately hitting each target, would for example be a crucial prerequisite to translating certain evolutionary medicine approaches to clinical settings (see SI Sec.~H for a fuller discussion).}  Thus, if we were enumerating the characteristics of an ideal control mechanism, at the very least it should be able to drive the system from one equilibrium distribution, $p(\bm{x},t_0) = \rho(\bm{x};\lambda_0)$, to another, $p(\bm{x},t_f) = \rho(\bm{x};\lambda_f)$, over a finite time $t_f - t_0$.  

In the context of quantum adiabatic computing~\cite{farhi2001quantum}, the typical focus is on the initial ground state (which has to be easy to realize experimentally) and the final ground state (since it encodes the solution to the computational problem).  
In the evolutionary case, we can imagine additional desired characteristics for our driving, beyond the start and end-point distributions.  
There are many ways to go from an initial fitness landscape, $\bm{s}(\lambda_0)$, to a final fitness landscape, $\bm{s}(\lambda_f)$, corresponding to different possible trajectories in the selection coefficient space of Fig.~\ref{f1}A that share initial and final values.  
Depending on how we empirically implement the control, many of these trajectories may be physically inaccessible.  
But among the remaining set of realizable trajectories, some may be more desirable than others (i.e. have different evolutionary consequences~\cite{nichol2019antibiotic}, or trade-offs~\cite{li2019single}).  
Each trajectory defines a continuous sequence of IE distributions $\rho\big(\bm{x};\lambda(t)\big)$, and for each distribution there is a mean genotype frequency $\overline{\bm{x}}\big(\lambda(t)\big)$, illustrated in the lower half of Fig.~\ref{f1}B.  
We may, for example, want protocols that minimize the chances of our system visiting certain problematic genotypes: in practice this could translate to demanding that the curve $\overline{\bm{x}}\big(\lambda(t)\big)$ for $t_0 < t < t_f$ stays far away from certain regions of the genotype frequency space.  
This in turn restricts the $\bm{s}\big(\lambda(t)\big)$ trajectories and hence the protocols $\lambda(t)$ of practical interest.  
In simpler terms, we would ideally like to control not just the distributions at the beginning and end of the driving, but also if possible along the way.

We formulate this ideal control problem in the following way: we demand that $p(\bm{x},t) = \rho\big(\bm{x};\lambda(t)\big)$ for some chosen control protocol $\lambda(t)$ between $t_0 < t < t_f$.  
The protocol $\lambda(t)$ is determined with the above considerations in mind, and thus defines a particular path through the space of genotype frequency distributions over which we would like to guide our system.  
Clearly we will not achieve success by just directly implementing $\lambda(t)$, since $p(\bm{x},t)$ obeying Eq.~\eqref{t1} will generally lag behind $\rho\big(\bm{x};\lambda(t)\big)$ \cite{vaikuntanathan2009lag}.  
The resolution of this problem in the quantum case through CD driving is to add a specially constructed auxiliary time-dependent Hamiltonian to the original Hamiltonian~\cite{demirplak2003adiabatic,demirplak2005assisted,berry2009transitionless}.  For a specific choice of this auxiliary Hamiltonian, we can guarantee that our new system always remains in the instantaneous ground state of the original Hamiltonian.
The evolutionary analogue of CD is to replace the Fokker-Planck operator ${\cal L}\big(\lambda(t)\big)$ in Eq.~\eqref{t1} with a different operator $\widetilde{\cal L}\big(\lambda(t),\bdot{\lambda}(t)\big)$, which depends on both $\lambda(t)$ and its time derivative $\bdot{\lambda}(t) \equiv d\lambda(t)/dt$.  
This CD operator satisfies

\begin{equation}\label{t3}
\partial_t \rho\big(\bm{x};\lambda(t)\big) = \widetilde{\cal L}\big(\lambda(t),\bdot{\lambda}(t)\big) \rho\big(\bm{x};\lambda(t)\big).
\end{equation}

\noindent Thus by construction, $p(\bm{x},t) = \rho\big(\bm{x};\lambda(t)\big)$ is a solution to the Fokker-Planck equation with the new operator.
Additionally, to be consistent with the slow adiabatic driving limit discussed above, $\widetilde{\cal L}(\lambda(t),0) = {\cal L}\big(\lambda(t)\big)$, so we recover the original Fokker-Planck operator when the speed of driving $\bdot{\lambda}(t) \to 0$ and $t_f \to \infty$.

Of course defining $\widetilde{\cal L}\big(\lambda(t), \bdot{\lambda}(t)\big)$ in this way is the easy part: figuring out how to implement a new control protocol to realize $\widetilde{\cal L}\big(\lambda(t),\bdot{\lambda}(t)\big)$ is more challenging.  
In the Methods, we show how the most general solution to go from ${\cal L}$ to $\widetilde{\cal L}$ is to replace the original selection coefficient trajectory $\bm{s}\big(\lambda(t)\big)$ with a frequency-dependent version, $\tilde{\bm{s}}\big(\bm{x};\lambda(t),\bdot{\lambda}(t)\big)$.  
Implementing a particular frequency dependent fitness seascape is a degree of control that is generally impossible in realistic scenarios.  
Fortunately, we show that in one important parameter regime the CD seascape becomes approximately frequency-independent, $\tilde{\bm{s}}\big(\bm{x};\lambda(t),\bdot{\lambda}(t)\big) \approx \tilde{\bm{s}}\big(\lambda(t),\bdot{\lambda}(t)\big)$.  
This occurs in the large population, frequent mutation regime: if the typical mutation rate scale is $\mu$, meaning $m_{\beta\alpha} \sim {\cal O}(\mu)$ for all nonzero mutation rates where $\alpha \ne \beta$, then this corresponds to $\mu N \gg 1$, $N \gg 1$~\cite{gillespie1983simple,gerrish1998fate,desai2007beneficial,sniegowski2010beneficial}.  
In this regime multiple genotypes can generally coexist in the population at equilibrium (though one may be quite dominant), which is particularly relevant for pathogenic populations, especially ones spreading through space~\cite{martens2011interfering,magdanova2013heterogeneity,krishnan2019range}.
Remarkably, there is a simple analytical expression that provides an excellent approximation to $\tilde{\bm{s}}\big(\lambda(t),\bdot{\lambda}(t)\big)$ in this case:

\begin{equation}\label{t4}
    \tilde{s}_i\big(\lambda(t),\bdot{\lambda}(t)\big) \approx s_i\big(\lambda(t)\big) + \frac{d}{dt} \ln \frac{\overline{x}_i\big(\lambda(t)\big)}{\overline{x}_M\big(\lambda(t)\big)}, \qquad i=1,\ldots,M-1,
\end{equation}

\noindent where $\overline{x}_M\big(\lambda(t)\big) \equiv 1- \sum_{i=1}^{M-1} \overline{x}_i\big(\lambda(t)\big)$. 
We see that the new selection coefficient protocol is defined through the target mean genotype frequency trajectory $\overline{\bm{x}}\big(\lambda(t)\big)$, and reduces to the original protocol when $\bdot{\lambda}(t) \to 0$.
Moreover, as we show in the examples below for specific systems, Eq.~\eqref{t4} can at least in certain cases be implemented through physically realistic manipulations of the environment, like time-varying drug dosages.  While we focus on the frequent mutation regime in the current work, the applicability of CD ideas is not limited to just this regime:  for the opposite case of infrequent mutations, $\mu N \ll 1$, where the evolutionary dynamics can be modeled as a sequence of mutant fixations, one can also formulate a CD theory based on a discrete Markov state description~\cite{Sella2005} (see our follow-up article~\cite{discreteCDpreprint}).

\section{Results}


\subsection{Two genotypes}

\begin{figure}[h]
\includegraphics[width=\textwidth]{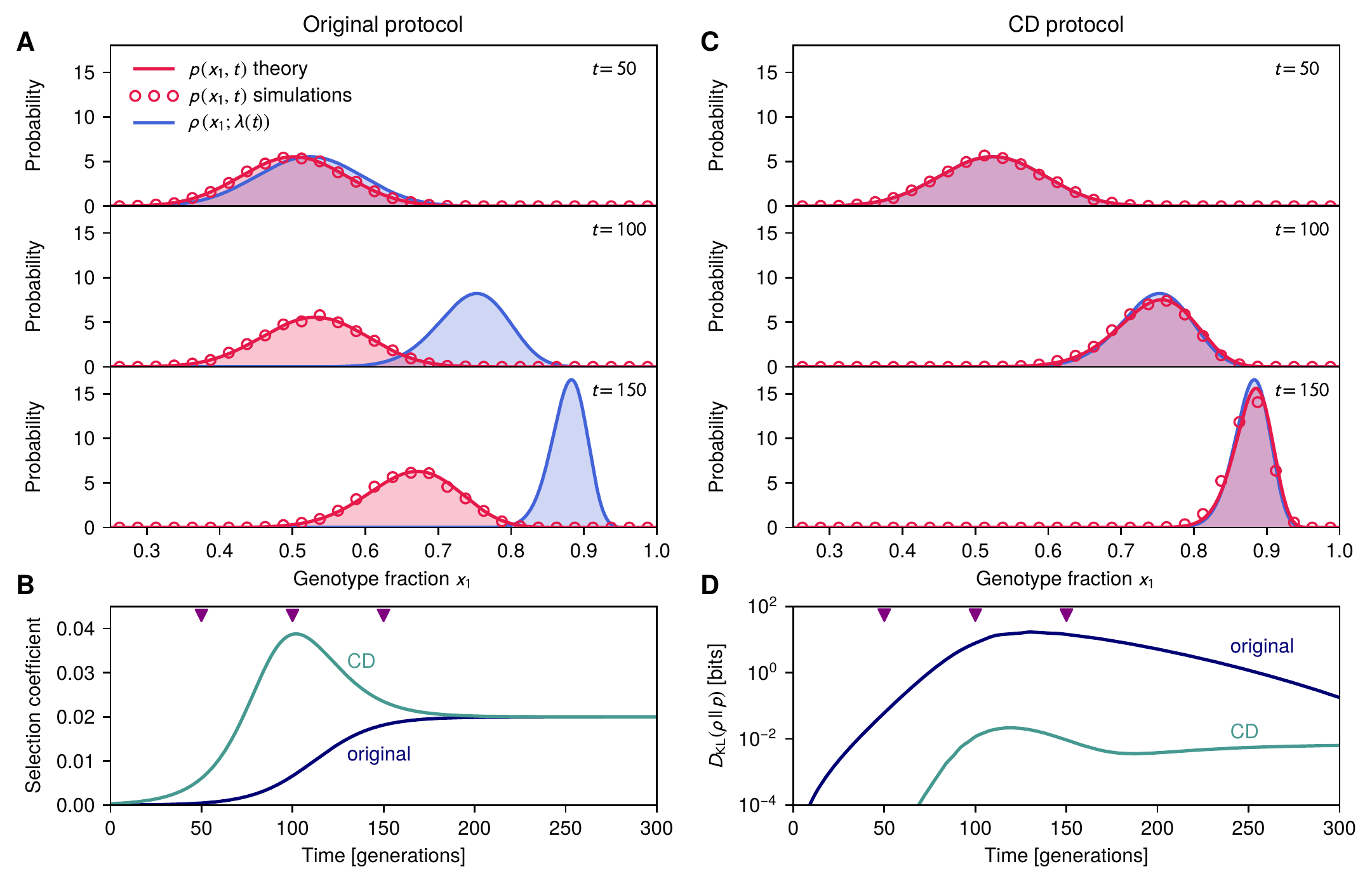}
\caption{\textbf{Results for two genotypes.} 
\textbf{(A,C)} We plot three time snapshots of the actual probability distribution $p(x_1,t)$ versus the IE distribution $\rho\big(x_1;\lambda(t)\big)$ for driving with the original control protocol \textbf{(A)} and with the CD driving protocol \textbf{(C)}, where $x_1$ is the fraction of genotype 1 in the population.  
Solid red curves are numerical solutions of the Fokker-Planck Eq.~\eqref{t1} for $p(x_1,t)$, red circles are agent-based simulations.
Without CD driving, the actual distribution always lags behind the IE. 
\textbf{B)} The selection coefficient trajectory $s_1$ for the original control protocol (dark blue) versus the corresponding CD prescription (green) $\tilde{s}_1$ from Eq.~\eqref{eq: 2 type prescription}.  The three snapshot times \revision{(50, 100, 150 generations)} are indicated by triangles. 
\textbf{(D)}  Kullback-Leibler divergence between actual and IE distributions versus time, with and without CD driving, calculated using the numerical Fokker-Planck solution.}
\label{fig: fig2}
\end{figure}

The simplest example of our CD theory is for a two genotype ($M=2$) system, where the dynamics are one dimensional, described by a single frequency $x_1$ and selection coefficient $s_1(\lambda(t))$.  
As shown in the Methods, Eq.~\eqref{t4} in this case can be evaluated analytically.  To illustrate driving,  we assume a control protocol $\lambda(t)$ such that the selection coefficient increases according to a smooth ramp (the original protocol in Fig.~\ref{fig: fig2}B).  This starts from zero at $t_0$ (both genotypes have equal fitness) and increases until reaching a plateau at a final selection coefficient that favors genotype 1. Fig.~\ref{fig: fig2}A shows $p(x_1,t)$ from a numerical solution of Eq.~\eqref{t1} using this protocol, compared against the IE distribution $\rho\big(x_1;\lambda(t)\big)$, solved using Eq.~\eqref{t2}, at three time snapshots.  
To validate the Fokker-Planck approach, we also designed an agent-based model (ABM), described in the Methods section~\ref{method: ABM}, which simulates the individual life trajectories of an evolving population of cells.  
Because there exists a mapping between the parameters of the ABM and the equivalent Fokker-Planck equation (Methods section~\ref{method: ABM mapping}), one can directly compare the $p(x_1,t)$ results from the ABM simulations (circles) to the Fokker-Planck numerical solution of Eq.~\eqref{t1} (curves), which show excellent agreement.  
In the absence of CD driving, as expected, $p(x_1,t)$ lags behind $\rho\big(x_1;\lambda(t)\big)$, with the latter shifting rapidly to larger $x_1$ frequencies as the fitness of genotype 1 increases.

To eliminate this lag, we implement the alternative selection coefficient trajectory of Eq.~\eqref{eq: 2 type prescription}.  
Fig.~\ref{fig: fig2}B shows a comparison between $\tilde{s}_1\big(\lambda(t),\bdot{\lambda}(t)\big)$ and the original $s_1\big(\lambda(t)\big)$.  
We see that the CD intervention requires a transient overshoot of the selection coefficient during the driving, nudging $p(x_1,t)$ to keep up with $\rho\big(x_1;\lambda(t)\big)$.  
Panel C shows the same snapshots as in panel A, but now with CD driving: we see the actual and IE distributions nearly perfectly overlap at all times.  
To quantify the effectiveness of the CD protocol, we measure the degree of overlap through the Kullback-Leibler (KL) divergence~\cite{kullback_information_1951,vaikuntanathan2009lag}, defined for any two probability distributions $p(\bm{x})$ and $q(\bm{x})$ as $D_\text{KL}(p||q) = \int d\bm{x}\,p(\bm{x}) \log_2 \big(p(\bm{x})/q(\bm{x})\big)$. Expressed in bits, the KL divergence is always $\ge 0$, and equals 0 for identical distributions.  Panel D shows $D_\text{KL}(\rho||p)$ for both the original and CD protocols, with the latter dramatically reducing the divergence across the time interval of driving.

\clearpage

\subsection{Multiple genotypes via agent-based modeling}

\begin{figure}[h]
\includegraphics[width=\textwidth]{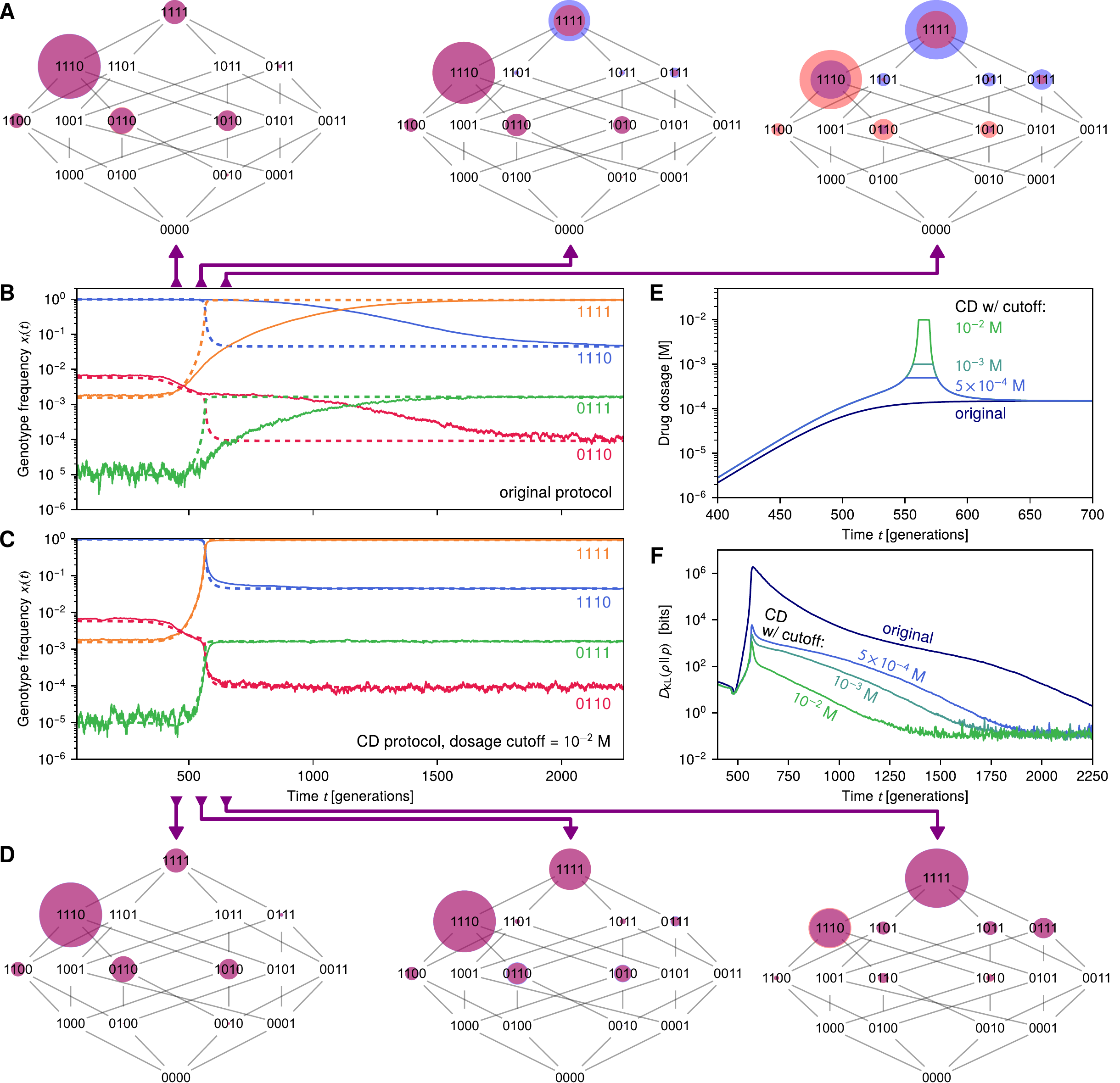}
\caption{\textbf{CD driving eliminates evolutionary lag in 16 genotype simulation.}  
\textbf{(A,D)} We plot three snapshots of our evolving agent-based population model without \textbf{(A)} and with \textbf{(D)} CD driving.
Each of the three $16$ genotype ($4$ binary alleles) hypercubic graphs (``tesseracts'') has vertices with log-scaled radii representing the fraction of each genotype in the total population at a given time.  Orange is the actual fraction, blue the IE fraction, and the overlap appears purple. The CD driving in this case is implemented approximately through a drug dosage protocol (panel \textbf{E}) with cutoff 10$^{-2}$ M.  \textbf{(B,C)} Corresponding sample simulation trajectories (solid lines) versus IE expectation (dashed lines) for the fraction of 4 representative genotypes without \textbf{(B)} and with \textbf{(C)} CD driving. The latter significantly reduces the nearly 1500 generation lag. \textbf{(E)} Original drug protocol versus the CD protocol with different possible dosage cutoffs.  \textbf{(F)} Kullback-Leibler divergence between actual and IE distributions versus time, with and without CD driving.  For the latter, increasing the dosage cutoff value makes the protocol more closely approximate the true CD solution, and hence decreases the divergence.}
\label{n_dim_fig}
\end{figure}

The ABM simulations also allow us to test the CD theory in more complex scenarios.  To this end we considered a system with 16 genotypes (4 alleles), with selection coefficients based on a well characterized experimental system: the fitness effects of the anti-malarial drug pyrimethamine at varying concentrations on all possible combinations of four different drug-resistance alleles~\cite{ogbunugafor2016adaptive, brown_compensatory_2010}.  Our control parameter $\lambda(t)$ is the drug concentration, and we implement the seascape by increasing the drug over time (after an initial equilibration period), eventually saturating at a concentration of $10^{-4}$ M (the protocol labeled ``original'' in Fig.~\ref{n_dim_fig}E).  With our choice of simulation parameters (Methods section~\ref{method: ABM simulation}), a number of the genotypes have sufficient resistance to survive even at higher drug dosages, so the overall population remains at carrying capacity.  What changes as the dosage increases is the distribution of genotypes.  Fig.~\ref{n_dim_fig}A and \ref{n_dim_fig}B show the results in the absence of CD driving, with each genotype labeled by a 4-digit binary sequence.  The population goes from being dominated by 1110 (with smaller fractions of other genotypes) to eventually becoming dominated by 1111.  However there is a dramatic lag behind the IE distribution, taking more than 1500 generations to resolve.  This is quantified in the KL divergence $D_\text{KL}(\rho||p)$ in Fig.~\ref{n_dim_fig}F, which rapidly increases by 5 orders of magnitude as the drug ramp starts showing its effects (around generation 500).  Equilibration to the higher drug dosages brings the divergence back down over time, but it only achieves relatively small amplitudes after generation 2000.  Note that the scale of the KL divergences for 15-dimensional probability distributions is larger than for the 1D example in the previous section:  this reflects both the greater sensitivity of the KL measure to small discrepancies in a 15-dimensional space, as well as the fact that distributions estimated from an ensemble of simulations (1000 indepedent runs in this case) will always have a degree of sampling error.  Thus it is more instructive to look at the relative change of the KL with driving rather than the absolute magnitudes.

To reduce the lag through CD driving, one should in principle implement the selection coefficients according to Eq.~\eqref{t4}.  However this involves guiding the system along a fitness trajectory in a 15-dimensional space, and in this case we have a single tuning knob (the concentration of pyrimethamine) to perturb fitnesses.  In such scenarios one then looks at the closest approximation to CD driving that can be achieved with the experimentally accessible control parameters.  In this particular case the genotypes which dominate the population at small and large drug concentrations are 1110 ($i=15$) and the wild-type 1111 ($i=16$), so the selection coefficient $s_{15}$ which encodes their fitness relative to each other plays the most important role in the dynamics.  We thus choose a CD drug dosage by numerically solving for the concentration that most closely approximates the $i=15$ component of Eq.~\eqref{t4} at each time.  Because in real-world scenarios there will be limits on the maximum allowable dosage, we constrain the CD concentrations to be below a certain cutoff.  \revision{The approximation described here,  where two different genotypes dominate at different times during the driving, is just a special case of a more general approximation approach where we seek to achieve the closest possible protocol to the one described by Eq.~\eqref{t4}, given the experimental constraints.  In SI Sec.~I we illustrate how this general strategy works in two additional 16-genotype seascape examples (including the empirical seascape for the drug cycloguanil~\cite{ogbunugafor2016adaptive}) where more than two genotypes dominate during driving.}

Fig.~\ref{n_dim_fig}E shows CD drug protocols with three different cutoffs:  $10^{-2}$, $10^{-3}$, $5\times 10^{-4}$ M (all within the experimentally measured dosage range).  The higher the cutoff, the better the approximation to CD driving.  \revision{We can directly quantify the overall reduction in lag time $\Delta t$ due to CD from the KL divergence results of Fig.~\ref{n_dim_fig}F, as explained in Methods section~\ref{method: KL}.}  For a cutoff of $10^{-2}$ M the lag is reduced by \revision{$\Delta t = 1210$} generations.  Notably, though the approximation is based on the top two genotypes (1110 and 1111), it reduces the lag time across the board for all genotypes (see Fig.~\ref{n_dim_fig}C for four representative genotype trajectories at $10^{-2}$ M cutoff, with other genotypes shown in the snapshots of Fig.~\ref{n_dim_fig}D).  This is because driving of the top two also entrains the dynamics of the subdominant genotypes whose populations are sustained by mutations out of and into the dominant ones.  Even with the more restrictive constraint of $5 \times 10^{-4}$ M there is still a substantial benefit, with the lag reduced by \revision{$\Delta t = 656$} generations.  This highlights the robustness of the CD approach:  even if one cannot implement the solution of Eq.~\eqref{t4} exactly, we can still arrive at the target distribution faster through an approximate CD protocol.

\clearpage

\section{Discussion and Conclusions}

Our demonstration of the CD driving approach in a population model with empirically-derived drug-dependent fitnesses shows that we can accelerate evolution toward a target distribution \textit{in silico}.  
As new technologies progressively allow us to assemble ever more extensive fitness landscapes for various organisms as a function of external perturbations like drugs~\cite{mira2015rational,ogbunugafor2016adaptive,brown_compensatory_2010}, the next step is implementing CD driving in the laboratory.  \revision{This would be a necessary milestone on the path to a range of potential applications (the latter discussed in more detail in SI Sec.~H).}  Thus it is worth considering the challenges and potential workarounds that will be involved in experimental applications.  

One salient issue is the range of control parameters available in laboratory settings.  Our examples have focused on the simplest cases of one-dimensional control, but to access the full power of the CD approach presented here we should explore a richer parameter space:  not only single drugs, but combinations, along with varying nutrients, metabolites, oxygen levels, osmotic pressure, and temperature.  The eventual goal would be \revision{to have for every system} a library of well-characterized interventions that could be applied in tandem, allowing us the flexibility to map out desired target trajectories through a multidimensional fitness landscape.  \revision{In other words for a given system we would have access to a selection coefficient function $\bm{s}(\bm{\lambda}(t))$, where $\bm{\lambda}(t) = (\lambda_1(t), \lambda_2(t), \ldots)$ is a multidimensional vector of control parameters at time $t$: $\lambda_1(t)$ the concentration of one drug, $\lambda_2(t)$ the concentration of another drug (or nutrient), and so on.} An interesting future extension of the theory would also investigate the role of spatial environments and restrictions as a potential control knob.  More generally, one could explore how fundamental differences among fitness landscapes (i.e. the difficulties in reaching local optima in so-called ``hard'' landscapes~\cite{kaznatcheev2019computational}) influences the types of interventions needed to achieve driving and their effectiveness.

Even given accurately measured fitnesses, one might be hampered by imperfect estimation of other system parameters, for example mutation rates.  To determine how large the margin for error is, we tested the CD driving prescription calculated using incorrect mutation rates, varying the degree of discrepancy  over two orders of magnitude (see SI for details). While such discrepancies do reduce the efficacy of CD driving, leading to deviations between actual and IE distributions at intermediate times, populations driven with an incorrect protocol still reached the target distribution faster than in the absence of driving.  As in the case of the dosage cutoff discussed above, the CD approach has a degree of robustness to errors in the protocol, which increases its chances of success in real-world settings.

But what if we lacked measurements of the underlying fitness seascape?  
Interestingly, there might still be some utility of the CD method even in this case.  
We could first do a preliminary quasi-adiabatic experimental trial:  vary external parameter(s) $\lambda$ extremely gradually, and use sequencing at regularly spaced intervals to determine the quasi-equilibrium mean genotype fractions $\overline{x}_i(\lambda)$ as a function of $\lambda$.  
If we now wanted to guide the system through the same sequence of evolutionary distributions but much faster, we have enough information to approximately evaluate the CD perturbation in Eq.~\eqref{t4}, which just depends on $\overline{x}_i(\lambda)$ and the rate $\bdot{\lambda}(t)$ that we would like to implement.  
So at the very least the CD prescription could be estimated, providing a blueprint, and the remaining challenge would be figuring out what combination of external perturbations would yield the right sorts of fitness perturbations to achieve CD driving.

``Nothing makes sense in biology except in the light of evolution'' is an oft-quoted maxim which was the title of a 1973 essay by Theodosius Dobzhansky~\cite{dobzhansky1973nothing}.  However evolution is not just the fundamental paradigm through which we can understand living systems, but also a framework by which we can shape and redesign nature at a variety of scales:  from engineering new proteins~\cite{romero2009exploring} and aptamers~\cite{tuerk1990systematic,ellington1990vitro} to combating drug resistance in pathogens and cancer~\cite{nichol2015steering} to the development of crops that can withstand climate stress~\cite{bita2013plant}.  
In all its manifestations, natural and synthetic, evolution is a stochastic process that occurs across a wide swath of timescales.  Our work represents a significant step toward more precise control of both the distribution of possible outcomes and the timing of this fundamental process.

\section{Methods}

\begin{tcolorbox}[
  colback=lightgreen_custom,
  colframe=Green,
  title={~\centering Box 1: Analogies between quantum physics and evolutionary dynamics}]
Here we summarize the connections between quantum and evolutionary concepts used in our theory.  Each numbered item in the quantum column on the left has its analogue in the evolutionary column on the right.
\begin{multicols*}{2}
\begin{center} $\quad$\underline{\bf Quantum physics} \end{center}

\begin{enumerate}
    \item {\bf Wavefunction:} describes the state of a quantum system.  For a simple quantum particle in a spatial region described by coordinates $\bm{x}$, this is a function $\psi(\bm{x},t)$ whose squared amplitude $|\psi(\bm{x},t)|^2$ is the probability density of finding the particle at $\bm{x}$ at time $t$.
    
    \item {\bf Hamiltonian operator:} a differential operator ${\cal H}(\lambda(t))$ depending on control parameters $\lambda(t)$ defined below.  It describes how the wavefunction changes in time through the {\bf time-dependent Schr\"odinger equation}, $i\hslash\partial_t \psi(\bm{x},t) = {\cal H}(\lambda(t)) \psi(\bm{x},t)$, where $\hslash$ is the reduced Planck constant.  ${\cal H}(\lambda(t))$ involves terms that correspond to the kinetic and potential energies of the quantum particle.
    
    \item {\bf Control parameters:} a set of parameters $\lambda(t)$ that can be manipulated over time by an experimentalist.  These parameters modify the kinetic/potential energy terms in the Hamiltonian, and thus influence the quantum dynamics.  An example of this would be the magnitude of an externally applied electromagnetic field.
    
    \item {\bf Ground state:} the lowest energy state of a quantum system.  In general, for a Hamiltonian ${\cal H}(\lambda)$ and given parameter values $\lambda$, the energy states (labeled by $n=0,1,2,\ldots$) correspond to solutions of the time-independent Schr\"odinger equation:  ${\cal H}(\lambda)\psi_n(\bm{x};\lambda) = E_n(\lambda) \psi_n(\bm{x};\lambda)$.  Here $E_n(\lambda)$ and $\psi_n(\bm{x};\lambda)$ are the energy and wavefunction respectively of the $n$th state.  The energies $E_0 < E_1 < \cdots$, and the ground state corresponds to $n=0$.  If $\lambda$ is fixed, a system whose wavefunction is $\psi_n(\bm{x};\lambda)$ will be stationary, with its wavefunction not changing in time.
    
    \item {\bf Adiabatic theorem:} if we start in the $n$th energy state, $\psi(\bm{x},t=0) = \psi_n(\bm{x};\lambda(0))$ for some initial control parameters $\lambda(0)$, and then vary $\lambda(t)$ infinitesimally slowly (adiabatically) the theorem states that the wavefunction at later times remains in $n$th energy state corresponding to the instantaneous value of the parameters, $\psi(\bm{x},t) = \psi_n(\bm{x}; \lambda(t))$.  This is true so long as there is always a nonzero difference between $E_n(\lambda(t))$ and any $E_m(\lambda(t))$ for $m\ne n$ at all $t$.
\end{enumerate} 
\vfill\columnbreak
\begin{center} $\qquad$\underline{\bf Evolutionary dynamics} \end{center}

\vspace{-2em}
\begin{enumerate}
    \item {\bf Genotype probability distribution:} the distribution $p(\bm{x},t)$ of genetic variants (genotypes) in a population of organisms at time $t$, where $\bm{x}$ is a vector of genotype fractions.
    
    \item {\bf Fokker-Planck operator:} a differential operator ${\cal L}(\lambda(t))$ depending on control parameters $\lambda(t)$ defined below.  It describes how the genotype probability $p(\bm{x},t)$ changes in time through the {\bf Fokker-Planck equation}, $\partial_t p(\bm{x},t) = {\cal L}(\lambda(t)) p(\bm{x},t)$.  The full form of ${\cal L}(\lambda(t))$ [Eqs.~\eqref{e1}-\eqref{e3}] involves terms that describe the mean change in genotype fractions due to mutations and selection, as well as the effects of genetic drift.
    
    \item {\bf Control parameters:} a set of parameters $\lambda(t)$ that can be manipulated over time by an experimentalist.  These parameters modify genotype fitnesses, and hence influence evolutionary dynamics through the selection terms in the Fokker-Planck operator.  An example would be the concentration of a drug applied to a microbial population, where different genotypes exhibit different degrees of resistance against the drug depending on the concentration.
    
    \item {\bf Equilibrium state:} for a given set of parameter values $\lambda$, this is the genotype probability $\rho(\bm{x};\lambda)$ which would remain unchanged in time (stationary) during evolutionary dynamics under fixed $\lambda$.  In general a Fokker-Planck operator ${\cal L}(\lambda)$ has a set of eigenfunctions $\psi_n(\bm{x};\lambda)$ and eigenvalues $-\kappa_n(\lambda) \le 0$ for $n=0,1,2\ldots$ defined through the following equation: ${\cal L}(\lambda) \psi_n(\bm{x};\lambda) = -\kappa_n(\lambda) \psi_n(\bm{x};\lambda)$.  The equilibrium state corresponds to $n=0$, with eigenvalue $-\kappa_0(\lambda) =0$ and $\rho(\bm{x};\lambda) \equiv \psi_0(\bm{x};\lambda)$.
    
    \item {\bf Adiabatic theorem:} If we start at equilibrium $p(\bm{x},t=0) = \rho(\bm{x};\lambda(0))$ for some initial control parameters $\lambda(0)$, and then vary $\lambda(t)$ infinitesimally slowly (adiabatically) the theorem (derived in SI Sec.~A) states that at later times we will always remain in the equilibrium state corresponding to the instantaneous value of the parameters, $p(\bm{x},t) = \rho(\bm{x};\lambda(t))$.
\end{enumerate}
\end{multicols*}

\label{box1}
\end{tcolorbox}

\subsection{Fokker-Planck description of Wright-Fisher evolutionary model}\label{method: FP of WF}

The underlying evolutionary dynamics of our model are based on the canonical haploid Wright-Fisher (WF) model with mutation and selection, and we adopt the formalism of recent approaches~\cite{Baxter2007,Mustonen2010} that generalized Kimura's original two-allele diffusion theory~\cite{Kimura1955} to the case of multiple genotypes. A convenient feature of the WF formalism is that other, more detailed descriptions of the population dynamics (for example agent-based models that track the life histories of individual organisms) can often be mapped onto an effective WF form, as we illustrate below.

The starting point of the Fokker-Planck diffusion approximation~\cite{Kimura1955,Baxter2007,Mustonen2010} for evolutionary population dynamics is the assumption that genotype frequencies change only by small amounts in each generation.  Thus we can take the genotype frequency vector $\bm{x}$ to be a continuous variable that follows a stochastic trajectory.  The key quantities describing these stochastic dynamics are the lowest order moments of $\delta \bm{x}$, the change in genotype frequency per generation.  
We will denote the mean of the change in the $i$th genotype, $\delta x_i$, taken over the ensemble of possible trajectories, as $v_i\big(\bm{x};\lambda(t)\big) \equiv \langle \delta x_i\rangle$. Note that in general $v_i\big(\bm{x};\lambda(t)\big)$ will be a function of the genotype frequencies $\bm{x}$ at the current time step, and also have a dependence on the control parameter $\lambda(t)$ through the selection coefficient vector $\bm{s}\big(\lambda(t)\big)$ (which influences $\langle \delta x_i\rangle$).

In non-evolutionary contexts $v_i\big(\bm{x};\lambda(t)\big)$ is called the drift function, but here we will call it a velocity function to avoid confusion with genetic drift.  Similarly we will introduce an $(M-1)\times(M-1)$ diffusivity matrix $D_{ij}(\bm{x})$ to describe the covariance of the genotype changes, defined through $2D_{ij}(\bm{x}) \equiv \langle \delta x_i \delta x_j\rangle - \langle \delta x_i \rangle \langle \delta x_j \rangle$.  As shown in the SI, to lowest order approximation $D_{ij}(\bm{x})$ is independent of $\bm{s}\big(\lambda(t)\big)$, and hence is not an explicit function of $\lambda(t)$.   If we are interested in the dynamics on time scales much larger than a single generation, the probability $p(\bm{x},t)$ to observe a genotype state $\bm{x}$ at time $t$ obeys a multivariate Fokker-Planck equation~\cite{Gillespie1996},

\begin{equation}
\begin{split}\label{e1}
\partial_t p(\bm{x},t) &= -\partial_i \left( v_i\big(\bm{x};\lambda(t)\big) p(\bm{x},t) \right) + \partial_i \partial_j \big(D_{ij}(\bm{x}) p(\bm{x},t) \big)\\
&\equiv {\cal L}\big(\lambda(t)\big) p(\bm{x},t),
\end{split}
\end{equation}
where $\partial_t \equiv \partial/\partial t$ and $\partial_i \equiv \partial / \partial x_i$.  Note that we use Einstein summation notation, where repeated indices are summed over, and furthermore designate Greek indices to range from $1$ to $M$ while Roman indices range from $1$ to $M-1$. So for example the term $\partial_i \partial_j \big(D_{ij}(\bm{x}) p(\bm{x},t) \big) \equiv \sum_{i=1}^{M-1} \sum_{j=1}^{M-1} \partial_i \partial_j \big(D_{ij}(\bm{x}) p(\bm{x},t) \big)$. The right-hand-side of Eq.~\eqref{e1} defines the Fokker-Planck differential operator ${\cal L}\big(\lambda(t)\big)$ in main text Eq.~\eqref{t1}.  In order to correspond to genotype fractions, the vectors $\bm{x}$ have to lie in the $(M-1)$-dimensional simplex $\Delta$ defined by the conditions $x_i \ge 0$ for all $i$ and $\sum_{j=1}^{M-1} x_j \le 1$.  If $\bm{x} \in \Delta$, then the wild type fraction $x_M = 1-\sum_{j=1}^{M-1}x_j$ lies between 0 and 1.  Normalization of $p(\bm{x},t)$ takes the form $\int_{\Delta} d\bm{x}\, p(\bm{x},t) = 1$, where the integral is over the volume of the simplex $\Delta$.

To complete the description of the model, we need expressions for the functions $v_i\big(\bm{x};\lambda(t)\big)$ and $D_{ij}(\bm{x})$. Given a Wright-Fisher evolutionary model, these take the following form (see SI for a detailed derivation):

\begin{equation}\label{e2}
v_i(\bm{x};\lambda(t)) = m_{i\mu} x_{\mu}  + g_{ij}(\bm{x}) s_j(\lambda(t)), \qquad D_{ij}(\bm{x},t) = \frac{g_{ij}(\bm{x})}{2N},
\end{equation}
where $m$ is the $M \times M$ mutation rate matrix defined in the main text, and $g(\bm{x})$ is an $(M-1) \times (M-1)$ matrix with elements given by

\begin{equation}\label{e3}
    g_{ij}(\bm{x}) \equiv \begin{cases} -x_i x_j & i\ne j\\
    x_i (1-x_i) & i = j \text{, no sum over } i \\\end{cases}.
\end{equation}

\subsection{Instantaneous equilibrium distributions}\label{method: IE}

The instantaneous equilibrium (IE) distribution $\rho(\bm{x};\lambda(t))$ is defined through main text Eq.~\eqref{t2}, ${\cal L}(\lambda(t)) \rho(\bm{x};\lambda(t)) = 0$.  Because we evaluate the effectiveness of our driving by comparing the actual distribution $p(\bm{x},t)$ to the IE distribution, it is useful to know the form of $\rho(\bm{x};\lambda(t))$.  Unfortunately it is generally not possible to find an IE analytical expression, except in some specific cases~\cite{Baxter2007,Mustonen2010}.  The two genotype system ($M=2$) is one example where an exact solution is known.  It has a form analogous to the Boltzmann distribution of statistical physics~\cite{Baxter2007,Mustonen2010},

\begin{equation}
    \label{e4}
    \rho(\bm{x};\lambda(t)) = \frac{e^{-\Phi(\bm{x};\lambda(t))}}{Z(\lambda(t))},
\end{equation}
where $\Phi(\bm{x};\lambda(t))$ is an effective ``potential'' given by

\begin{equation}\label{e4b}
    \Phi(\bm{x};\lambda(t)) = -2N \left(m_{12} \log x_1 + m_{21} \log (1-x_1) + s_1(\lambda(t))x_1\right) + \log{\det g(\bm{x})}
\end{equation}
and $Z(\lambda(t))$ is a normalization constant.

To estimate the IE distribution for general $M$, we take advantage of the large population, frequent mutation regime: $m_{\beta\alpha} \sim {\cal O}(\mu)$ for all nonzero matrix entries where $\alpha \ne \beta$, with $\mu N \gg 1$, $N \gg 1$.  In this case we know that $\rho(\bm{x};\lambda(t))$ is approximately a multivariate normal distribution of the form

\begin{equation}
    \label{e5}
    \rho(\bm{x};\lambda(t)) \approx \left( (2\pi)^{M-1} \det \Sigma(\lambda(t))\right)^{-1/2} \exp\left(-\frac{1}{2}\bigl(x_i - \overline{x}_i(\lambda(t))\bigr) \Sigma^{-1}_{ij}(\lambda(t)) \bigl(x_j - \overline{x}_j(\lambda(t))\bigr) \right).
\end{equation}
Here $\overline{x}_i(\lambda) = \int_{\Delta} d\bm{x}\, x_i \rho(\bm{x};\lambda)$ is the $i$th mean genotype fraction for the IE distribution, and $\Sigma^{-1}(\lambda)$ is the inverse of the covariance matrix $\Sigma(\lambda)$ for this distribution.  The latter has entries $\Sigma_{ij}  \equiv \overline{x_i x_j} - \overline{x}_i \overline{x}_j$.  In order to make practical use of Eq.~\eqref{e5}, we need a method to estimate $\overline{x}_i(\lambda)$ and $\Sigma(\lambda)$.  As shown in the SI, this can be done through an approximate numerical solution to a set of exact equations involving the moments of $\rho(\bm{x};\lambda)$.  

\subsection{Counterdiabatic driving protocol}\label{method: CD}

To implement CD driving, we need to solve for the CD Fokker-Planck operator $\widetilde{\cal L}\big(\lambda(t),\bdot{\lambda}(t)\big)$ that satisfies main text Eq.~\eqref{t3}.  We posit that $\widetilde{\cal L}$ should be in the Fokker-Planck form of Eq.~\eqref{e1}, but with some CD version of the selection coefficient, $\tilde{\bm{s}}\big(\bm{x};\lambda(t),\bdot{\lambda}(t)\big)$, instead of the original $\bm{s}(\lambda(t))$.  The necessary perturbation to the fitness seascape to achieve CD driving, $\delta \tilde{\bm{s}}\big(\bm{x};\lambda(t),\bdot{\lambda}(t)) \equiv \tilde{\bm{s}}\big(\bm{x};\lambda(t),\bdot{\lambda}(t))- \bm{s}(\lambda(t))$, we take for now to be frequency-dependent for generality.
Thus main text Eq.~\eqref{t3} takes the form

\begin{equation}\label{e6}
\begin{split}
\partial_t \rho\big(\bm{x};\lambda(t)\big) &= \widetilde{\cal L}\big(\lambda(t),\bdot{\lambda}(t)\big) \rho\big(\bm{x};\lambda(t)\big)\\
&=-\partial_i \left( \tilde{v}_i\big(\bm{x};\lambda(t),\bdot{\lambda}(t)\big) \rho\big(\bm{x};\lambda(t)\big) \right) + \partial_i \partial_j \big(D_{ij}(\bm{x}) \rho\big(\bm{x};\lambda(t)\big) \big)
\end{split}
\end{equation}
with a modified velocity function:

\begin{equation}\label{e7}
\begin{split}
\tilde{v}_i\big(\bm{x};\lambda(t),\bdot{\lambda}(t)\big) &= m_{i\mu} x_{\mu}  + g_{ij}(\bm{x}) \tilde{s}_j\big(\bm{x};\lambda(t),
\bdot{\lambda}(t)\big)\\
&= v_i\big(\bm{x};\lambda(t)\big) + g_{ij}(\bm{x}) \delta \tilde{s}_j\big(\bm{x};\lambda(t),
\bdot{\lambda}(t)\big).
\end{split}
\end{equation}
Using the fact that ${\cal L}(\lambda(t)) \rho\big(\bm{x};\lambda(t)\big) =0$, since $\rho\big(\bm{x};\lambda(t)\big)$ is the IE distribution of the original operator ${\cal L}$, we can rewrite Eq.~\eqref{e6} as

\begin{equation}\label{e8}
\begin{split}
\partial_t \rho\big(\bm{x};\lambda(t)\big) &=- \partial_i \left(\rho\big(\bm{x};\lambda(t)\big) g_{ij}(\bm{x}) \delta \tilde{s}_j\big(\bm{x};\lambda(t),\bdot{\lambda}(t)\big) \right)\\
&\revision{\equiv -\partial_i \mathcal{J}_i.}
\end{split}
\end{equation}
\revision{where $\mathcal{J}_i \equiv \rho\big(\bm{x};\lambda(t)\big) g_{ij}(\bm{x}) \delta \tilde{s}_j\big(\bm{x};\lambda(t),\bdot{\lambda}(t)\big)$ is a probability current.  In this form Eq.~\eqref{e8} looks like a continuity equation, describing the local transport of probability density due to the current field $\bm{\mathcal{J}}$. In order for this equation to conserve total probability over the simplex $\Delta$, we also require the condition that $\mathcal{J}_i n_i = 0$ at any point on the boundary of the simplex, where the vector $\bm{n}$ is normal to the boundary at the point.  The perturbation $\delta \tilde{\bm{s}}\big(\bm{x};\lambda(t),\bdot{\lambda}(t))$ that satisfies Eq.~\eqref{e8} and the boundary condition defines an exact CD protocol for the evolutionary system.}

\revision{Given an arbitrary continuous time sequence of IE distributions $\rho\big(\bm{x};\lambda(t)\big)$, such a perturbation always exists.  In fact, from a formal mathematical standpoint~\cite{sahoo2008}, any perturbation of the following form is a solution (note that for clarity we do not use Einstein summation in this case):}

\revision{\begin{equation}\label{e8b}
\delta \tilde{\bm{s}}\big(\bm{x};\lambda(t),\bdot{\lambda}(t)) = \bm{g}^{-1}(\bm{x})\left[\frac{1}{\rho\big(\bm{x};\lambda(t)\big)}\left(-\sum_{i=1}^{M-1}\hat{\bm{x}}_i w_i \int_{0}^{x_i} dx_i^\prime\, \partial_t \rho\big(x_1,\ldots,x_i^\prime,\ldots,x_{M-1};\lambda(t)\big) + \bm{f}\big(\bm{x};\lambda(t),\bdot{\lambda}(t)\big)\right) \right].
\end{equation}
Here $\bm{g}^{-1}(\bm{x})$ is the inverse of the matrix $\bm{g}(\bm{x})$, $\hat{\bm{x}}_i$ is the unit vector along the $i$th axis, and the integral in the $i$th term of the sum is carried out only over the $i$th genotype fraction, keeping all other components $x_{j}$, $j\ne i$, fixed.  There are two quantities in Eq.~\eqref{e8b} that make the solution potentially non-unique:  the weights $w_i$ can be any real numbers, so long as $\sum_{i=1}^{M-1} w_i = 1$; and $\bm{f}\big(\bm{x};\lambda(t),\bdot{\lambda}(t)\big)$ is an $(M-1)$-dimensional vector function which has zero divergence, $\partial_i f_i =0$.  However we an have additional constraints on this function $\bm{f}$: it has to be compatible with the vanishing of the current orthogonal to boundary, $\mathcal{J}_i n_i =0$.  For $M=2$, where necessarily $w_1 =1$, these constraints mean that only $\bm{f}=0$ is allowed, and we get a unique exact CD solution.  For $M>2$, the partial differential equation $\partial_i f_i =0$ and the boundary condition do not specify $\bm{f}$ uniquely, and hence we get many possible allowable CD solutions all of which satisfy Eq.~\eqref{e8}.  This in turn means that we can always find CD Fokker-Planck operators $\widetilde{\cal L}\big(\lambda(t),\bdot{\lambda}(t)\big)$ that satisfy main text Eq.~\eqref{t3}.}

\revision{However the formal existence of such perturbations $\delta \tilde{\bm{s}}$ is not the end of the story, because many of the solutions described by Eq.~\eqref{e8b} may not be physically realizable.  In order to get at a more practical (though approximate) CD solution, we proceed as follows.} As discussed Sec.~\ref{method: IE}, in the regime of interest it is easier to work with moments of the IE distribution, so it is useful to convert \revision{Eq.~\eqref{e8}} into a relation involving the IE first moment $\overline{x}_i(\lambda)$.  Multiply both sides of \revision{Eq.~\eqref{e8}} by $x_k$, and notice that $x_k \partial_{i}\mathcal{J}_{i} = \partial_{i}\left(x_k \mathcal{J}_{i}\right) - \delta_{ik}\mathcal{J}_{i}$, where $\delta_{ik}$ is the Kronecker delta function. Integrating over the entire simplex gives

\begin{equation}\label{e10}
    \int_{\Delta} d\bm{x}\,x_k \partial_t\rho(\bm{x};\lambda(t))=  - \int_{\Delta} d\bm{x}\, \partial_{i}\left(x_k \mathcal{J}_{i}\right) + \int_{\Delta}d\bm{x}\,\mathcal{J}_{k}.
\end{equation}
By Gauss's theorem, $\int_{\Delta}d\bm{x}\, \partial_{i}\left(x_k \mathcal{J}_{i}\right) = \int_{\partial \Delta} d\sigma\, x_k \mathcal{J}_{i}n_i$, where the integral involves area elements $d\sigma$ of the simplex boundary $\partial \Delta$, and $n_i$ are the components of the normal vector to this boundary. By conservation of probability, the component of $\mathcal{J}$ normal to $\partial \Delta$ vanishes, i.e. $\mathcal{J}_i n_i = 0$, so the first term in Eq.~\eqref{e10} is zero.  Plugging the definition of $\mathcal{J}_k$ into the second term, we get

\begin{equation}\label{e11}
    \int_{\Delta}d\bm{x}\, x_k \partial_t {\rho}(\bm{x};\lambda(t))= \int_{\Delta} d\bm{x}\,{\rho}(\bm{x};\lambda(t)) g_{kj}(\bm{x}) \delta \tilde{s}_{j}(\bm{x};\lambda(t),\bdot{\lambda}(t)),
\end{equation}
or equivalently

\begin{equation}\label{e12}
\partial_t \overline{\bm{x}}(\lambda(t)) = \left\langle \bm{g}(\bm{x}) \delta \tilde{\bm{s}}\big(\bm{x};\lambda(t),\bdot{\lambda}(t)\big)\right\rangle,
\end{equation}
where the brackets $\langle \: \rangle$ denote an average over the simplex with respect to ${\rho}(\bm{x};\lambda(t))$.

So far both Eq.~\eqref{e8} and \eqref{e12} are exact relations satisfied by the CD perturbation $\delta\tilde{\bm{s}}$.  However we can simplify the results in the large population, frequent mutation regime, where $\rho(\bm{x};\lambda(t))$ has the approximate normal form of Eq.~\eqref{e5}.  As argued in the SI, in this case the leading contribution to $\delta \tilde{\bm{s}}$ is frequency-independent, $\delta \tilde{\bm{s}}(\bm{x};\lambda(t),\bdot{\lambda}(t)) \approx \delta \tilde{\bm{s}}(\lambda(t),\bdot{\lambda}(t))$, with corrections that vanish in the large $N$ limit.  The leading contribution $\delta \tilde{\bm{s}}(\lambda(t),\bdot{\lambda}(t))$ satisfies a version of Eq.~\eqref{e12} with $\bm{x}$ on the right-hand side replaced by the IE mean $\overline{\bm{x}}(\lambda(t))$,

\begin{equation}\label{e13}
\partial_t \overline{\bm{x}}(\lambda(t)) =  \bm{g}\big(\overline{\bm{x}}(\lambda(t))\big) \delta \tilde{\bm{s}}\big(\lambda(t),\bdot{\lambda}(t)\big).
\end{equation}
This equation can be directly solved for $\delta \tilde{\bm{s}}(\lambda(t),\bdot{\lambda}(t))$ in terms of $\overline{\bm{x}}(\lambda(t))$, yielding the approximate CD solution of main text Eq.~\eqref{t4}.  Thus knowing the IE first moment $\overline{\bm{x}}(\lambda(t))$ over the duration of the protocol (via the numerical procedure described in the SI) allows us to estimate a CD driving prescription.

\subsection{CD driving for the two genotype example}

For the $M=2$ system, the exact IE distribution is given by Eqs.~\eqref{e4}-\eqref{e4b}.  In the large population, frequent mutation limit we can estimate the mean frequency $\overline{x}_1(\lambda(t))$ corresponding to this distribution as:

\begin{equation}\label{j6}
    \overline{x}_{1}(\lambda(t)) \approx \frac{-m_{12}-m_{21}+s_1(\lambda(t)) + \sqrt{\big(m_{12}+m_{21}-s_1(\lambda(t))\big)^2+ 4 m_{12} s_1(\lambda(t))}}{2 s_1(\lambda(t))}.
\end{equation}
This allows the CD prescription in Eq.~\eqref{t4} to be evaluated analytically, yielding

\begin{equation}\label{eq: 2 type prescription}
\tilde{s}_1\big(\lambda(t),\bdot{\lambda}(t)\big)= s_1\big(\lambda(t)\big) + \frac{\partial_t {s}_1\big(\lambda(t)\big)}{\sqrt{\left(m_{12}+m_{21}-s_1\big(\lambda(t)\big)\right)^2 + 4 m_{12} s_1\big(\lambda(t)\big)}}.
\end{equation}
For the results in Fig.~\ref{fig: fig2}, we assume the following ramp for the selection coefficient:  $s_1\big(\lambda(t)\big) = \sigma/(1+a e^{-kt}) - \sigma/(1+a)$, with $\sigma = 0.02$, $a = 817$, $k = 0.06$.  The other model parameters are set to:  $N=10^4$, $m_{12} = m_{21} = 2.5 \times 10^{-3}$.

\subsection{Agent based model}\label{method: ABM}

\subsubsection{Model description}

For the agent based model (ABM) simulations, we track a population of single-celled organisms that undergo birth (through binary division), death, and mutations.  There are $M$ genotypes, and the fitness of genotype $i<M$ relative to the $M$th one (the wildtype) is $1+s_i(\lambda(\tau))$, which depends on the drug dosage $\lambda(\tau)$ at the current simulation time step $\tau$.  (The mapping between simulation time steps $\tau$ and Wright-Fisher generations $t$ will be discussed below.) At each simulation time step, every cell in the population undergoes the following process:  i) with probability $d$ it dies; ii) if it survives, the cell divides with a genotype-dependent probability

\begin{equation}\label{1}
b_i(\tau) = \begin{cases} \text{min}\left(b_0(1+s_i(\lambda(\tau)) \left(1-\frac{N_\text{cell}(\tau)}{K}\right),1\right) & N_\text{cell}(\tau) \le K\\
0 & N_\text{cell}(\tau)>K,
\end{cases}
\end{equation}
where $i$ is the cell's genotype, $b_0$ is a baseline birth rate, $N_\text{cell}(\tau)$ is the current number of cells in the population, and $K$ is the carrying capacity.  Upon division, the daughter cell mutates to another genotype $j$ with probability $\hat{m}_{ji}$, $j \ne i$.

\subsubsection{In silico implementation}
\label{method: ABM simulation}

\begin{figure}[h]
\centering
\includegraphics[width=0.7\textwidth]{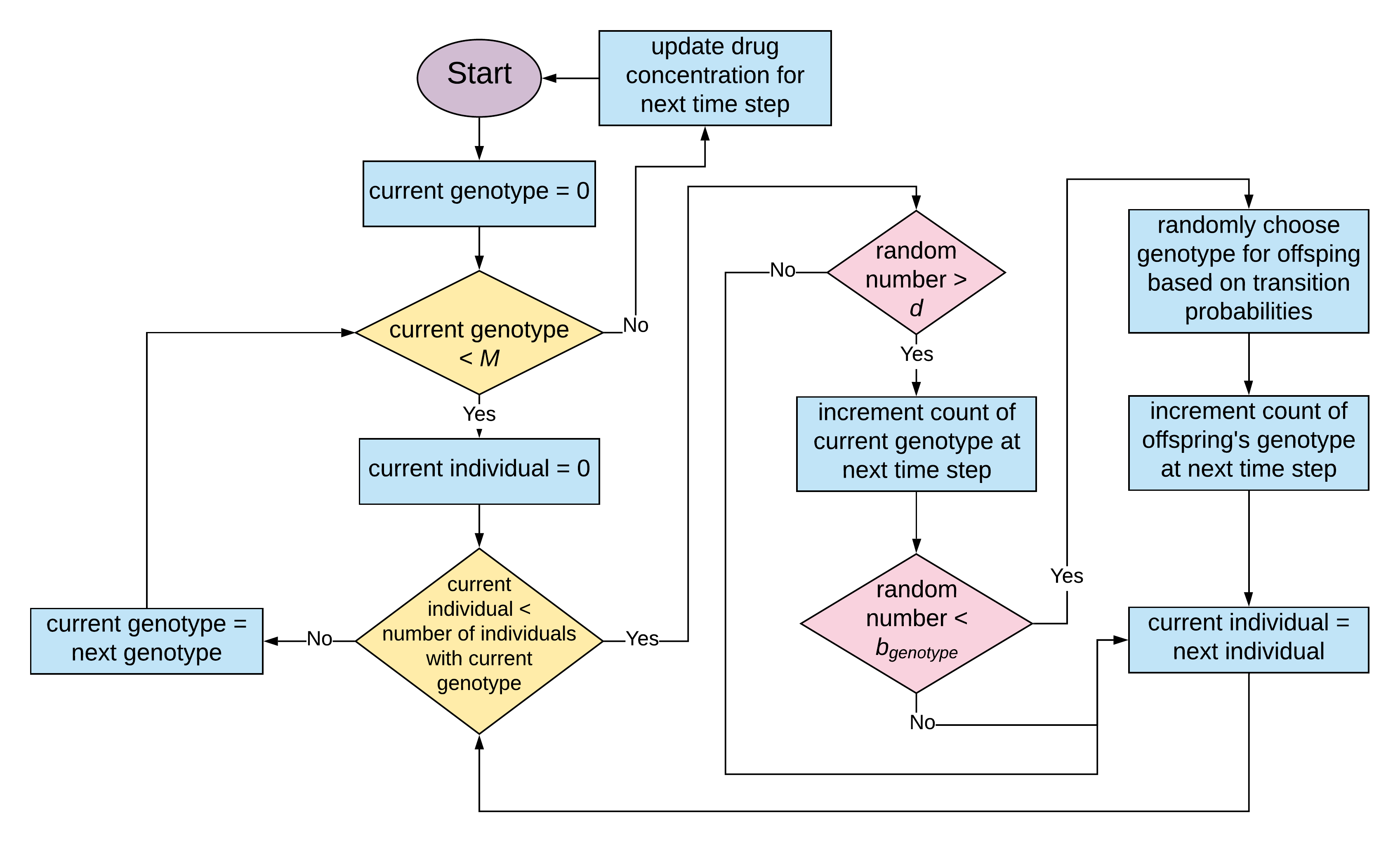}
\caption{Flow chart describing the code executed at each time step by ABMs in this paper. 
$M$ refers to the total number of genotypes in the model, $d$ to the death rate, and $b_{genotype}$ to the birth rate of the current genotype.  Random numbers in the chart are drawn uniformly from the range 0 to 1.}
\label{abm_flowchart}
\end{figure}

The ABM  was implemented for the $M=2$ and $M=16$ examples described in the main text using code written in the C++ programming language.  Code, configuration files, and analysis scripts for these models can be found on \url{https://github.com/Peyara/Evolution-Counterdiabatic-Driving}. The code directly implements the model of the previous section, and is summarized in the flowchart of Fig.~\ref{abm_flowchart}.  The $M=2$ selection coefficient $s_1(\lambda(t))$ and other model parameters are as described in the main text.

For $M=16$, the simulations were run for $4.5 \times 10^4$ time steps with a death rate $d = 0.05$, a baseline birth rate $b_0 = 2$, and a carrying capacity $K = 5 \times 10^6$. The mutation probability $\hat{m}_{ji}$, $j \ne i$ is zero unless the Hamming distance between the binary string representation of $i$ and $j$ is 1.  This gives the ``tesseract'' connectivity seen in Fig.~\ref{n_dim_fig}A,D.  Where nonzero, the probability $\hat{m}_{ji} = 2.5\times 10^{-4}$, giving a total mutation probability $\sum_{j\ne i} \hat{m}_{ji} = 10^{-3}$ for all offspring.  To give the population time to reach an initial equilibrium, the drug concentration $\lambda(\tau)$ is initially small, increases substantially around time step $\tau \sim 10^4$, and then plateaus at later times.  The dosage follows the equation,

\begin{equation}
\lambda(\tau) = \frac{a}{1 + \exp\big(-b(\tau - c)\big)},
\label{eq: drug_ramp}
\end{equation}
with parameters: $a = 1.5 \times 10^{-4}$ M, $b = 2\times 10^{-3}$, $c = 10,110$.  The selection coefficients $s_i(\lambda(\tau))$ were varied with concentration $\lambda(\tau)$ in accordance with the experimentally measured dose-fitness curves of 16 genotypes for the anti-malarial drug pyrimethamine~\cite{ogbunugafor2016adaptive, brown_compensatory_2010}.  To calculate distributions of genotype frequencies, every simulation is repeated 1000 times.  

\subsubsection{Mapping the ABM simulations to a Fokker-Planck equation}\label{method: ABM mapping}

In order to implement the CD driving protocol, derived for Wright-Fisher Fokker-Planck dynamics, in the context of the ABM, we need a mapping between the ABM parameters and the corresponding Fokker-Planck parameters.  As shown in the SI, this can be done by describing the ABM simulation population updating at each time step as an effective Langevin equation, and then using the connection between the Langevin and Fokker-Planck descriptions~\cite{Gillespie1996,gillespie2000chemical}.  The resulting approximate correspondence is summarized as follows:  i) a duration of $\tau$ ABM simulation time steps maps to $t \approx \tau d$ Wright-Fisher generations, where $d$ is the ABM death rate.  ii) The Fokker-Planck mutation matrix entries $m_{i\nu}$, $i \ne \nu$, are given by $m_{i\nu} \approx \hat{m}_{i\nu}(1+s_\nu)$, where $\hat{m}_{i\nu}$ are the ABM mutation probabilities. iii) The effective population $N$ in the Fokker-Planck model is given by $N \approx \frac{1}{2}K(1-d b_0^{-1}(1-d)^{-1})$, where $K$ is the ABM carrying capacity and $b_0$ the baseline birth rate.  The accuracy of this mapping is illustrated in Fig.~\ref{fig: fig2}A,C, where the distributions from ABM simulations for $M=2$ (red circles) are compared against numerical Fokker-Planck solutions with parameters calculated using the mapping (red curves).

\subsubsection{Numerical estimation of the KL divergence \revision{and reduction in lag time}}\label{method: KL}

To quantify the effectiveness of the CD driving, we use the KL divergence between the actual distribution, $p(\bm{x},t)$ and the IE one, $\rho(\bm{x};\lambda(t))$, defined as $D_\text{KL}(\rho||p) = \int d\bm{x}\,\rho(\bm{x};\lambda(t)) \log_2 \big(\rho(\bm{x};\lambda(t))/p(\bm{x},t)\big)$.  For $M=2$, the Fokker-Planck equation can be solved numerically for $p(x_1,t)$, while $\rho(x_1;\lambda(t))$ is known analytically (Eqs.~\eqref{e4}-\eqref{e4b}).  Hence the one-dimensional integral for $D_\text{KL}(\rho||p)$ can be numerically evaluated.  For $M=16$ the situation becomes more complicated.  There is no analytical solution for $\rho(\bm{x};\lambda(t))$, but we do have a good approximation in terms of the multivariate normal distribution of Eq.~\eqref{e5}, expressed in terms of the mean vector $\overline{\bm{x}}(\lambda(t))$ and covariance matrix $\Sigma(\lambda(t))$ that are calculated using the moment approach described in the SI.  The ABM simulation results are also normally distributed in this parameter regime, and hence there is a corresponding simulation mean $\overline{\bm{x}}_\text{sim}(t)$ and covariance $\Sigma_\text{sim}(t)$ that can be calculated at each time $t$.  These are calculated from the ensemble of 1000 simulations that are run for each parameter set.  The integral for the KL divergence $D_\text{KL}(\rho||p)$ between the simulation and IE multivariate normal distributions can then be evaluated directly, yielding

\begin{equation}
D_\text{KL}(\rho||p) = \frac{1}{2 \ln 2}\left[\ln\frac{\det \Sigma_\text{sim}(t)}{\det \Sigma(\lambda(t))} - M + 1 +\text{tr}\,\left( \Sigma_\text{sim}^{-1}(t) \Sigma(\lambda(t))\right) + \left(\overline{\bm{x}}_\text{sim}(t) - \overline{\bm{x}}(\lambda(t))\right)^T \Sigma_\text{sim}^{-1}(t)\left(\overline{\bm{x}}_\text{sim}(t) - \overline{\bm{x}}(\lambda(t))\right) \right]
\end{equation}
Since $\Sigma_\text{sim}(t)$ will have some degree of sampling errors due to the finite size of the simulation ensemble, it can in some cases be badly conditioned.  In these scenarios the Moore-Penrose pseudo-inverse is used to estimate $\Sigma_\text{sim}^{-1}(t)$.

\revision{We can use the curves of $D_\text{KL}(\rho||p)$ as a function of time, for example those of Fig.~\ref{n_dim_fig}F, to estimate how much lag time $\Delta t$ is being eliminated using a given approximate CD protocol, relative to the original one.  This lag time savings $\Delta t = t^\text{orig}_\text{eq} - t^\text{CD}_\text{eq}$, where $t^\text{orig}_\text{eq}$ and $t^\text{CD}_\text{eq}$ are respectively the times at which probability distributions in the original and CD protocols reach their final IE target values.  In terms of $D_\text{KL}(\rho||p)$, there is minimum value $D^\text{eq}_\text{KL}$ attained at long times when $p(\bm{x},t)$ has converged with $\rho(\bm{x};\lambda(t))$.  Note this value is not precisely zero because of numerical noise associated with the estimation of the distribution $p(\bm{x},t)$ from a finite number of simulations.  At long times when $D_\text{KL}(\rho||p)$ approaches $D^\text{eq}_\text{KL}$, the final approach can be fit well by the following exponential decay function,}

\revision{
\begin{equation}\label{fit}
D_\text{KL}(\rho||p)\approx \begin{cases} D^\text{eq}_\text{KL} e^{-(t_\text{eq}- t)/\tau} & t\le t_\text{eq}\\
D^\text{eq}_\text{KL} & t > t_\text{eq}\end{cases}.
\end{equation}
Since we know $D^\text{eq}_\text{KL}$ from the long-time behavior of the KL divergence curves, we then can estimate $\tau$ and $t_\text{eq}$ by fitting Eq.~\eqref{fit} to the final decay portion of each $D_\text{KL}(\rho||p)$ curve (the time range where $D_\text{KL}(\rho||p)$ is within two orders of magnitude of $D^\text{eq}_\text{KL}$).  After finding $t_\text{eq}$ for the original and CD protocols, the difference gives us the $\Delta t$ values quoted in the main text and SI.}



\section*{Acknowledgements}

MH would like to thank the U.S. National Science Foundation for support through the CAREER grant (BIO/MCB 1651560).  JGS would like to thank the NIH Loan Repayment Program for their generous support and the Paul Calabresi Career Development Award for Clinical Oncology (NIH K12CA076917).
SD acknowledges support from the U.S. National Science Foundation under Grant No. CHE-1648973.

\section*{Author contributions statement}
\begin{figure}[h!]
    \centering
    \includegraphics[width=0.55\textwidth]{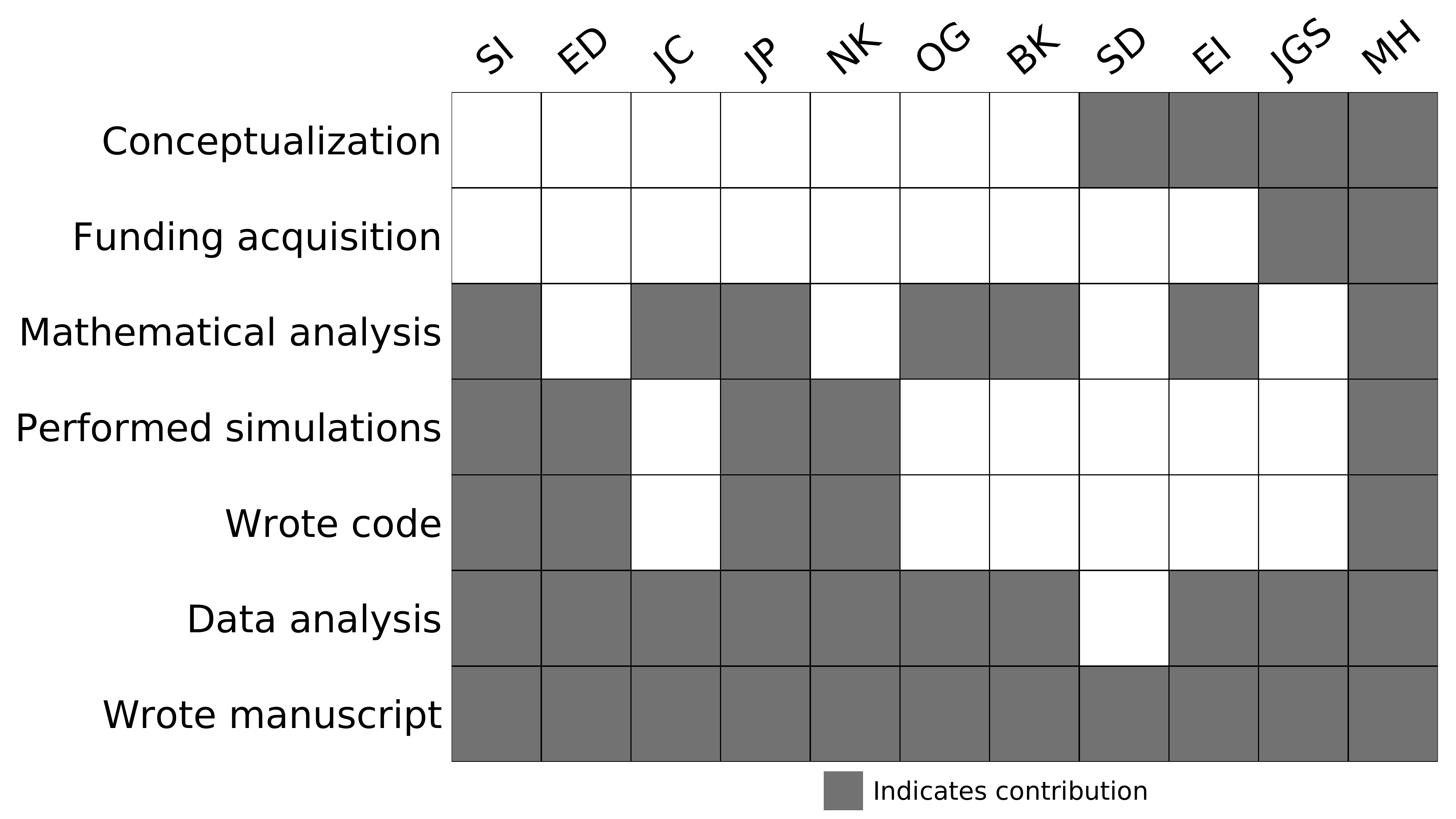}
    \caption{\textbf{Author Contributions} {
    }}
    \label{fig:authors}
\end{figure}

SI and JP performed mathematical analysis, wrote the two-allele code, peformed simulations, analyzed the data and wrote the manuscript.
JC, EI, OG, BK performed mathematical analysis, analyzed the data and wrote the manuscript.
ED and NK wrote the multidimensional ABM code, performed the simulations, analyzed data and wrote the manuscript.
JGS analyzed the data and wrote the manuscript. 
MH performed the mathematical analysis and simulations, wrote code, analyzed the data and wrote the manuscript.
SD wrote the manuscript, and SD, EI, JGS, MH contributed to developing the overall theoretical framework.  These contributions are graphically illustrated in Figure~\ref{fig:authors}.

{}

\section*{Data availability}

The raw numerical data for the figures in the main text and SI, as well as the code to generate the figures, is available via github at
\noindent
\parbox{\linewidth}{\url{https://github.com/Peyara/Evolution-Counterdiabatic-Driving}.}
\section*{Code availability}

The code to perform the numerical simulations and the specific driving protocols is available via github at 

\noindent
\parbox{\linewidth}{\url{https://github.com/Peyara/Evolution-Counterdiabatic-Driving}.}

\setcounter{section}{0}
\setcounter{figure}{0}
\setcounter{equation}{0}
\renewcommand{\theequation}{S\arabic{equation}}
\renewcommand{\thetable}{S\arabic{table}}
\renewcommand{\thefigure}{S\arabic{figure}}
\let\addcontentsline\oldaddcontentsline

\appendix

\newpage

{\raggedright\sffamily\bfseries\fontsize{20}{25}\selectfont Supplementary Information:\\ Controlling the speed and trajectory of evolution with counterdiabatic driving\par}

\tableofcontents

\section{Fokker-Planck analogue of quantum adiabatic theorem}

The Fokker-Planck dynamics of our model obeys a classical, stochastic analogue of the quantum adiabatic theorem~\cite{born1928adiabatic,griffiths2018introduction}.  As described in the main text, this means that if the system starts at $t=t_0$ in the equilibrium distribution $\rho(\bm{x};\lambda(0))$ corresponding to control parameter $\lambda(0)$, it will remain in the corresponding instantaneous equilibrium (IE) distribution  $\rho(\bm{x};\lambda(t))$ at all $t>t_0$ if $\lambda(t)$ is varied infinitesimally slowly.  Note that for the quantum case, if a system starts in {\it any} eigenstate of the Hamiltonian, it will remain in that eigenstate if the Hamiltonian is varied adiabatically, assuming the eigenvalues never become degenerate.  The classical version that we demonstrate here only applies to one eigenstate, the IE distribution, which corresponds to the quantum ground state.

\subsection{Fokker-Planck eigenfunction expansion, adjoint operator}
Before considering adiabatic driving, we start with some preliminaries for a Fokker-Planck (FP) system with a time-independent control parameter $\lambda(t) = \lambda$.  The FP operator for a given $\lambda$ is defined in Eq.~(5) of the Methods,
\begin{equation}\label{s1}
{\cal L}\big(\lambda\big) p(\bm{x},t) \equiv -\partial_i \left( v_i\big(\bm{x};\lambda\big) p(\bm{x},t) \right) + \partial_i \partial_j \big(D_{ij}(\bm{x}) p(\bm{x},t) \big).
\end{equation}
Throughout the Supplementary Information (SI) we will use the same Einstein summation notation that we described in the Methods, with repeated Roman indices summed from $1$ to $M-1$ and repeated Greek indices summed from $1$ to $M$.  The two exceptions for clarity will be: i) the eigenfunction indices $n$ and $m$ used in this section, where the summation convention will not apply and explicit sums will be always be indicated; ii) the final section describing the mapping between the agent-based model and the Fokker-Planck equation, where it will be more convenient to write out all sums explicitly.

The operator ${\cal L}\big(\lambda\big)$ has an associated set of eigenfunctions $\psi_n(\bm{x};\lambda)$ and eigenvalues $-\kappa_n(\lambda)$, satisfying~\cite{risken1996fokker}
\begin{equation}\label{s2}
{\cal L}(\lambda) \psi_n(\bm{x};\lambda) = - \kappa_n(\lambda) \psi_n(\bm{x};\lambda)
\end{equation}
for $n=0,1,2,\ldots$.  We assume an equilibrium distribution exists for every value of the control parameter $\lambda$, in which case we know one eigenvalue is zero.  By convention we choose this to be $n=0$, so $\kappa_0(\lambda) =0$, and the corresponding eigenfunction $\psi_0(\bm{x};\lambda) \equiv \rho(\bm{x};\lambda)$.  Eigenvalues for $n>0$ can be in general complex, but have positive real parts, $\operatorname{Re}(\kappa_n(\lambda)) > 0$, which guarantees that the system eventually equilibrates, as discussed below~\cite{risken1996fokker,ryter1987eigenfunctions}.

The Fokker-Planck equation for constant $\lambda$,
\begin{equation}\label{s3}
\partial_t p(\bm{x},t) = {\cal L}(\lambda)p(\bm{x},t)
\end{equation}
has a solution that can be expressed as a linear combination of the eigenfunctions,
\begin{equation}\label{s4}
p(\bm{x},t) = \sum_{n=0}^\infty c_n \psi_n(\bm{x};\lambda) e^{-\kappa_n(\lambda) t},
\end{equation}
for some constants $c_n$ where $c_0 = 1$.  The fact that $\operatorname{Re}(\kappa_n(\lambda)) > 0$ for $n>0$ ensures eventual equilibration:  $p(\bm{x},t) \to \psi_0(\bm{x};\lambda) = \rho(\bm{x};\lambda)$ as $t\to \infty$.

It is convenient to introduce the adjoint ${\cal L}^\dagger\big(\lambda\big)$ of the FP operator~\cite{risken1996fokker,ryter1987eigenfunctions},
\begin{equation}\label{s5}
{\cal L}^\dagger\big(\lambda\big) p(\bm{x},t) \equiv v_i\big(\bm{x};\lambda\big) \partial_i  p(\bm{x},t) +D_{ij}(\bm{x})  \partial_i \partial_j p(\bm{x},t),
\end{equation}
with corresponding eigenfunctions $\xi_n(\bm{x};\lambda)$,
\begin{equation}\label{s6}
{\cal L}^\dagger(\lambda) \xi_n(\bm{x};\lambda) = - \kappa_n^\ast(\lambda) \xi_n(\bm{x};\lambda).
\end{equation}
where the asterisk denotes complex conjugation.  Let us define the scalar product of two functions $f(\bm{x};\lambda)$ and $h(\bm{x};\lambda)$ through
\begin{equation}\label{s7}
    \langle f,h\rangle_\lambda  \equiv \int_{\Delta} d\bm{x}\,f(\bm{x};\lambda) h^\ast(\bm{x};\lambda)
\end{equation}
where the integral is over the $M-1$ dimensional simplex $\Delta$ defined in the Methods, and the $\lambda$ subscript denotes the dependence on $\lambda$.  Then ${\cal L}^\dagger\big(\lambda\big)$ has the conventional property of an adjoint:
\begin{equation}\label{s8}
\langle {\cal L} f , h \rangle_\lambda = \langle f,{\cal L}^\dagger h\rangle_\lambda.
\end{equation}
A consequence of this property, using $f = \psi_n$ and $h=\xi_m$, is that the eigenfunctions can be chosen to ensure biorthonormality of the form
\begin{equation}\label{s9}
\langle \psi_n , \xi_m \rangle_\lambda = \delta_{nm},
\end{equation}
where $\delta_{nm}$ is the Kronecker delta.  By inspection of Eq.~\eqref{s5} one can see that the $n=0$ adjoint eigenfunction, with eigenvalue $\kappa_{0}^\ast(\lambda) = 0$, is $\xi_0(\bm{x};\lambda) = 1$.  The biorthonormality relation in this case,
\begin{equation}\label{s10}
    1 = \langle \psi_0,\xi_0 \rangle_\lambda = \int_\Delta d\bm{x}\, \rho(\bm{x};\lambda)
\end{equation}
corresponds to the normalization of the equilibrium distribution $\rho(\bm{x};\lambda)$.  We also know that for $n>0$,
\begin{equation}\label{s11}
    0 = \langle \psi_n,\xi_0 \rangle_\lambda = \int_\Delta d\bm{x}\, \psi_n(\bm{x};\lambda).
\end{equation}
This property ensures that $p(\bm{x},t)$ from Eq.~\eqref{s4}, with $c_0 = 1$, is also properly normalized, $\int_\Delta d\bm{x}\,p(\bm{x},t) =1$.

\subsection{Fokker-Planck adiabatic driving}

We now allow the control parameter $\lambda(t)$ to vary with time.  At any given time $t$, the definitions of the previous section generalize to give instantaneous operators ${\cal L}(\lambda(t))$, ${\cal L}^\dagger(\lambda(t))$, and corresponding instantaneous eigenfunctions/eigenvalues $\psi_n(\bm{x};\lambda(t))$, $\xi_n(\bm{x};\lambda(t))$ and $\kappa_n(\lambda(t))$.  The dynamics of the system are now described by the FP equation
\begin{equation}\label{s12}
\partial_t p(\bm{x},t) = {\cal L}(\lambda(t)) p(\bm{x},t).
\end{equation}
Working by analogy with the standard proof of the quantum adiabatic theorem~\cite{griffiths2018introduction}, let us posit a solution to this equation of the form
\begin{equation}\label{s13}
p(\bm{x},t) = \sum_{n=0}^\infty c_n(t) \psi_n(\bm{x};\lambda(t)) e^{-\Theta_n(t)},
\end{equation}
where $c_n(t)$ are some functions to be determined and $\Theta_n(t) \equiv \int_{t_0}^t dt^\prime\,\kappa_n(\lambda(t^\prime))$.  Plugging Eq.~\eqref{s13} into Eq.~\eqref{s12}, and using the fact that ${\cal L}(\lambda(t)) \psi_n(\bm{x};\lambda(t)) = -\kappa_n(\lambda(t))\psi_n(\bm{x};\lambda(t))$, we see the $p(\bm{x},t)$ form satisfies the FP equation assuming the following relation is true:
\begin{equation}\label{s14}
    \sum_{n=0}^\infty \bdot{c}_n(t) \psi_n(\bm{x};\lambda(t)) e^{-\Theta_n(t)} = -\sum_{n=0}^\infty c_n(t) \bdot\lambda(t) \partial_\lambda \psi_n(\bm{x};\lambda(t))  e^{-\Theta_n(t)},
\end{equation}
where $\bdot{f}(t) \equiv df/dt$ for a function $f(t)$.  Taking the scalar product of both sides with respect to $\xi_m(\bm{x};\lambda(t))$, and using the biorthonormality relations we find a set of coupled differential equations for $m=0,1,2,\ldots$ that can in principle be used to solve for the functions $c_n(t)$ if we knew the eigenfunctions/eigenvalues:
\begin{equation}\label{s15}
    \bdot{c}_m(t) = -\sum_{n=0}^\infty c_n(t) \bdot\lambda(t) \langle \partial_\lambda \psi_n, \xi_m\rangle_{\lambda(t)}  e^{-(\Theta_n(t)-\Theta_m(t))}.
\end{equation}
For $m=0$ the relations in Eqs.~\eqref{s10}-\eqref{s11} allow us to simplify the above to $\bdot{c}_0(t) = 0$, which means $c_0(t)$ is time-independent (and has to be equal to 1 to ensure normalization).  Thus we can write Eq.~\eqref{s13} as
\begin{equation}\label{s16}
p(\bm{x},t) = \rho(\bm{x};\lambda(t)) + \sum_{n=1}^\infty c_n(t) \psi_n(\bm{x};\lambda(t)) e^{-\Theta_n(t)}.
\end{equation}

Let us imagine that we start at $t_0$ in equilibrium, so $c_n(0) = 0$ for $n>0$.  For a general control protocol $\lambda(t)$, Eq.~\eqref{s15} implies that $c_n(t)$, $n>0$, would not necessarily stay zero at later times $t> t_0$.  Hence $p(\bm{x},t)$ in Eq.~\eqref{s16} would gain contributions from higher eigenfunctions in the second term, and no longer remain in the IE distribution.  This is the classical analogue of the observation that for a general time-dependent Hamiltonian driven at a finite rate, a quantum system that started in a ground state will evolve into a superposition of instantaneous ground and excited states.

However if the driving was infinitesimally slow, $\bdot\lambda(t) \to 0$, then the right-hand side of Eq.~\eqref{s15} becomes negligible, and hence $c_n(t) \approx 0$ for $n>0$.  Thus for adiabatically slow driving, $p(x,t) \approx \rho(\bm{x};\lambda(t))$ at all times $t>t_0$.  The system remains in the IE distribution, just like the corresponding quantum system remains in the instantaneous ground state.

\section{Derivation of $v_i$ and $D_{ij}$ for Fokker-Planck Wright-Fisher model with mutation and selection}

To derive the expressions for $v_i$ and $D_{ij}$ in Eqs.~(6)-(7) of the Methods, we start with haploid WF evolutionary dynamics defined as follows:  each time step corresponds to a generation, and every new generation is created by each child randomly ``choosing'' a parent in the previous generation and copying the parental genotype.  Let $N$ be the total population, assumed fixed between generations (i.e. at carrying capacity).  Let $\phi_i(\bm{x};\lambda(t))$ be the probability of choosing a parent of genotype $i$, which may depend on the control parameter $\lambda(t)$ through its influence on selection coefficients.  In the absence of mutation and selection, $\phi_i(\bm{x};\lambda(t)) = x_i$, the fraction of that genotype in the parental generation.  However we will keep $\phi_i$ general, in order to incorporate mutation / selection effects later on.  The probability of the new generation having a set of genotype populations $\bm{n} \equiv (n_1,n_2,\ldots,n_{M-1})$, where $n_i$ is the number of type $i$ individuals, is given by the multinomial distribution,
\begin{equation}\label{w1}
{\cal P}(\bm{n};\bm{x},\lambda(t))= \frac{N!}{n_1!n_2! \cdots n_{M-1}!n_M!} \phi_1^{n_1}(\bm{x};\lambda(t)) \phi_2^{n_2}(\bm{x};\lambda(t)) \cdots \phi_{M-1}^{n_{M-1}}(\bm{x};\lambda(t))\phi_M^{n_M}(\bm{x};\lambda(t))
\end{equation}
where $n_M = N - \sum_{j=1}^{M-1} n_j$ is the number of type $M$ individuals.  Note that Eq.~\eqref{w1} is defined for all allowable configurations of types, or in other words for any $\bm{n}$ where $N - \sum_{j=1}^{M-1} n_j \ge 0$.  If we denote $\sum_{\bm{n}}$ as a sum over these allowable configurations, then $\sum_{\bm{n}} {\cal P}(\bm{n};\bm{x};\lambda(t)) =1$.  The genotype fraction $x_i^\prime$ in the new generation is just $x_i^\prime = n_i/N$, and hence the mean difference $v_i(\bm{x};\lambda(t))$ in genotype fractions per generation can be expressed as
\begin{equation}\label{w2}
v_i(\bm{x};\lambda(t)) = \langle \delta x_i \rangle = \langle x_i^\prime - x_i\rangle = \sum_{\bm{n}} \left( \frac{n_i}{N} - x_i\right){\cal P}(\bm{n};\bm{x},\lambda(t)) = \phi_i(\bm{x};\lambda(t)) - x_i.
\end{equation}
Similarly $D_{ij}(\bm{x};\lambda(t))$ can be calculated through the covariance as:
\begin{equation}\label{w3}
\begin{split}
2 D_{ij}(\bm{x};\lambda(t)) &= \langle \delta x_i \delta x_j \rangle- \langle \delta x_i \rangle \langle \delta x_j \rangle\\ &= \left \langle \left(\frac{n_i}{N}-x_i \right)\left(\frac{n_j}{N}-x_j \right) \right \rangle
- \left(\phi_i(\bm{x};\lambda(t)) - x_i\right)\left(\phi_j(\bm{x};\lambda(t))- x_j\right)\\
&= \frac{g_{ij}(\bm{\phi}(\bm{x};\lambda(t)))}{N},
\end{split}
\end{equation}
where $g_{ij}(\bm{\phi}(\bm{x};\lambda(t)))$ is given by Methods Eq.~(7) with $\bm{\phi}(\bm{x};\lambda(t)) = (\phi_1(\bm{x};\lambda(t)),\ldots,\phi_{M-1}(\bm{x};\lambda(t)))$ replacing $\bm{x}$.  In order to derive Eqs.~\eqref{w2}-\eqref{w3}, we have used the mean and covariance properties of the multinomial distribution of Eq.~\eqref{w1}:
\begin{equation}\label{w3b}
\begin{split}
    \langle n_i \rangle &= N \phi_i(\bm{x};\lambda(t)),\\
    \langle n_i n_j \rangle - \langle n_i \rangle \langle n_j \rangle &= \begin{cases} N \phi_i(\bm{x};\lambda(t)) (1-\phi_i(\bm{x};\lambda(t))) & j = i\\
    -N \phi_i(\bm{x};\lambda(t)) \phi_j(\bm{x};\lambda(t)) & j \ne i
    \end{cases}.
\end{split}
\end{equation}

To complete the derivation, we need an expression for $\phi_i(\bm{x};\lambda(t))$ when mutation and selection are included in the model.  Consider first the probability $\phi^0_i(\bm{x})$ of picking a parent of type $i$, assuming only mutation was allowed, but no selection.  Accounting for the gain and loss of possible type $i$ parents through mutation, we have $\phi^0_i(\bm{x}) = x_i + m_{i\mu} x_{\mu}$, for $1\le i \le M-1$. We can express the probability to choose a parent of type $M$ as $\phi_M^0(\bm{x}) = 1-\sum_{j=1}^{M-1} \phi_j^0(\bm{x})$ by normalization.  To include selection and obtain the final expressions for $\phi_i(\bm{x};\lambda(t))$, we note that we can write
\begin{equation}\label{w4}
\frac{\phi_i(\bm{x};\lambda(t))}{\phi_M(\bm{x};\lambda(t))} = (1+s_i(\lambda(t)))\frac{\phi^0_i(\bm{x})}{\phi^0_M(\bm{x})} \qquad \text{[no summation over $i$],}
\end{equation}
Eq.~\eqref{w4}, for $i=1,\ldots,M-1$, states that the ratio of $\phi_i(\bm{x};\lambda(t))$, the chance of picking a type $i$ parent, to $\phi_M(\bm{x};\lambda(t))$, the chance of picking a wild type parent, is modified by a factor $1+s_i(t)$ due to selection, compared to the case without selection.  In other words, if $s_i(t)$ is positive because type $i$ has a greater fitness than the wild type, the chance of getting a type of $i$ parent relative to a wild type parent increases by $1+s_i(t)$.  Recalling the definition of $\phi_M(\bm{x};\lambda(t))$ in terms of $\bm{\phi}(\bm{x};\lambda(t))$'s components, we can solve the systems of equations in Eq.~\eqref{w4} to find
\begin{equation}\label{w5}
    \phi_i(\bm{x};\lambda(t)) = \frac{(1+s_i(\lambda(t)))\phi^0_i(\bm{x})}{1+ s_j(\lambda(t))\phi^0_j(\bm{x})} \quad \text{[no summation over $i$]}.
\end{equation}
Let us substitute in the expression for $\phi_i^0(\bm{x})$, and make the typical assumption that $|s_i(\lambda(t))|$, $|m_{ij}| \ll 1$.  After Taylor expanding to first order in these quantities, Eq.~\eqref{w5} becomes
\begin{equation}\label{w6}
    \phi_i(\bm{x};\lambda(t)) \approx x_i + m_{i\mu}x_{\mu} +g_{ij}(\bm{x})s_{j}(\lambda(t)).
\end{equation}
Plugging this into Eq.~\eqref{w2} gives the expression in Methods Eq.~(6) for $v_i$.  Similarly, if we plug Eq.~\eqref{w6} into Eq.~\eqref{w3}, and keep only the leading order contribution, we get $D_{ij}(\bm{x}) \approx g_{ij}(\bm{x})/(2N)$, which is the expression in Methods Eq.~(6) for $D_{ij}$.

\section{Estimating the mean and covariance of the instantaneous equilibrium distribution}

As discussed in the Methods, for general $M$ we do not have an analytical expression for IE distribution.  However in the large population, frequent mutation regime we can use the multivariate normal approximation of Methods Eq.~(10).  This requires us to be able to calculate the mean genotype frequencies $\overline{x}_i(\lambda)$ and covariance matrix $\Sigma(\lambda)$ associated with the IE distribution $\rho(\bm{x};\lambda)$ at a given value of the control parameter $\lambda$.  In this section we outline how to do this calculation.  The first step is deriving a set of coupled equations for the first and second moments of the IE distribution (part of a larger hierarchy of moment equations).  The second step will be to approximately solve these equations using a moment closure technique.  We consider each step in turn.

\subsection{Deriving equations for the first and second moments of the IE distribution}

In order to characterize the moments of the IE distribution $\rho(\bm{x};\lambda)$, let us consider an auxiliary problem:  imagine a system described by a genotype frequency probability $\tilde{p}(\bm{x},\tau)$ evolving under a Fokker-Planck equation with constant $\lambda$:
\begin{equation}\label{m1}
\begin{split}
    \partial_\tau \tilde{p}(\bm{x},\tau) &= {\cal L}(\lambda) \tilde{p}(\bm{x},\tau)\\
    &= -\partial_i \left[ v_i(\bm{x};\lambda) \tilde{p}(\bm{x},\tau) + \partial_j \big(D_{ij}(\bm{x}) \tilde{p}(\bm{x},\tau) \big)\right]\\
    &\equiv - \partial_i J_i(\bm{x},\tau),
\end{split}
\end{equation}
where for later convenience we have introduced the probability current $J_i(\bm{x},\tau)$.  We know that in the limit $\tau \to \infty$ the system equilibrates, $\tilde{p}(\bm{x},\tau) \to \rho(\bm{x};\lambda)$.  We use the auxiliary time $\tau$ here for clarity, since it is distinct from the actual time $t$ in the original system.  Let us define the first and second moments of the genotype frequencies with respect to $\tilde{p}(\bm{x},\tau)$:
\begin{equation}\label{m2}
    \begin{split}
    \langle x_i \rangle_\tau &\equiv \int_{\Delta}d\bm{x}\,x_i \tilde{p}(\bm{x},\tau), \qquad
    \langle x_i x_j \rangle_\tau &\equiv \int_{\Delta}d\bm{x}\,x_i x_j \tilde{p}(\bm{x},\tau),\\
    \end{split}
\end{equation}
where the $\tau$ subscript denotes the dependence of the moments on $\tau$.  Since $\tilde{p}(\bm{x},\tau) \to \rho(\bm{x};\lambda)$ as $\tau \to\infty$, the IE distribution quantities we are interested in are just the following limiting values:
\begin{equation}\label{m3}
    \overline{x}_i(\lambda) = \lim_{\tau\to\infty} \langle x_i \rangle_\tau, \qquad     \Sigma_{ij}(\lambda) = \lim_{\tau\to\infty} \left(\langle x_i x_j \rangle_\tau - \langle x_i \rangle_\tau \langle x_j \rangle_\tau\right).
\end{equation}

We will derive the following exact moment relationships:
\begin{eqnarray}\label{mm0}
0 &=& m_{i\mu} \overline{x}_\mu(\lambda) + \big[\overline{x}_i(\lambda)\delta_{ik} - \Sigma_{ik}(\lambda) -\overline{x}_i(\lambda)\overline{x}_k(\lambda)\big]s_k(\lambda),\\
0 &=& m_{i\mu} \Sigma_{j\mu}(\lambda) + m_{j\mu} \Sigma_{i\mu}(\lambda) + \Sigma_{ij}(\lambda)\big(s_i(\lambda)+s_j(\lambda)\big)\label{mm5}\\
&\quad& -\big[2T_{ijk}(\lambda)-\overline{x}_j(\lambda) \Sigma_{ik}(\lambda)-\overline{x}_i(\lambda) \Sigma_{jk}(\lambda) -2 \overline{x}_i(\lambda)\overline{x}_j(\lambda)\overline{x}_k(\lambda) \big]s_k(\lambda)\nonumber\\ &\quad& +\frac{1}{N}\big[\overline{x}_i(\lambda)\delta_{ij} - \Sigma_{ij}(\lambda) -\overline{x}_i(\lambda)\overline{x}_j(\lambda)\big],\nonumber
\end{eqnarray}
where $i,\,j=1,\ldots,M-1$ (the indices $i$, $j$ are not summed over).  Here $T_{ijk}(\lambda) \equiv \lim_{\tau \to \infty} \langle x_i x_j x_k\rangle_\tau$ is a third moment of the IE distribution.  The derivation of these equations is shown below.  For readers not interested in the details, they can skip ahead to Sec.~\ref{mom:app}.

To find Eq.~\eqref{mm0}, let us start with the first moment $\langle x_i\rangle_\tau$ from Eq.~\eqref{m2}, take the derivative with respect to $\tau$, and plug in the Fokker-Planck equation from Eq.~\eqref{m1}:
\begin{equation}\label{h1}
\frac{d}{d\tau}\langle x_i \rangle_\tau = \int_\Delta d
\bm{x}\, x_i \partial_{\tau} \tilde{p}(\bm{x},\tau) = - \int_\Delta d
\bm{x}\, x_i \partial_j J_j(\bm{x},\tau),
\end{equation}
Notice that $x_i \partial_j J_j = \partial_j(x_iJ_j) - \delta_{ij}J_j$. By Gauss's theorem, the integral over the first term is $\int_\Delta d\bm{x}\,\partial_j(x_i J_j) = \int_{\partial \Delta} d\sigma\,x_i J_j n_j$.  The latter integral is expressed in terms of area elements $d\sigma$ of the simplex boundary $\partial \Delta$, and $n_j$  is the $j$th component of the normal vector to this boundary. Since probability is conserved within the simplex, $J_j n_j = 0$, and hence the integral vanishes. Thus only the integral over the second term contributes, and Eq.~\eqref{h1} can be rewritten:
\begin{equation}\label{b1}
\frac{d}{d\tau}\langle x_i \rangle_\tau = \int d
\bm{x}\, J_i(\bm{x},\tau) = \int d
\bm{x}\, \big[ v_i(\bm{x};\lambda) \tilde{p}(\bm{x},\tau) - \partial_j\left(D_{ij}(\bm{x})\tilde{p}(\bm{x},\tau) \right)\big].
\end{equation}
Focusing on the second term of the integrand in Eq.~\eqref{b1}, we can rewrite this term using Gauss's law as:
\begin{equation}\label{b2}
-\int_{\Delta} d\bm{x}\,\partial_j\left(D_{ij}(\bm{x})\tilde{p}(\bm{x},\tau)\right) = -\int_{\partial \Delta} d\sigma\, \tilde{p}(\bm{x},\tau) D_{ij}(\bm{x}) n_j = -\frac{1}{2N}\int_{\partial \Delta} d\sigma\, \tilde{p}(\bm{x},\tau) g_{ij}(\bm{x}) n_j.
\end{equation}
The term $g_{ij}(\bm{x})n_j =0$ for $\bm{x} \in \partial \Delta$, which makes the integral vanish.  We prove this in the following lemma.

\begin{enumerate}
    \item[] {\bf Lemma:} For $g_{ij}(\bm{x})$ as defined in Methods Eq.~(7) and for any normal vector $\bm{n}$ to the simplex boundary $\partial \Delta$, $g_{ij}(\bm{x})n_j = 0$ for all $\bm{x} \in \partial V$. .

    \item[] {\it Proof:} A simplex is defined by $\Delta = \{ \bm{x} \,|\, x_i \geq 0, \, x_ie_i \leqslant 1\}$, where $\bm{e}$ is an $M-1$ dimensional vector with all components equal to 1. There are two classes of hypersurfaces on the simplex boundary (where $x_i e_i = 1$) to consider:

{\bf Case 1}: One type of hypersurface on the simplex boundary is $S_k$, defined by the conditions $x_k = 0$ and $x_j e_j = 1$.  There are $M-1$ such hypersurfaces, one for each $k=1,\ldots,M-1$.  The components of the normal vector $\bm{n}_k$ to $S_k$ are given by $n_{kj} = -\delta_{kj}$ (the minus sign ensures $\bm{n}_k$ faces away from the simplex volume). Then, for all $\bm{x} \in S_k$, $g_{ij}(\bm{x})n_{kj} = -g_{ij}(\bm{x})\delta_{kj} = -g_{ik}(\bm{x})$, and we know from the definition of $g$ that $g_{ik}(\bm{x}) \propto x_k = 0$.  The last result follows because $x_k=0$ for $\bm{x} \in S_k$.

{\bf Case 2}: The only other type of hypersurface on the simplex boundary is $S^\prime$, defined by the conditions $x_i > 0$ for all $i$, and $x_i e_i = 1$.  We find the normal to this surface as follows. Define $F(\bm{x}) = x_i e_i - 1$.  Then $\bm{n} = \nabla F = \bm{e}$.  For all $\bm{x}\in S^\prime$, we have $g_{ij}(\bm{x})n_ j =  g_{ij}(\bm{x})e_j = x_i (1- x_j e_j)$.  Since $x_j e_j = 1$ for all $\bm{x} \in S^\prime$, we see that $g_{ij}(\bm{x})n_ j = 0$.
\end{enumerate}

Using the lemma, and the definition of $v_i(\bm{x};\lambda)$ from Methods Eq.~(6), we can rewrite Eq.~\eqref{b1} as:
\begin{equation}\label{m4}
\begin{split}
\frac{d}{d\tau}\langle x_i \rangle_\tau = \int d
\bm{x}\, v_i(\bm{x};\lambda) \tilde{p}(\bm{x},\tau) &= \langle v_i(\bm{x};\lambda) \rangle_\tau\\
&= m_{i\mu}\langle x_\mu \rangle_\tau + \langle g_{ij}(\bm{x}) \rangle_\tau s_j(\lambda).
\end{split}
\end{equation}
Taking the $\tau \to \infty$ limit on sides of the equation, we note that the left-hand side vanishes because $\langle x_i \rangle_\tau \to \overline{x}_i(\lambda)$, a constant independent of $\tau$.  On the right-hand side we can subsitute in Eq.~\eqref{m2} and use the definition of $g$ in Methods Eq.~(7).  The end result is Eq.~\eqref{mm0}.  This is the first moment equation we will be interested in.  

To derive Eq.~\eqref{mm5}, we start analogously to Eq.~\eqref{h1}, but now with the second moment:
\begin{equation}\label{mm1}
\frac{d}{d\tau}\langle x_i x_j \rangle_\tau = \int_\Delta d
\bm{x}\, x_i x_j \partial_{\tau} \tilde{p}(\bm{x},\tau) = - \int_\Delta d
\bm{x}\, x_i x_j \partial_k J_k(\bm{x},\tau).
\end{equation}
Notice, by the product rule, that $x_i x_j \partial_k J_k = \partial_k(x_i x_j J_k) - J_k \partial_k(x_i x_j) = \partial_k(x_i x_j J_k) - J_i x_j - J_j x_i$. By Gauss' law, we have $\int_\Delta d\bm{x}\,\partial_k(x_i x_j J_k) = \int_{\partial \Delta} d\sigma\,x_i x_j J_k n_k$. Since probability is conserved, $J_k n_k = 0$, and we can thus rewrite Eq.~\eqref{mm1} as:
\begin{equation}\label{mm2}
\begin{split}
\frac{d}{d\tau}\langle x_i x_j \rangle_\tau &= \int d
\bm{x} \left(J_i x_j + J_j x_i \right) \\ &= \int d
\bm{x} \Bigl[\left(v_i(\bm{x};\lambda) \tilde{p}(\bm{x},\tau) - \partial_k\big(D_{ik}(\bm{x})\tilde{p}(\bm{x},\tau) \big)\right)x_j + \left(v_j(\bm{x};\lambda) \tilde{p}(\bm{x},\tau) - \partial_k\big(D_{jk}(\bm{x})\tilde{p}(\bm{x};\tau) \big)\right)x_i\Bigr].
\end{split}
\end{equation}
Note that $\partial_k(D_{ik}\tilde{p})x_j = \partial_k(D_{ik}\tilde{p}x_j) - D_{ik}\tilde{p}\partial_k x_j = \partial_k(D_{ik}\tilde{p}x_j) - D_{ik}\tilde{p}\delta_{jk} = \partial_k(D_{ik}\tilde{p}x_j) - D_{ij}\tilde{p}$.  By Gauss's theorem,
\begin{equation}\label{mm3}
-\int_{\Delta}d\bm{x}\, \partial_k\left(D_{ik}(\bm{x})\tilde{p}(\bm{x},\tau)x_j\right) = -\int_{\partial \Delta} d\sigma\, \tilde{p}(\bm{x},\tau) x_j D_{ik}(\bm{x})n_k = -\frac{1}{2N}\int_{\partial \Delta} d\sigma\, \tilde{p}(\bm{x},\tau) x_j g_{ik}(\bm{x})n_k.
\end{equation}
By the lemma, $g_{ik}(\bm{x})n_k = 0$ for $\bm{x} \in \partial \Delta$, so the integral vanishes.  Using this and the fact that $D_{ij} = D_{ji}$, Eq.~\eqref{mm2} becomes
\begin{equation}\label{mm4}
\begin{split}
   \frac{d}{d\tau}\langle x_i x_j \rangle_\tau &= \int_\Delta d
\bm{x}\, \bigl[x_j v_i(\bm{x};\lambda) \tilde{p}(\bm{x},\tau)  + x_i v_j(\bm{x};\lambda) \tilde{p}(\bm{x},\tau) + 2 D_{ij}(\bm{x})\tilde{p}(\bm{x},\tau) \bigr] \\ &=  \langle x_j v_i(\bm{x};\lambda) \rangle_\tau + \langle x_i v_j(\bm{x};\lambda) \rangle_\tau + 2\langle D_{ij}(\bm{x}) \rangle_\tau.
\end{split}
\end{equation}
In the $\tau \to \infty$ limit this equation can be written in the form of Eq.~\eqref{mm5}.

\subsection{Approximate solution of moment equations}\label{mom:app}

Eqs.~\eqref{mm0} and \eqref{mm5} constitute the first two of a hierarchy of coupled moment equations, with each set of equations involving moments of one higher order (i.e. Eq.~\eqref{mm0} involves $\Sigma_{ij}$, Eq.~\eqref{mm5} involves $T_{ijk}$).  In the limit $N \to \infty$ Eq.~\eqref{mm5} for the second moments can be trivially satisfied, because the IE distribution becomes a delta function with zero spread.  In this case $\Sigma_{ij}(\lambda) = 0$, $T_{ijk}(\lambda) = \overline{x}_i(\lambda)\overline{x}_j(\lambda)\overline{x}_k(\lambda)$ is a solution to Eq.~\eqref{mm5}. For large but finite $N$ (assuming $\mu N \gg 1$ is still satisfied, where $\mu$ is the order of magnitude of the nonzero mutation rates) the distribution becomes spread out by a small amount, and we will approximate it by a multivariate Gaussian.  As a result we assume third and higher order cumulants are negligible, which will allow us to approximately solve Eqs.~\eqref{mm0} and \eqref{mm5} for $\overline{x}_i(\lambda)$ and $\Sigma_{ij}(\lambda)$.  This approach, known as moment closure, effectively truncates the moment hierarchy after the second order.  It is justified by the fact that while $\Sigma_{ij}(\lambda)$ scales like $N^{-1}$, the third cumulant scales like $N^{-2}$, etc., which allows the higher order cumulants to be neglected for large $N$.  Letting the third cumulant be zero means the third moment can be approximated as follows:
\begin{equation}\label{a1}
T_{ijk}(\lambda) \approx \overline{x}_i(\lambda) \Sigma_{jk}(\lambda) + \overline{x}_j(\lambda) \Sigma_{ik}(\lambda) + \overline{x}_k(\lambda) \Sigma_{ij}(\lambda) + \overline{x}_i(\lambda)\overline{x}_k(\lambda)\overline{x}_k(\lambda).
\end{equation}
Plugging this into Eq.~\eqref{mm5} gives:
\begin{equation}\label{a2}
\begin{split}
0 &\approx m_{i\mu} \Sigma_{j\mu}(\lambda) + m_{j\mu} \Sigma_{i\mu}(\lambda) + \Sigma_{ij}(\lambda)\big(s_i(\lambda)+s_j(\lambda)\big)\\
&\quad -\big[2\overline{x}_k(\lambda) \Sigma_{ij}(\lambda)+\overline{x}_i(\lambda) \Sigma_{jk}(\lambda) +\overline{x}_j(\lambda) \Sigma_{ik}(\lambda) \big]s_k(\lambda)\\ &\quad +\frac{1}{N}\big[\overline{x}_i(\lambda)\delta_{ij} - \Sigma_{ij}(\lambda) -\overline{x}_i(\lambda)\overline{x}_j(\lambda)\big].
\end{split}
\end{equation}
For Eq.~\eqref{mm0}, since the term $\Sigma_{ik}(\lambda) \sim {\cal O}(N^{-1})$ becomes negligible relative to the other terms for large $N$, we can approximate the equation as
\begin{equation}\label{a3}
0 \approx m_{i\mu} \overline{x}_\mu(\lambda) + \big[\overline{x}_i(\lambda)\delta_{ik} -\overline{x}_i(\lambda)\overline{x}_k(\lambda)\big]s_k(\lambda).
\end{equation}
The following procedure can then be used to solve Eqs.~\eqref{a2}-\eqref{a3}:  i) numerically solve the nonlinear set of equations in Eq.~\eqref{a3} for $\overline{x}_i(\lambda)$, $i=1,\ldots,M-1$.  ii) Plug this solution into Eq.~\eqref{a2}, and then numerically solve the resulting set of linear equations for $\Sigma_{ij}(\lambda)$, $i,\,j=1,\ldots,M-1$.  Because the $\Sigma$ indices in Eq.~\eqref{a2} run up to $M$, we use the following identities to express those elements in terms of lower indices:  $\Sigma_{MM} = \Sigma_{k\ell}e_k e_\ell$, $\Sigma_{iM} = -\Sigma_{ik}e_k$.  Both of these identities follow from the fact that $x_M = 1 - x_k e_k$.

In principle we can iterate this procedure to progressively add small corrections to the solution, converging to self-consistency between Eq.~\eqref{mm0} and Eq.~\eqref{a2}:  plug the $\Sigma_{ij}(\lambda)$ values obtained from the first iteration into Eq.~\eqref{mm0}, solve for the updated $\overline{x}_i(\lambda)$, plug these into Eq.~\eqref{a2}, and so on.  However for all the cases we examined in the main text the corrections resulting from multiple iterations are negligible, so we use one iteration only.

\section{Approximating the counterdiabatic driving protocol in the large population, frequent mutation regime}

As derived in Methods Sec. 4, the selection coefficient perturbation $\delta \tilde{\bm{s}}\big(\bm{x};\lambda(t),\bdot\lambda(t)) = \tilde{\bm{s}}\big(\bm{x};\lambda(t),\bdot\lambda(t))- \bm{s}(\lambda(t))$ needed to implement the CD protocol satisfies Methods Eq.~(13):
\begin{equation}\label{ap1}
\partial_t \rho\big(\bm{x};\lambda(t)\big) =- \partial_i \left(\rho\big(\bm{x};\lambda(t)\big) g_{ij}(\bm{x}) \delta \tilde{s}_j\big(\bm{x};\lambda(t),\bdot\lambda(t)\big) \right),
\end{equation}
which in turn implies the relation in Methods Eq.~(17),
\begin{equation}\label{ap2}
\partial_t \overline{\bm{x}}(\lambda(t)) = \left\langle \bm{g}(\bm{x}) \delta \tilde{\bm{s}}\big(\bm{x};\lambda(t),\bdot{\lambda}(t)\big)\right\rangle.
\end{equation}
Here the brackets $\langle \: \rangle$ denote an average over the simplex with respect to the IE distribution ${\rho}(\bm{x};\lambda(t))$.  Let us define a function $\bm{F}(\bm{x}) \equiv \bm{g}(\bm{x}) \delta \tilde{\bm{s}}\big(\bm{x};\lambda(t),\bdot{\lambda}(t)\big)$.  For simplicity of notation we do not explicitly show the $\lambda(t)$, $\bdot{\lambda}(t)$ dependence in $\bm{F}(\bm{x})$.  We can then Taylor expand the right-hand side of Eq.~\eqref{ap2} around $\overline{\bm{x}}(\lambda(t))$ up to second order.  In component form, this looks like
\begin{equation}\label{ap3}
\begin{split}
\partial_t \overline{x}_i(\lambda(t)) &= \left\langle F_i(\bm{x})\right\rangle\\
&=  F_i\big(\overline{\bm{x}}(\lambda(t))\big) + \partial_j F_i\big(\overline{\bm{x}}(\lambda(t))\big) \left\langle x_j - \overline{x}_j(\lambda(t))\right\rangle + \frac{1}{2} \partial_j\partial_k F_i\big(\overline{\bm{x}}(\lambda(t))\big) \left\langle \big(x_j - \overline{x}_j(\lambda(t))\big)\big(x_k - \overline{x}_k(\lambda(t))\big)\right\rangle + \cdots\\
&=  F_i\big(\overline{\bm{x}}(\lambda(t))\big) + \frac{1}{2} \partial_j\partial_k F_i\big(\overline{\bm{x}}(\lambda(t))\big) \Sigma_{jk}(\lambda(t)) + \cdots
\end{split}
\end{equation}
In the last line we have used the definition of the IE covariance matrix $\Sigma_{jk}(\lambda(t))$, and the fact that $\left\langle x_j - \overline{x}_j(\lambda(t))\right\rangle = 0$ since $\overline{x}_j(\lambda(t))$ is the mean of the IE distribution.  From the discussion in the previous section we know that $\Sigma_{jk}(\lambda(t))$ scales like $N^{-1}$ when $N$ is large (and $\mu N \gg 1$).  This means that the $\Sigma$ term in Eq.~\eqref{ap3} becomes small compared to the leading term for large $N$.  Keeping only the leading term, and substituting in the definition of $\bm{F}(\bm{x})$, we find the approximate relation
\begin{equation}\label{ap4}
\partial_t \overline{\bm{x}}(\lambda(t)) \approx  \bm{g}\big(\overline{\bm{x}}(\lambda(t))\big) \delta \tilde{\bm{s}}\big(\lambda(t),\bdot{\lambda}(t)\big),
\end{equation}
where $\delta \tilde{\bm{s}}\big(\lambda(t),\bdot{\lambda}(t)\big) \equiv \delta \tilde{\bm{s}}\big(\overline{\bm{x}}(\lambda(t));\lambda(t),\bdot{\lambda}(t)\big)$.  This is what is shown in Methods Eq.~(18).

\section{Counterdiabatic driving under time-varying total populations}

\begin{figure}[t!]
\centering\includegraphics[width=0.6\textwidth]{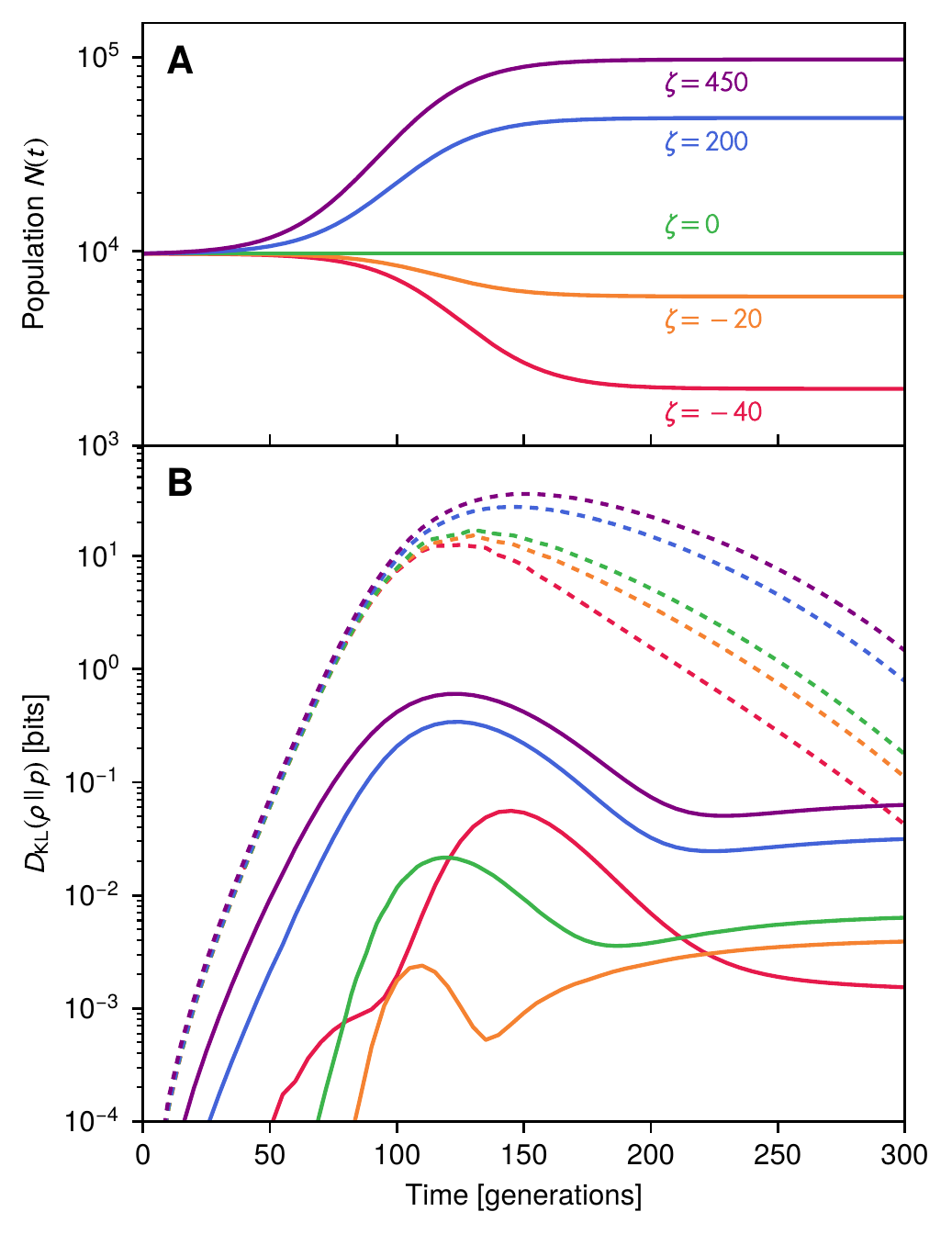}
\caption{\textbf{(A)} Five examples of time-varying total populations $N(t) = N_0(1+\zeta s_i(\lambda(t))$ for the two genotype system described in Methods Sec.~4.4.  Here $N_0 = 10^4$ and $\zeta = -40$, $-20$, $0$, $200$, and $450$. \textbf{(B)} For each case in panel \textbf{A}, the Kullback-Leibler divergence between the IE and actual genotype frequency distributions while driving the system according to the original protocol (dashed curves) or the approximate CD protocol (solid curves). Colors are matched to the $\zeta$ values indicated in panel \textbf{A}.  The KL divergences are calculated using numerical solutions to the associated Fokker-Planck equations.}\label{fig:varying_N}
\end{figure}

The derivation of the CD protocol discussed in Methods Sec.~4 and the previous section of the SI remains valid even when the total population $N(t)$ varies in time as a result of the control protocol, i.e. the carrying capacity of the system changes along with the control parameters.  In this case we can effectively absorb $N(t)$ into the set of control parameters $\lambda(t)$ during the derivation, and we end up with the same approximate CD protocol defined through Eq.~\eqref{ap4}, and whose explicit solution is shown in main text Eq.~(4).  This assumes the conditions for the validity of the approximation hold at all times of interest, $N(t) \gg 1$ and $\mu N(t) \gg 1$.  It is interesting to note that main text Eq.~(4) depends only on the selection coefficients under the protocol $s_i(\lambda(t))$ and the corresponding instantaneous equilibrium mean genotype frequencies $\overline{x}_i(\lambda(t))$.  As can be seen from Eq.~\eqref{a3}, to leading order for large $N(t)$ the means $\overline{x}_i(\lambda(t))$ are independent of $N(t)$.  If the selection coefficients are also independent of $N(t)$ then the entire CD driving protocol becomes (at least to leading order) independent of $N(t)$.  Thus the same CD protocol should work for a variety of $N(t)$ behaviors.

To illustrate this, we have redone the two genotype CD driving results from main text Fig.~2 using time-varying $N(t)$.  All parameters are as described in Methods Sec.~4.4, except that $N(t)$ varies with the control parameter according to: $N(t) = N_0(1+\zeta s_1(\lambda(t)))$, where $N_0 = 10^4$ and $\zeta$ is a constant.  The original two genotype results for constant $N$ are recovered when $\zeta = 0$.  Fig.~\ref{fig:varying_N}A shows five forms for $N(t)$ for different $\zeta$, and Fig.~\ref{fig:varying_N}B shows the corresponding driving results, calculated using numerical solutions of the Fokker-Planck dynamics (main text Eq.~(1)).  The dashed curves show the Kullback-Leibler divergence between the IE and actual genotype frequency distributions using the original protocol, and the solid curves using the approximate CD protocol (which is independent of $N(t)$).  In all cases CD driving dramatically reduces lag, improving the agreement between IE and actual distributions:  the KL divergences under CD remain below 1 bit throughout the whole protocol, in contrast to the original cases where the divergence peaks above 10 bits.

\section{Robustness of counterdiabatic driving to errors in the protocol}

\begin{figure}[t!]
\centering\includegraphics[width=0.6\textwidth]{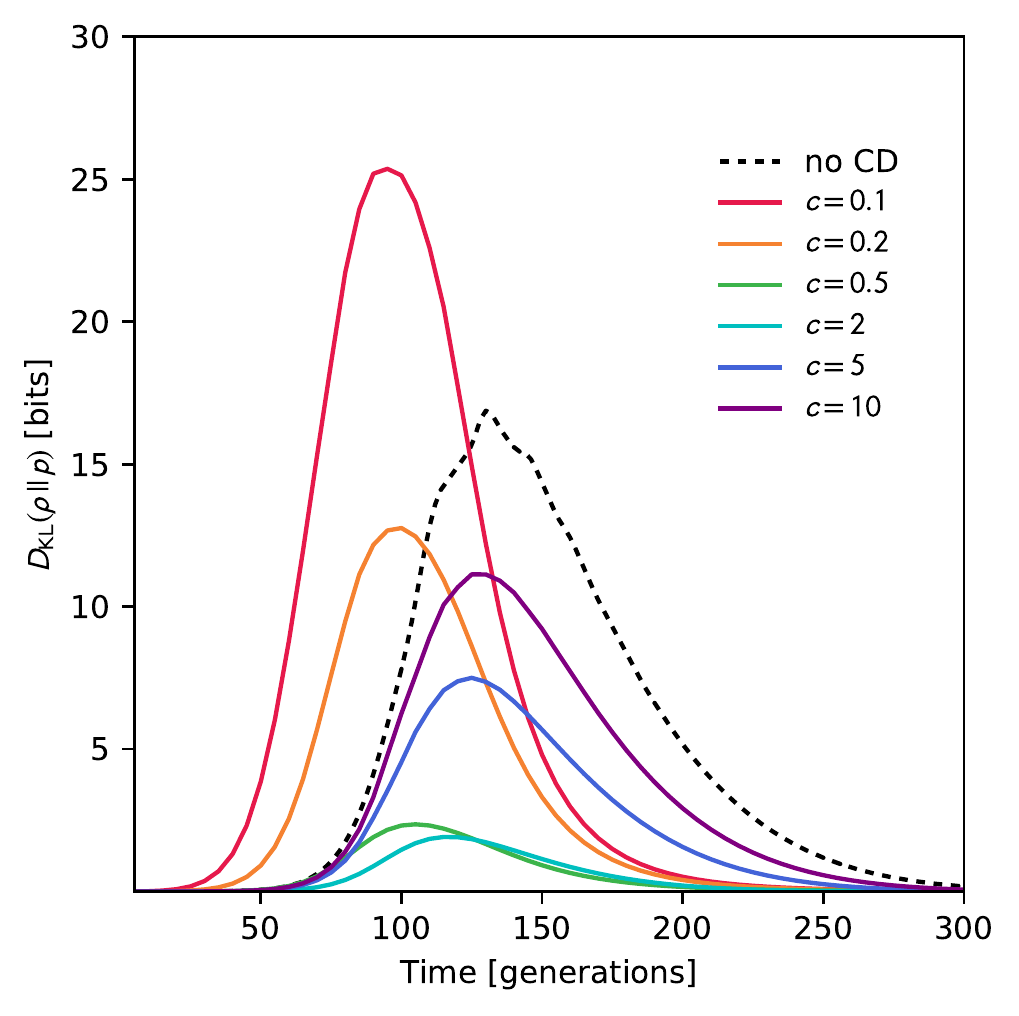}
\caption{The KL divergence between the actual and IE genotype frequency distributions for the $M=2$ system described in Methods Sec.~4.4, calculated using  numerical solutions to the Fokker-Planck equation.  The dashed curve shows the results of the original protocol, while the colored curves show deliberately inaccurate CD protocols:  the mutation rates that enter into the CD solution of Methods Eq.~(20) are scaled by a factor $c$ to mimic experimental measurement errors, where $c=1$ corresponds to the true CD protocol.  Six different inaccurate protocols are shown, with $c$ ranging from 0.1 to 10.}\label{fig:robustness}
\end{figure}

In the main text we explored one way in which counterdiabatic driving may still be effective even in scenarios where precise implementation of the protocol described by main text Eq.~(4) is hampered by practical constraints.  In the $M=16$ example direct realization of the full $\tilde{\bm{s}}(\lambda(t),\bdot\lambda(t))$ solution is constrained by the fact that there is one control variable (drug concentration) and this control variable may have a limited range (no larger than a certain maximum allowable dosage).  However we showed protocols that approximate the CD solution could still produce excellent results in terms of driving the system near the desired trajectory of genotype distributions over finite times.

Here we look at another potential hindrance:  what if there were errors in the quantities we use to calculate CD driving via main text Eq.~(4)?  The latter involves the mean IE genotypes frequencies $\overline{\bm{x}}(\lambda(t))$ and selection coefficients $\bm{s}(\lambda(t))$ in the original protocol as a function of the control parameter.  Using the moment relationship of Eq.~\eqref{a3}, $\overline{\bm{x}}(\lambda(t))$ can in turn be calculated with knowledge of the mutation rate matrix $m$ and $\bm{s}(\lambda(t))$.  We can imagine different potential sources of error:  i) One possibility is that our estimate for $\bm{s}(\lambda(t))$ of the original protocol contains some inaccuracies.  For example, in the case where the control parameter is a drug concentration, the genotype fitness versus drug dosage curve from earlier experiments might have measurement artifacts.  Assuming that we had a good estimate of the mutation rates $m_{\mu\nu}$, we could still calculate the correct $\overline{\bm{x}}(\lambda(t))$ associated with each value of our $\bm{s}(\lambda(t))$ trajectory through Eq.~\eqref{a3}.  Hence main text Eq.~(4) would still yield a valid CD protocol, in the sense that the system would be approximately guided through a series of IE distributions corresponding to $\bm{s}(\lambda(t))$,  assuming we could implement the calculated $\tilde{\bm{s}}(\lambda(t),\bdot\lambda(t))$.  It would not be precisely the series of IE distributions actually associated with $\lambda(t)$, but nevertheless Eq.~(4) would still provide a recipe for following a path in the space of IE distributions.  ii) Another possibility is that our estimate for the mutation rates $m_{\mu\nu}$ is flawed.  This would propagate into errors in our calculation of $\overline{\bm{x}}(\lambda(t))$ associated with $\bm{s}(\lambda(t))$.  Because of these errors the results of Eq.~(4) would no longer necessarily be a CD protocol for $\bm{s}(\lambda(t))$.  

To understand the effects of errors in the mutation rates, we investigated the $M=2$ example with deliberately inaccurate protocols.  For $M=2$ the expression for the mean genotype frequency $\overline{x}_1(\lambda(t))$ is given by Methods Eq.~(19), which in turn yields the CD protocol solution in Methods Eq.~(20).  We modified the protocol by scaling the mutation rates in Eq.~(20) by a factor of $c$, so $m_{12} = m_{21} = c \mu_0$, where $\mu_0 = 2.5\times 10^{-3}$ is the actual mutation rate of the $M=2$ system.  All other parameters as described in Methods Sec.~4.4.  The value $c=1$ corresponds to the true CD protocol, while other values correspond to inaccurate protocols.  As seen in main text Fig.~2D, the true CD protocol works effectively at all times, with the KL divergence between the IE and actual genotype distributions getting no larger than about 0.02 bits.  In contrast, the original protocol leads to severe lag at intermediate times, with the KL divergence peaking around 17 bits.  Fig.~\ref{fig:robustness} shows KL divergence results for inaccurate protocols, with $c$ ranging from 0.1 to 10, calculated using a numerical Fokker-Planck approach.  As expected, we see a breakdown of CD driving, with the IE and actual distributions differing dramatically at intermediate times.  For the most inaccurate protocols, at $c=0.1$ and $c=10$, the peaks in the KL divergence are comparable to or greater than those without CD.  However milder errors, like $c=0.5$ and $c=2$, perform much better than the original protocol, with the divergence peaking around 2 bits.

There is one silver lining in the error analysis:  even though we lose a degree of control over the genotype distributions at intermediate times (they no longer follow precisely the desired series of IE distributions over time), we still retain some benefits in arriving at our destination faster.  All the inaccurate protocols analyzed, particularly those where the mutation rates were underestimated ($c<1$) showed faster convergence to the final equilibrium distribution at long times (i.e. faster decay of the KL divergence) than the original protocol (dashed curve in Fig.~\ref{fig:robustness}).  Thus some practical benefit of the calculated protocol remains for errors up to an order of magnitude in the mutation rates.  Clearly having as good an estimate as possible of the mutation rates is beneficial for complete control, but the approach has some tolerance for the inevitable inaccuracies that will enter into experimentally estimated system parameters.

\section{Mapping the agent-based model to a Fokker-Planck equation}

In order to derive the mapping summarized in Methods Sec. 4.5.3, the first step is describing the dynamics of the agent-based model (ABM) over each time step in terms of a chemical Langevin equation~\cite{gillespie2000chemical}.  This in turns allows us to map the ABM to an effective Fokker-Planck equation, using the standard relationship between Langevin and Fokker-Planck dynamics~\cite{Gillespie1996}.  At simulation time step $\tau$ we have $n_\mu(\tau)$ organisms with genotype $\mu$, and the ABM code will update this to $n_\mu(\tau+1)$ at the next time step.  So long as we are interested in time scales much larger than a single simulation time step, and the updates per step are small compared to $n_\mu(\tau) \gg 1$, we can treat $n_\mu(\tau)$ as a continuous population variable and $\tau$ as a continuous time variable, so that $n_\mu(\tau+1) - n_\mu(\tau) \approx dn_\mu(\tau)/d\tau$.  In this continuum description, the rate of change $dn_\mu(\tau)/d\tau$ will be related to a series of stochastic ``reactions'', which we will label with an index $k=1,\ldots,R$, where $R$ is the total number of possible reactions.  Each reaction represents an aspect of the code that contributes to changes in the genotype populations.  The amount by which the population of genotype $\mu$ changes in the $k$th reaction is denoted as $\Delta_{k\mu}$, and the probability of this reaction occurring at time step $\tau$ is $\pi_k(\tau)$.  The corresponding chemical Langevin approximation describing the change in the population of genotype $\mu$ is given by~\cite{gillespie2000chemical}:
\begin{equation}\label{c1}
\frac{dn_\mu}{d\tau} = \sum_{k=1}^R \Delta_{k\mu}\pi_k(\tau) + \sum_{k=1}^R \Delta_{k\mu} \sqrt{\pi_k(\tau)} \Gamma_k(\tau).
\end{equation}
Note that throughout this section we will not be using the Einstein summation convention, so all sums will be explicitly indicated.  The first term represents the deterministic contribution of all the reactions, while the second term is the corresponding noise introduced by the stochasticity of the reactions.  The $\Gamma_k(\tau)$ functions are independent Gaussian white noise functions with zero mean that satisfy $\langle \Gamma_k(\tau) \Gamma_{k^\prime}(\tau^\prime) \rangle = \delta_{kk^\prime} \delta(\tau-\tau^\prime)$.  The average $\langle \: \rangle$ is taken over an ensemble of realizations of the system (i.e. an ensemble of simulation trajectories).

To complete the Langevin description, we have to identify $\Delta_{k\mu}$ and $\pi_k(\tau)$ for each reaction.  Based on the ABM procedure described in Methods Sec.~4.5.3, we can enumerate the different types of reactions as follows:
\begin{itemize}
    \item There are $M$ possible cell death reactions, one for each genotype.  If reaction $k$ corresponds to the death of a genotype $\mu$ organism, then $\Delta_{k\mu} = -1$, $\Delta_{k\nu} = 0$ for all $\nu \ne \mu$, and $\pi_k(\tau) = n_\mu(\tau)d$, where $d$ is the death probability at each simulation time step.

    \item There are $M$ possible cell division reactions, one for each genotype.  If reaction $k$ corresponds to the division of a genotype $\mu$ organism, then $\Delta_{k\mu} = 1$, $\Delta_{k\nu} = 0$ for all $\nu \ne \mu$, and $\pi_k(\tau) = b_\mu(\tau)(1-d)n_\mu(\tau)$.  
    Here $b_\mu(\tau)$ is the cell division probability given by Methods Eq.~(21), which due to the carrying capacity in the system depends on the current total cell population $N_\text{cell}(\tau) = \sum_{\mu=1}^M n_\mu(\tau)$.  The factor $(1-d)$ accounts for the fact that cell division can only occur if the cell survives the initial culling at the death step of the code.
    
    \item There are $M(M-1)$ possible mutation reactions (assumed to occur instantaneously after the cell division reactions) where newly born cells of one genotype mutate into a different genotype.  If reaction $k$ corresponds a new born cell of type $\nu$ mutating into a different type $\mu$, then $\Delta_{k\mu} = 1$, $\Delta_{k\nu}=-1$, $\Delta_{k\sigma} = 0$ for all $\sigma \ne \mu,\nu$, and $\pi_k(\tau) = \hat{m}_{\mu\nu} b_\nu(\tau)(1-d)n_\nu(\tau)$ where $\hat{m}_{\mu\nu}$ is the mutation probability per time step in the ABM simulation.
    
\end{itemize}

Thus there are altogether $R = M(M+1)$ possible reactions.  Before going further, we will narrow our focus to the system after it reaches equilibrium at some constant set of selection coefficients $s_i(\lambda)$.  The total  population will then fluctuate around some steady state mean value $\overline{N}_\text{cell} = \sum_{\mu=1}^M \langle n_\mu(\tau)\rangle$.  The latter can be calculated by summing both sides of Eq.~\eqref{c1} over all $\mu$, and then taking the mean.  The result is:
\begin{equation}\label{c2}
0 = \sum_{\mu=1}^{M} \sum_{k=1}^R \Delta_{k\mu}\langle\pi_k(\tau)\rangle = -d \overline{N}_\text{cell} + (1-d) \sum_{\mu=1}^M \langle b_\mu(\tau) n_\mu(\tau)\rangle.
\end{equation}
In the second equality we have explicitly plugged in all the different possible reactions and their probabilities.  From Methods Eq.~(21) we know that for $N_\text{cell}(\tau) < K$ the division probability is given by $b_\mu(\tau) = b_0(1+s_i(\lambda))(1-N_\text{cell}(\tau)/K)$.  For simplicity, let us ignore $N_\text{cell}(\tau)$ fluctuations in this expression and write
\begin{equation}\label{c3}
b_\mu(\tau) \approx b_0(1+s_\mu(\lambda))\left(1-\frac{\overline{N}_\text{cell}}{K}\right).
\end{equation}
Substituting this into Eq.~\eqref{c2} we get:
\begin{equation}\label{c4}
0 \approx -d \overline{N}_\text{cell} + (1-d) b_0 \left(1-\frac{\overline{N}_\text{cell}}{K}\right) \sum_{\mu=1}^M (1+s_\mu(\lambda))\langle n_\mu(\tau)\rangle.
\end{equation}
If we make the typical assumption that $|s_\mu(\lambda)| \ll 1$, then the sum on the right is approximately equal to $\overline{N}_\text{cell}$.  This approximation still remains valid even if a subset of genotypes satisfies $|s_\mu(\lambda)| \ll 1$, while the others have substantial negative selection coefficients (as is the case for the fitness landscapes in our simulations).  Then the latter group of less fit genotypes will have negligible populations relative to the former group, and the sum will still approximately evaluate to $\overline{N}_\text{cell}$.  Using this simplification, we can solve Eq.~\eqref{c4} for $\overline{N}_\text{cell}$:
\begin{equation}\label{c5}
\overline{N}_\text{cell} \approx K\left(1-\frac{d}{b_0(1-d)} \right).
\end{equation}
Plugging this into Eq.~\eqref{c3} we find
\begin{equation}\label{c5b}
b_\mu(\tau) \approx (1+s_\mu(\lambda))\frac{d}{1-d}.
\end{equation}
This approximate form for $b_\mu(\tau)$ will prove useful below.

In order to connect the ABM dynamics described through Eq.~\eqref{c1} to our Fokker-Planck formalism, we need to express these dynamics in terms of genotype frequencies, $x_\mu(\tau) = n_\mu(\tau)/N_\text{cell}(\tau)$.  Using the chain rule, the derivative of $x_\mu(\tau)$ with respect to $\tau$ can be written as:
\begin{equation}\label{c6}
\begin{split}
    \frac{dx_\mu(\tau)}{d\tau} &= \frac{1}{N_\text{cell}(\tau)} \frac{dn_\mu(\tau)}{d\tau} - \frac{n_\mu(\tau)}{N_\text{cell}^2(\tau)} \sum_{\nu=1}^{M} \frac{dn_\nu(\tau)}{d\tau}\\
    &=\frac{(1-x_\mu(\tau))}{N_\text{cell}(\tau)} \frac{dn_\mu(\tau)}{d\tau} - \frac{x_\mu(\tau)}{N_\text{cell}(\tau)} \sum_{\nu \ne \mu}^{M} \frac{dn_\nu(\tau)}{d\tau}.
\end{split}
\end{equation}
Using Eq.~\eqref{c1} we can rewrite this as follows:
\begin{equation}\label{c7}
\frac{dx_\mu(\tau)}{d\tau} = \frac{1}{N_\text{cell}(\tau)} \sum_{k=1}^R \pi_k(\tau)\left[(1-x_\mu(\tau)) \Delta_{k\mu}- x_\mu(\tau) \sum_{\nu\ne\mu}^M \Delta_{k\nu} \right] + \eta_\mu(\tau)d,
\end{equation}
where $\eta_\mu(\tau)$ incorporates all the noise contributions,
\begin{equation}\label{c8}
\eta_\mu(\tau) \equiv \frac{1}{N_\text{cell}(\tau)d} \sum_{k=1}^R \Gamma_k(\tau) \sqrt{\pi_k(\tau)}\left[(1-x_\mu(\tau)) \Delta_{k\mu}- x_\mu(\tau) \sum_{\nu\ne\mu}^M \Delta_{k\nu} \right].
\end{equation}
Let us plug in the details of all the reaction types into Eq.~\eqref{c7}, and use Eq.~\eqref{c5b} for $b_\mu(\tau)$ to simplify.  For $\mu = i =1,\ldots,M-1$, we find:
\begin{equation}\label{c9}
\frac{dx_i(\tau)}{d\tau} = \sum_{\nu\ne i}^M \hat{m}_{i\nu}(1+s_{\nu}(\lambda))x_\nu(\tau)d - \sum_{\nu\ne i}^M \hat{m}_{\nu i}(1+s_{i}(\lambda))x_i(\tau)d + \sum_{j=1}^{M-1}  g_{ij}(\bm{x}(\tau)) s_j(\lambda)d + \eta_i(\tau)d,
\end{equation}
where $g$ is the matrix defined in Methods Eq.~(7) and we note that $s_M(\lambda) = 0$ from the definition of the selection coefficients (since genotype $M$ is the wild type).  Let us define a rescaled time $t \equiv \tau d$, which allows Eq.~\eqref{c9} to be written as:
\begin{equation}\label{c10}
\frac{dx_i(t)}{dt} = v_i(\bm{x}(t);\lambda) + \eta_i(t),
\end{equation}
where
\begin{equation}\label{c13}
v_i(\bm{x}(t);\lambda) = \sum_{\nu=1}^M \hat{m}_{i\nu}(1+s_{\nu}(\lambda))x_\nu(t) - \sum_{\nu\ne i}^M \hat{m}_{\nu i}(1+s_{i}(\lambda))x_i(t) + \sum_{j=1}^{M-1}  g_{ij}(\bm{x}(t)) s_j(\lambda).
\end{equation}
Using Eqs.~\eqref{c8}, \eqref{c5b}, the details of the reaction types, and the relation $\langle \Gamma_k(\tau) \Gamma_{k^\prime}(\tau^\prime) \rangle = \delta_{kk^\prime} \delta(\tau-\tau^\prime)$, we can write the correlations of the noise terms for $i,j = 1,\ldots,M-1$ as:
\begin{equation}\label{c11}
\langle \eta_i(t) \eta_j(t^\prime) \rangle = \frac{2\delta(t-t^\prime)}{N_\text{cell}(t)} \left[g_{ij}(\bm{x}(t))+ {\cal O}(\hat{m},\bm{s}(\lambda))\right].
\end{equation}
Note that we have used the fact that Dirac delta functions change under a rescaling of variables as $\delta(\tau-\tau^\prime) = \delta(t-t^\prime) d$. The correction terms not explicitly shown in the bracket of Eq.~\eqref{c11} involve elements of the matrix $\hat{m}$ and selection coefficient vector $\bm{s}(\lambda)$.  Assuming $\hat{m}_{i\mu},\,|s_i(\lambda)| \ll 1$, these can be ignored relative to the leading contribution $g_{ij}(\bm{x}(\tau))$, and we will also approximate $N_\text{cell}(t) \approx \overline{N}_\text{cell}$.  Finally, since the noise functions $\eta_i(t)$ have zero mean, Eq.~\eqref{c11} represents the covariance of the noise, and hence can be used to define a diffusivity matrix:
\begin{equation}\label{c12}
\langle \eta_i(t) \eta_j(t^\prime) \rangle = 2 D_{ij}(\bm{x}(t)) \delta(t-t^\prime), \qquad \text{where} \: D_{ij}(\bm{x}(t)) \approx \frac{g_{ij}(\bm{x}(t))}{\overline{N}_\text{cell}}.
\end{equation}
Eqs.~\eqref{c10}-\eqref{c13}, together with the noise covariance result of Eq.~\eqref{c12}, constitute a system of Langevin equations for the genotype frequencies $x_i(t)$.  Using the standard relation between Langevin and Fokker-Planck equations~\cite{Gillespie1996}, we know that the corresponding Fokker-Planck operator has the form of Eq.~\eqref{s1}, with diffusivity matrix $D_{ij}(\bm{x})$ defined through Eq.~\eqref{c12}, and the velocity function $v_i(\bm{x};\lambda)$ defined through Eq.~\eqref{c13}.

Comparing Eqs.~\eqref{c13} and \eqref{c12} to the velocity and diffusivity definitions in our Wright-Fisher Fokker-Planck formalism, main text Eq.~(6), we see that they match under the mapping:
\begin{equation}\label{c14}
m_{i\nu} = \hat{m}_{i\nu}(1+s_{\nu}(\lambda)) \quad \text{for}\: \nu \ne i, \qquad N = \frac{1}{2}\overline{N}_\text{cell} = \frac{K}{2}\left(1-\frac{d}{b_0(1-d)} \right).
\end{equation}
To derive this mapping, we have used the fact that the diagonal entries of $m$ are defined as $m_{\alpha\alpha} \equiv -\sum_{\beta \ne \alpha} m_{\beta\alpha}$.

\section{\revision{Potential future applications of CD driving in evolutionary systems}}

\revision{
The CD driving examples in the main text and SI were based on anti-malarial drug resistance fitness data from Refs.~\citen{ogbunugafor2016adaptive,brown_compensatory_2010}.  
This experimental system is particularly attractive for our purposes, since it is the only one we are aware of to date for which a full ``seascape'' has been measured:  the set of fitnesses for all possible combinations of a set of alleles of interest over a range of environmental conditions, i.e. drug dosages.
However there are no reasons why similar experimental techniques cannot be applied to construct seascapes for other systems.   
For example, single-environment (single dosage) drug-resistance landscapes have been measured in \textit{E. coli} for 16 antibiotics~\cite{mira2015rational}, and these could be expanded to include fitnesses across a range of drug doses along the same lines as in Ref.~\citen{ogbunugafor2016adaptive}. 

One incentive for measuring such seascapes would be applications of CD in the treatment of drug resistance in infectious disease and cancer.  
Let us consider, for instance, a pathogen and two different drugs that can potentially kill or hinder its growth.  
Prior to drug treatment the pathogen population might consist of a heterogeneous mixture of genetic variants.  
With exposure to one of the drugs, that mixture changes as the pathogen mutates, with resistant variants coming to dominate due to natural selection.  
However suppose that these variants resistant to the first drug are in turn extremely susceptible to the second drug.  
Thus by steering of the evolution of the population by use of the first drug, we have created a situation where the second drug can be particularly effective, improving our chances of eradicating the disease using a two-stage sequence of drugs.  
Clearly for this to work the initial steering stage has to be accomplished quickly~\cite{basanta2012exploiting,gerlee2017extinction}, before the overall population of resistant variants under the first drug rebounds sufficiently to endanger the patient.  
The aforementioned scenario is not merely hypothetical, but describes a phenomenon known as ``collateral sensitivity'' between the pathogen and the two drugs in question.  
It has been observed \textit{in vitro} and in clinical isolates for a variety of infectious diseases, as well as cancer~\cite{imamovic2013use,dhawan2017collateral}.  
Theoretical and experimental work has demonstrated the prevalence of this phenomenon and the potential for carefully crafted drug sequences to exploit this relationship~\cite{nichol2015steering,zhao2016exploiting,barbosa2018antibiotic,nichol2019antibiotic,maltas2019pervasive,acar2020exploiting}.  
However a key question in applying such a strategy clinically is how long it would take for the ``steering'' stage under the first drug to be completed, achieving the desired target distribution of genetic variants ready for the application of the second drug~\cite{basanta2012exploiting,gerlee2017extinction,kaznatcheev2019computational,barbosa2019evolutionary}.  
Given the slow nature of evolution, having the first stage drag out into weeks or months would severely limit the clinical usefulness of this approach.  
CD speed-up could be useful in overcoming this challenge, by providing a dosage protocol to hasten arrival at the target distribution.  
This could be one of the key steps in translating this theoretical treatment idea into a clinically actionable therapeutic strategy. 

There are also systems outside the context of evolutionary medicine where we expect this protocol to be useful despite the required seascape details.
In the context of agriculture, CD driving could potentially be used to speed of up the process of crop breeding. 
As in \textit{E. coli}, fitness landscapes have been successfully measured in maize~\cite{chenu_simulating_2009, messina_yieldtrait_2011}, although not full seascapes. 
Note that these are technically trait-yield landscapes rather than true fitness landscapes, but that by performing artificial selection based on yield we can effectively use them as fitness landscapes. 
Of course, if we had sufficient information to use CD driving on a crop, we would also have sufficient information to use genetic engineering instead.
It is unclear whether it would ever be advantageous to use CD driving over genetic engineering, but we suspect instances exist where it would be (perhaps due to CD diving's ability to act on an entire population at once, or due to anti-GMO sentiment). 
A similar argument holds for the potential applicability of CD driving to synthetic biology.  
  
Measuring seascapes in additional systems would be labor-intensive, but tractable if it were the only obstacle standing between us and substantial advances in applications like those mentioned above.  
If there do exist cases where collecting such full seascapes may be prohibitively difficult, we can try the alternative strategy described in the Discussion section of the main text:  quasi-adiabatic {\it in vitro} experimental trials where we vary external parameters (i.e. drug dosages) extremely gradually, and use sequencing at regular intervals to determine the quasi-equilibrium mean genotype fractions as a function of dosage.
This would be generally faster than trying to determine the fitnesses of each genotype at each dosage level, and would provide enough information to construct our CD protocol through Eq.~(4) of the main text.
}

\section{\revision{Generality of the CD approach:  deriving approximate protocols in additional landscape examples}}

\revision{
For the empirical seascape used in the main text, describing the response of 16 genotypes to the anti-malarial drug pyrimethamine~\cite{ogbunugafor2016adaptive,brown_compensatory_2010}, we noted that two of those genotypes dominate at different times during the driving:  initially at small concentrations the genotype 1110 ($i=15$) has the largest population, but is eventually overtaken by the wild-type genotype 1111 ($i=16$) at large concentrations as the drug is ramped up.  This allowed a particularly simple approach to approximate the CD protocol of main text Eq.~(4) by focusing on the the selection coefficient $s_{15}$ that describes the relative fitness of 1110 versus 1111:  at every time step we numerically find the drug concentration $\tilde{\lambda}(t)$ where the corresponding selection coefficient $s_{15}(\tilde{\lambda}(t))$ is closest to the value of the CD solution $\tilde{s}_{15}(\lambda(t),\dot\lambda(t))$ from Eq.~(4), with the constraint that $\tilde{\lambda}(t)$ must be lower than a certain maximum cutoff $\tilde\lambda_\text{max}$.  This approximate CD dosage protocol $\tilde{\lambda}(t)$ is compared to the original dosage protocol $\lambda(t)$ in main text Fig.~3E for different cutoffs $\tilde\lambda_\text{max}$.  While we saw the resulting CD driving worked quite well for this example, how would we generalize this approach to more complex scenarios?

In this section we describe such a general approach for finding the set of experimental control parameters $\tilde{\lambda}(t)$ that best approximates main text Eq.~(4) given an arbitrary seascape.   We show how the two-genotype approximation described above is a special case of this general approach, and illustrate the latter through two additional driving examples:  one using a modified version of the pyrimethamine seascape, and the other using the empirical seascape for cycloguanil, another anti-malarial drug~\cite{ogbunugafor2016adaptive}.  In both of these cases driving takes us through three different regimes, with different genotypes dominating in each regime.  

Let us start with Eq.~(18) of the Methods,
\begin{equation}\label{g1}
\begin{split}
\partial_t \overline{\bm{x}}(\lambda(t)) &=  \bm{g}\big(\overline{\bm{x}}(\lambda(t))\big) \delta \tilde{\bm{s}}\big(\lambda(t),\bdot{\lambda}(t)\big)\\
&= \bm{g}\big(\overline{\bm{x}}(\lambda(t))\big) \left( \tilde{\bm{s}}\big(\lambda(t),\bdot{\lambda}(t)\big) - \bm{s}(\lambda(t))\right).
\end{split}
\end{equation}
The CD solution $\tilde{\bm{s}}\big(\lambda(t),\bdot{\lambda}(t)\big)$ to this equation is shown in main text Eq.~(4).  Now let us imagine we want to use the control parameters accessible to us and implement a protocol $\tilde\lambda(t)$ that approximates this CD solution.  To do this, let us replace $\tilde{\bm{s}}\big(\lambda(t),\bdot{\lambda}(t)\big)$ with $\bm{s}(\tilde\lambda(t))$ on the right-hand side of Eq.~\eqref{g1}.  Subtracting the left and right-hand sides of the resulting equation from each other, and squaring the difference, gives us the following loss function: 
\begin{equation}\label{g2}
L(\tilde\lambda(t)) = \left[\partial_t \overline{\bm{x}}(\lambda(t)) - \bm{g}\big(\overline{\bm{x}}(\lambda(t))\big) \left( \bm{s}\big(\tilde\lambda(t)\big) - \bm{s}(\lambda(t))\right)\right]^2.
\end{equation}
If we could find a $\tilde\lambda(t)$ such that $L(\tilde\lambda(t)) = 0$ at all $t$ then we would have exactly implemented the CD solution of main text Eq.~(4).  However in practice the control parameters $\tilde\lambda(t)$ that we can externally manipulate may not allow this to be perfectly satisfied.  Hence we do the next best thing and find $\tilde\lambda(t)$ that minimizes the loss function.  Note that the loss function can also be rewritten in the following form, where we substitute in the right-hand side of Eq.~\eqref{g1} for $\partial_t \overline{\bm{x}}(\lambda(t))$:
\begin{equation}\label{g3}
\begin{split}
L(\tilde\lambda(t)) &= \left[\bm{g}\big(\overline{\bm{x}}(\lambda(t))\big) \left( \tilde{\bm{s}}\big(\lambda(t),\bdot{\lambda}(t)\big) - \bm{s}(\lambda(t))\right) - \bm{g}\big(\overline{\bm{x}}(\lambda(t))\big) \left( \bm{s}\big(\tilde\lambda(t)\big) - \bm{s}(\lambda(t))\right)\right]^2\\
&=\left[\bm{g}\big(\overline{\bm{x}}(\lambda(t))\big) \left( \tilde{\bm{s}}\big(\lambda(t),\bdot{\lambda}(t)\big) - \bm{s}\big(\tilde\lambda(t)\big)\right)\right]^2.
\end{split}
\end{equation}
In this form we can explicitly see that minimizing the loss function is the same as minimizing the difference between $\tilde{\bm{s}}\big(\lambda(t),\bdot{\lambda}(t)\big)$ from main text Eq.~(4) and the experimentally accessible selection coefficient trajectory $\bm{s}\big(\tilde\lambda(t)\big)$.  This difference is weighted by the response matrix $\bm{g}\big(\overline{\bm{x}}(\lambda(t))\big)$, which governs how much each component of the selection coefficient contributes to the velocity term in the Fokker-Planck equation through Methods Eq.~(6).  This is not the only possible form of the loss function (one can imagine alternative ways of weighting the selection coefficient differences) but in the numerical examples we investigated this form gave us the best driving results.

Minimizing Eq.~\eqref{g3} (or equivalently Eq.~\eqref{g2}) to find $\tilde\lambda(t)$ is a general prescription for getting an approximate CD protocol, regardless of the system.  How well that protocol will work (how close we can get to true CD driving) is of course a system-specific question, which will depend on the nature and number of control parameters that we can vary and the ranges they can explore (the dimensionality and extent of the space of possible $\tilde\lambda(t)$ curves).  In general we expect that the more degrees of freedom we have for $\tilde\lambda(t)$, the closer we can approach the CD solution.  Because simultaneously controlling a large number of parameters is experimentally more challenging, in practice one will seek out the smallest number that can still give a reasonable approximation to CD driving.

\begin{figure}[t!]
\centering\includegraphics[width=0.6\textwidth]{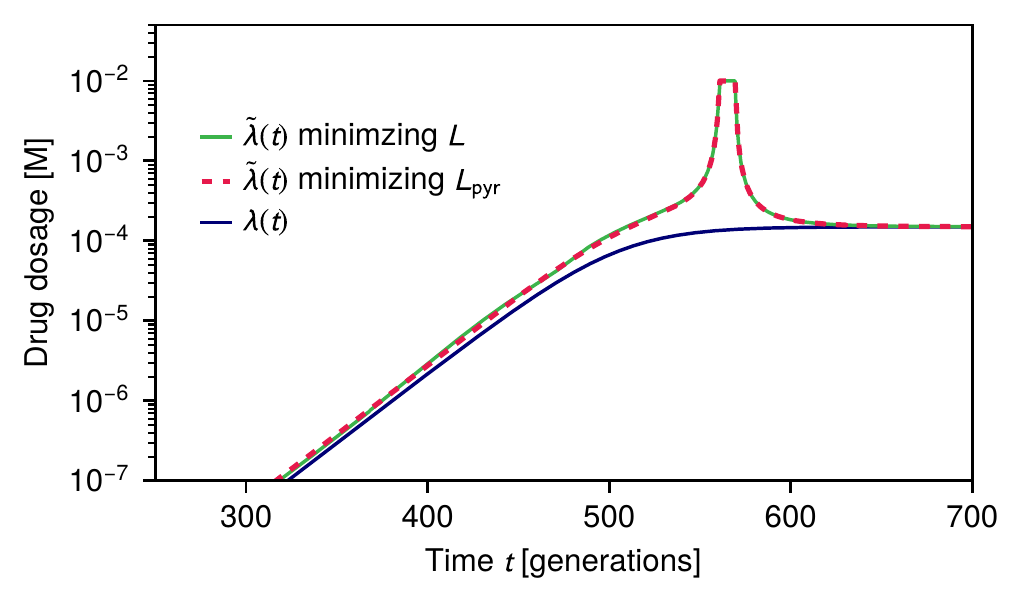}
\caption{\revision{Comparison of the original drug dosage $\lambda(t)$ (blue curve) for the pyrimethamine seascape example of the main text versus the approximate CD dosage protocol $\tilde\lambda(t)$ calculated using two methods:  minimizing the full loss function $L(\tilde\lambda(t))$ of Eq.~\eqref{g3} (green curve) versus minimzing the simplified loss function $L_\text{pyr}(\tilde\lambda(t))$ of Eq.~\eqref{g4} (dashed red curve).}}\label{fig:app1}
\end{figure}

\begin{figure}[t]
\centering\includegraphics[width=\textwidth]{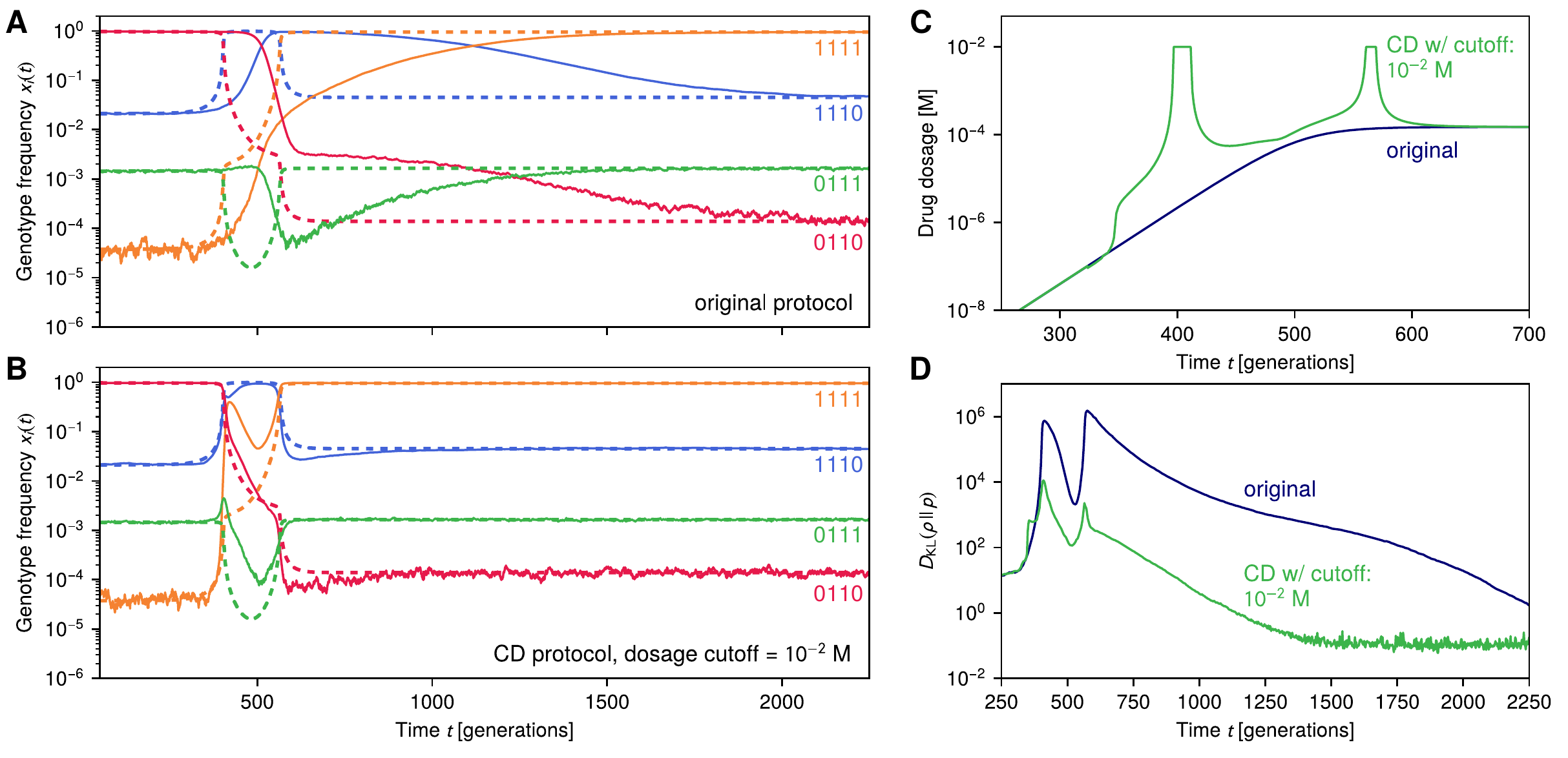}
\caption{\revision{\textbf{CD driving for an altered 16-genotype pyrimethamine seascape.}  This is the same seascape as in main text Fig.~3, using the experimental data of Ref.~\citen{ogbunugafor2016adaptive}, except that genotype 0110 has been modified to have a 5\% larger base growth rate under no drug conditions. \textbf{(A,B)} Sample simulation trajectories (solid lines) versus IE expectation (dashed lines) for the fraction of 4 representative genotypes without \textbf{(A)} and with \textbf{(B)} CD driving.  The CD driving is implemented approximately through the drug dosage protocol (green curve) shown in panel (\textbf{C}) with cutoff 10$^{-2}$ M.  The original protocol (blue curve) is shown for comparison. \textbf{(D)} Kullback-Leibler divergence between actual and IE distributions versus time, with and without CD driving.}}\label{fig:alt_driving}
\end{figure}

\begin{figure}[t]
\centering\includegraphics[width=\textwidth]{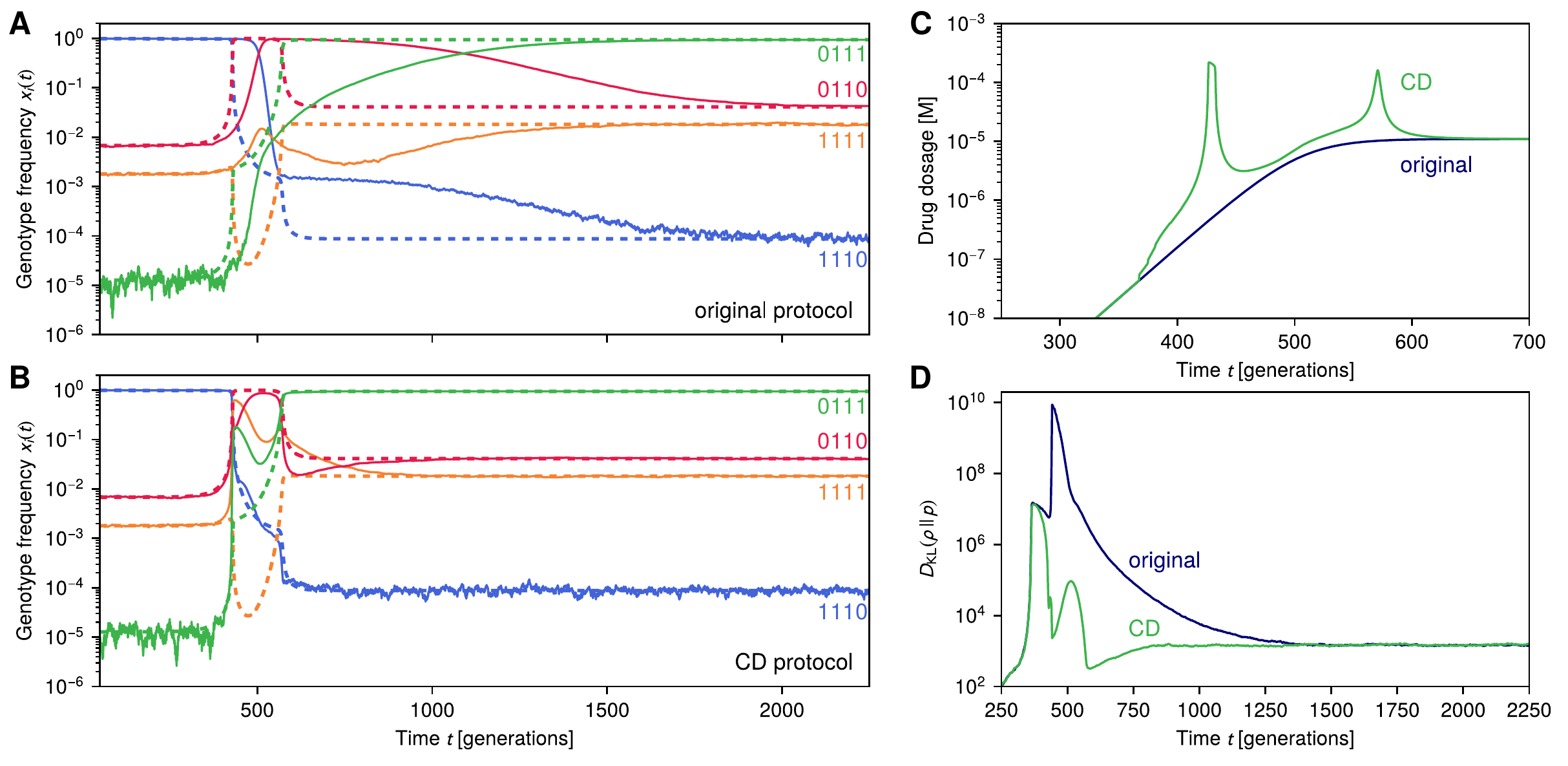}
\caption{\revision{\textbf{CD driving for a 16-genotype cycloguanil seascape.}  
This is the same 16-genotype system as in the examples of main text Fig.~3 and Fig.~\ref{fig:alt_driving}, except using the anti-malarial drug cycloguanil instead of pyrimethamine.  
The seascape is based on the experimental data of Ref.~\citen{ogbunugafor2016adaptive}, without any modifications. 
\textbf{(A,B)} Sample simulation trajectories (solid lines) versus IE expectation (dashed lines) for the fraction of 4 representative genotypes without \textbf{(A)} and with \textbf{(B)} CD driving.  
The CD driving is implemented approximately through the drug dosage protocol (green curve) shown in panel (\textbf{C}).  The original protocol (blue curve) is shown for comparison. \textbf{(D)} Kullback-Leibler divergence between actual and IE distributions versus time, with and without CD driving.}}\label{fig:cyc_driving}
\end{figure}

To get some intuition into this approximation scheme, let us consider several examples.  For the pyrimethamine seascape used in the main text, minimizing Eq.~\eqref{g3} turns out to be nearly the same as minimizing the simpler alternative loss function that involves only the $i=15$ selection coefficient component:
\begin{equation}\label{g4}
    L_\text{pyr}(\tilde\lambda(t)) = \left( \tilde{s}_{15}\big(\lambda(t),\bdot{\lambda}(t)\big) - s_{15}\big(\tilde\lambda(t)\big)\right)^2.
\end{equation}
This is because $i=15$ difference makes the major contribution to the right-hand side of Eq.~\eqref{g3}, given that the 1110 and 1111 genotypes dominate the population at different stages of the driving.  Fig~\ref{fig:app1} shows the drug protocol $\tilde\lambda(t)$ calculated by minimizing Eq.~\eqref{g3} versus that found from minimizing the simpler loss function of Eq.~\eqref{g4}.  The two are nearly identical throughout the entire time range.  As can be seen in main text Fig.~3, for this system a single control parameter (pyrimethamine drug dosage) is sufficient to get an excellent approximation to the CD protocol over the entire time range.  The genotype distributions under the protocol, $p(\bm{x},t)$, are kept close to the IE target distributions $\rho(\bm{x};\lambda(t))$, measured via their KL divergence (main text Fig.~3F).  Since the distributions are narrow for the large $N$ case, an individual simulation trajectory (solid curves in main text Fig.~3C) will closely follow the mean IE genotype frequencies $\overline{\bm{x}}\big(\lambda(t)\big)$ (dashed curves in main text Fig.~3C).  The excellent performance of the approximate CD protocol in this case likely stems from the fact that the loss function is dominated by a single component ($i=15$), and varying a one-dimensional control parameter allows us to effectively minimize this component.  

Let us now make the situation more complicated in the following way.  
Take the same pyrimethamine seascape, based on the empirical data of Ref.~\citen{ogbunugafor2016adaptive}, and make a single parameter alteration:  increase the base growth rate (at zero drug concentration) of genotype 0110 by 5\%.  Everything else in the system remains the same.  
Under the original drug protocol (blue curve in Fig.~\ref{fig:alt_driving}C) the mean IE genotype fractions (dashed curves in Fig.~\ref{fig:alt_driving}A,B) show a more complex behavior over time:  0110 dominates at small times, followed by a period of 1110 domination, until at large times 1111 takes the lead.  
As time is varied there are now multiple components on the right-hand side of Eq.~\eqref{g3} that shift in their relative significance.  
The result of minimizing $L(\tilde\lambda(t))$ is the approximate CD protocol $\tilde\lambda(t)$ shown in green in Fig.~\ref{fig:alt_driving}C.  
This now has two dosage peaks, one around the 0110-1110 transition and the other around the 1110-1111 transition.  Under the original protocol the simulation trajectory lags behind the IE expectation (Fig.~\ref{fig:alt_driving}A), but  
with the approximate CD driving this lag is largely eliminated (Fig.~\ref{fig:alt_driving}B).  The overall reduction in lag time (quantified as described in Methods Sec.~4.5.4) is $\Delta t = 1128$ generations, comparable to the reduction achieved in the main text example.  Note however that the agreement between simulation and IE curves during the 1110-dominated regime is only partial. 
Clearly, manipulating a single control parameter (pyrimethamine dosage) is not sufficient to achieve a close enough approximation to the CD protocol to get fine-grained control in this regime. 
The drug dosage peak near generation 400 is sufficient to drive the interchange between 0110 and 1110, but as a side-effect causes the overshoot of 1111 with respect to its IE expectation.  
However this overshoot is resolved quickly as 1111 approaches its IE curve, and the subsequent driving pushes the system toward the final IE values with little delay.  
Thus if one were interested only in getting to the final target distribution of genotypes quickly (likely the most common real-world application), this approximate CD protocol would suffice.  
On the other hand there could be situations where following the exact IE path through genotype space at all times was important, for example if we wanted to steer the system away from problematic intermediate genotypes in some evolutionary medicine scenario.  
In this case we could benefit from having additional control knobs.  
For example the possibility of using a second drug alongside pyrimethamine might allow for a better CD approximation.

Fig.~\ref{fig:cyc_driving} tells a qualitatively similar story to Fig.~\ref{fig:alt_driving}, but using the seascape from cycloguanil, a different anti-malarial drug.  
The cycloguanil data was also experimentally measured in Ref.~\citen{ogbunugafor2016adaptive}, and we do not make any modifications to the empirical values.  
We choose an original dosage protocol $\lambda(t)$ with the same form as before (Methods Eq.~(22)), but with saturating dosage parameter $a = 1.1 \times 10^{-5}$ M. 
Under this protocol, we see three regimes during the driving, but in a different combination from the previous example:  the dominance shifts from 1110 to 0110 to 0111 as the drug is increased.  
Again the approximate CD protocol involves two dosage peaks, and gets us to the final target distribution with less lag than the original protocol ($\Delta t = 373$ generations).  
Though we currently only have data in this system for one drug applied at any given time, we see that despite this limitation we can still do a fairly decent job of eliminating lag, whether that drug be pyrimethamine or cycloguanil. 
The imperfect control during the intermediate regime could potentially be resolved by a protocol that involves both drugs applied simultaneously (or another external intervention).  
Thus, while our theory provides the basic framework to understand CD driving in such evolutionary systems, there are a variety of pragmatic questions about balancing driving accuracy versus the complexity of the control protocol that will provide interesting subjects for future work.
}


\begin{thebibliography}{10}
\urlstyle{rm}
\expandafter\ifx\csname url\endcsname\relax
  \def\url#1{\texttt{#1}}\fi
\expandafter\ifx\csname urlprefix\endcsname\relax\def\urlprefix{URL }\fi
\expandafter\ifx\csname doiprefix\endcsname\relax\def\doiprefix{DOI: }\fi
\providecommand{\bibinfo}[2]{#2}
\providecommand{\eprint}[2][]{\url{#2}}

\bibitem{mira2015rational}
\bibinfo{author}{Mira, P.~M.} \emph{et~al.}
\newblock \bibinfo{journal}{\bibinfo{title}{Rational design of antibiotic
  treatment plans: a treatment strategy for managing evolution and reversing
  resistance}}.
\newblock {\emph{\JournalTitle{PloS One}}} \textbf{\bibinfo{volume}{10}},
  \bibinfo{pages}{e0122283} (\bibinfo{year}{2015}).

\bibitem{ogbunugafor2016adaptive}
\bibinfo{author}{Ogbunugafor, C.~B.}, \bibinfo{author}{Wylie, C.~S.},
  \bibinfo{author}{Diakite, I.}, \bibinfo{author}{Weinreich, D.~M.} \&
  \bibinfo{author}{Hartl, D.~L.}
\newblock \bibinfo{journal}{\bibinfo{title}{Adaptive landscape by environment
  interactions dictate evolutionary dynamics in models of drug resistance}}.
\newblock {\emph{\JournalTitle{PLoS Comp. Biol.}}}
  \textbf{\bibinfo{volume}{12}}, \bibinfo{pages}{e1004710}
  (\bibinfo{year}{2016}).

\bibitem{brown_compensatory_2010}
\bibinfo{author}{Brown, K.~M.} \emph{et~al.}
\newblock \bibinfo{journal}{\bibinfo{title}{Compensatory mutations restore
  fitness during the evolution of dihydrofolate reductase}}.
\newblock {\emph{\JournalTitle{Mol. Biol. Evol.}}}
  \textbf{\bibinfo{volume}{27}}, \bibinfo{pages}{2682--2690}
  (\bibinfo{year}{2010}).

\bibitem{WHO2014resistance}
\bibinfo{author}{{World Health Organization}}.
\newblock \bibinfo{title}{{Antimicrobial resistance: global report on
  surveillance.}} (\bibinfo{year}{2014}).
\newblock \bibinfo{note}{ISBN: 978 92 4 156474 8}.

\bibitem{holohan2013resistance}
\bibinfo{author}{Holohan, C.}, \bibinfo{author}{Van~Schaeybroeck, S.},
  \bibinfo{author}{Longley, D.~B.} \& \bibinfo{author}{Johnston, P.~G.}
\newblock \bibinfo{journal}{\bibinfo{title}{Cancer drug resistance: an evolving
  paradigm}}.
\newblock {\emph{\JournalTitle{Nat. Rev. Cancer}}}
  \textbf{\bibinfo{volume}{13}}, \bibinfo{pages}{714--726}
  (\bibinfo{year}{2013}).

\bibitem{WHO2019resistance}
\bibinfo{author}{{World Health Organization}}.
\newblock \bibinfo{title}{{HIV drug resistance report 2019}}
  (\bibinfo{year}{2019}).
\newblock \bibinfo{note}{Licence: CC BY-NC-SA 3.0 IGO.}

\bibitem{nichol2015steering}
\bibinfo{author}{Nichol, D.} \emph{et~al.}
\newblock \bibinfo{journal}{\bibinfo{title}{Steering evolution with sequential
  therapy to prevent the emergence of bacterial antibiotic resistance}}.
\newblock {\emph{\JournalTitle{PLoS Comp. Biol.}}}
  \textbf{\bibinfo{volume}{11}}, \bibinfo{pages}{e1004493}
  (\bibinfo{year}{2015}).

\bibitem{maltas2018pervasive}
\bibinfo{author}{Maltas, J.} \& \bibinfo{author}{Wood, K.~B.}
\newblock \bibinfo{journal}{\bibinfo{title}{Pervasive and diverse collateral
  sensitivity profiles inform optimal strategies to limit antibiotic
  resistance}}.
\newblock {\emph{\JournalTitle{bioRxiv}}} \bibinfo{pages}{241075}
  (\bibinfo{year}{2018}).

\bibitem{bason2012high}
\bibinfo{author}{Bason, M.~G.} \emph{et~al.}
\newblock \bibinfo{journal}{\bibinfo{title}{High-fidelity quantum driving}}.
\newblock {\emph{\JournalTitle{Nat. Phys.}}} \textbf{\bibinfo{volume}{8}},
  \bibinfo{pages}{147} (\bibinfo{year}{2012}).

\bibitem{zhou2017accelerated}
\bibinfo{author}{Zhou, B.~B.} \emph{et~al.}
\newblock \bibinfo{journal}{\bibinfo{title}{Accelerated quantum control using
  superadiabatic dynamics in a solid-state lambda system}}.
\newblock {\emph{\JournalTitle{Nat. Phys.}}} \textbf{\bibinfo{volume}{13}},
  \bibinfo{pages}{330} (\bibinfo{year}{2017}).

\bibitem{walther2012controlling}
\bibinfo{author}{Walther, A.} \emph{et~al.}
\newblock \bibinfo{journal}{\bibinfo{title}{Controlling fast transport of cold
  trapped ions}}.
\newblock {\emph{\JournalTitle{Phys. Rev. Lett.}}}
  \textbf{\bibinfo{volume}{109}}, \bibinfo{pages}{080501}
  (\bibinfo{year}{2012}).

\bibitem{farhi2001quantum}
\bibinfo{author}{Farhi, E.} \emph{et~al.}
\newblock \bibinfo{journal}{\bibinfo{title}{A quantum adiabatic evolution
  algorithm applied to random instances of an np-complete problem}}.
\newblock {\emph{\JournalTitle{Science}}} \textbf{\bibinfo{volume}{292}},
  \bibinfo{pages}{472--475} (\bibinfo{year}{2001}).

\bibitem{torrontegui2013shortcuts}
\bibinfo{author}{Torrontegui, E.} \emph{et~al.}
\newblock \bibinfo{title}{Shortcuts to adiabaticity}.
\newblock In \emph{\bibinfo{booktitle}{Adv. At. Mol. Opt. Phys.}},
  vol.~\bibinfo{volume}{62}, \bibinfo{pages}{117--169}
  (\bibinfo{publisher}{Elsevier}, \bibinfo{year}{2013}).

\bibitem{deffner2014classical}
\bibinfo{author}{Deffner, S.}, \bibinfo{author}{Jarzynski, C.} \&
  \bibinfo{author}{del Campo, A.}
\newblock \bibinfo{journal}{\bibinfo{title}{Classical and quantum shortcuts to
  adiabaticity for scale-invariant driving}}.
\newblock {\emph{\JournalTitle{Phys. Rev. X}}} \textbf{\bibinfo{volume}{4}},
  \bibinfo{pages}{021013} (\bibinfo{year}{2014}).

\bibitem{deffner2015njp}
\bibinfo{author}{Deffner, S.}
\newblock \bibinfo{journal}{\bibinfo{title}{Shortcuts to adiabaticity:
  suppression of pair production in driven dirac dynamics}}.
\newblock {\emph{\JournalTitle{New Journal of Physics}}}
  \textbf{\bibinfo{volume}{18}}, \bibinfo{pages}{012001}
  (\bibinfo{year}{2015}).

\bibitem{acconcia2015pre}
\bibinfo{author}{Acconcia, T.~V.},
  \bibinfo{author}{Bonan\ifmmode~\mbox{\c{c}}\else \c{c}\fi{}a, M. V.~S.} \&
  \bibinfo{author}{Deffner, S.}
\newblock \bibinfo{journal}{\bibinfo{title}{Shortcuts to adiabaticity from
  linear response theory}}.
\newblock {\emph{\JournalTitle{Phys. Rev. E}}} \textbf{\bibinfo{volume}{92}},
  \bibinfo{pages}{042148} (\bibinfo{year}{2015}).

\bibitem{campbell2017trade}
\bibinfo{author}{Campbell, S.} \& \bibinfo{author}{Deffner, S.}
\newblock \bibinfo{journal}{\bibinfo{title}{Trade-off between speed and cost in
  shortcuts to adiabaticity}}.
\newblock {\emph{\JournalTitle{Phys. Rev. Lett.}}}
  \textbf{\bibinfo{volume}{118}}, \bibinfo{pages}{100601}
  (\bibinfo{year}{2017}).

\bibitem{guery2019sta}
\bibinfo{author}{Gu\'ery-Odelin, D.} \emph{et~al.}
\newblock \bibinfo{journal}{\bibinfo{title}{Shortcuts to adiabaticity:
  Concepts, methods, and applications}}.
\newblock {\emph{\JournalTitle{Rev. Mod. Phys.}}}
  \textbf{\bibinfo{volume}{91}}, \bibinfo{pages}{045001}
  (\bibinfo{year}{2019}).

\bibitem{demirplak2003adiabatic}
\bibinfo{author}{Demirplak, M.} \& \bibinfo{author}{Rice, S.~A.}
\newblock \bibinfo{journal}{\bibinfo{title}{Adiabatic population transfer with
  control fields}}.
\newblock {\emph{\JournalTitle{J. Phys. Chem. A}}}
  \textbf{\bibinfo{volume}{107}}, \bibinfo{pages}{9937--9945}
  (\bibinfo{year}{2003}).

\bibitem{demirplak2005assisted}
\bibinfo{author}{Demirplak, M.} \& \bibinfo{author}{Rice, S.~A.}
\newblock \bibinfo{journal}{\bibinfo{title}{Assisted adiabatic passage
  revisited}}.
\newblock {\emph{\JournalTitle{J. Phys. Chem. B}}}
  \textbf{\bibinfo{volume}{109}}, \bibinfo{pages}{6838--6844}
  (\bibinfo{year}{2005}).

\bibitem{berry2009transitionless}
\bibinfo{author}{Berry, M.~V.}
\newblock \bibinfo{journal}{\bibinfo{title}{Transitionless quantum driving}}.
\newblock {\emph{\JournalTitle{J. Phys. A: Math. Theor.}}}
  \textbf{\bibinfo{volume}{42}}, \bibinfo{pages}{365303}
  (\bibinfo{year}{2009}).

\bibitem{patra2017shortcuts}
\bibinfo{author}{Patra, A.} \& \bibinfo{author}{Jarzynski, C.}
\newblock \bibinfo{journal}{\bibinfo{title}{Shortcuts to adiabaticity using
  flow fields}}.
\newblock {\emph{\JournalTitle{New J. Phys.}}} \textbf{\bibinfo{volume}{19}},
  \bibinfo{pages}{125009} (\bibinfo{year}{2017}).

\bibitem{li2017shortcuts}
\bibinfo{author}{Li, G.}, \bibinfo{author}{Quan, H.} \& \bibinfo{author}{Tu,
  Z.}
\newblock \bibinfo{journal}{\bibinfo{title}{Shortcuts to isothermality and
  nonequilibrium work relations}}.
\newblock {\emph{\JournalTitle{Phys. Rev. E}}} \textbf{\bibinfo{volume}{96}},
  \bibinfo{pages}{012144} (\bibinfo{year}{2017}).

\bibitem{martinez2016engineered}
\bibinfo{author}{Mart{\'\i}nez, I.~A.}, \bibinfo{author}{Petrosyan, A.},
  \bibinfo{author}{Gu{\'e}ry-Odelin, D.}, \bibinfo{author}{Trizac, E.} \&
  \bibinfo{author}{Ciliberto, S.}
\newblock \bibinfo{journal}{\bibinfo{title}{Engineered swift equilibration of a
  brownian particle}}.
\newblock {\emph{\JournalTitle{Nat. Phys.}}} \textbf{\bibinfo{volume}{12}},
  \bibinfo{pages}{843} (\bibinfo{year}{2016}).

\bibitem{le2016fast}
\bibinfo{author}{Le~Cunuder, A.} \emph{et~al.}
\newblock \bibinfo{journal}{\bibinfo{title}{Fast equilibrium switch of a micro
  mechanical oscillator}}.
\newblock {\emph{\JournalTitle{Appl. Phys. Lett.}}}
  \textbf{\bibinfo{volume}{109}}, \bibinfo{pages}{113502}
  (\bibinfo{year}{2016}).

\bibitem{schmiedl2007optimal}
\bibinfo{author}{Schmiedl, T.} \& \bibinfo{author}{Seifert, U.}
\newblock \bibinfo{journal}{\bibinfo{title}{Optimal finite-time processes in
  stochastic thermodynamics}}.
\newblock {\emph{\JournalTitle{Phys. Rev. Lett.}}}
  \textbf{\bibinfo{volume}{98}}, \bibinfo{pages}{108301}
  (\bibinfo{year}{2007}).

\bibitem{aurell2012refined}
\bibinfo{author}{Aurell, E.}, \bibinfo{author}{Gaw\k{e}dzki, K.},
  \bibinfo{author}{Mej{\'\i}a-Monasterio, C.}, \bibinfo{author}{Mohayaee, R.}
  \& \bibinfo{author}{Muratore-Ginanneschi, P.}
\newblock \bibinfo{journal}{\bibinfo{title}{Refined second law of
  thermodynamics for fast random processes}}.
\newblock {\emph{\JournalTitle{J. Stat. Phys.}}}
  \textbf{\bibinfo{volume}{147}}, \bibinfo{pages}{487--505}
  (\bibinfo{year}{2012}).

\bibitem{wright1932roles}
\bibinfo{author}{Wright, S.}
\newblock \emph{\bibinfo{title}{The roles of mutation, inbreeding,
  crossbreeding, and selection in evolution}}, vol.~\bibinfo{volume}{1}
  (\bibinfo{publisher}{na}, \bibinfo{year}{1932}).

\bibitem{Mustonen2010}
\bibinfo{author}{Mustonen, V.} \& \bibinfo{author}{L{\"a}ssig, M.}
\newblock \bibinfo{journal}{\bibinfo{title}{Fitness flux and ubiquity of
  adaptive evolution}}.
\newblock {\emph{\JournalTitle{Proc. Natl. Acad. Sci.}}}
  \textbf{\bibinfo{volume}{107}}, \bibinfo{pages}{4248--4253}
  (\bibinfo{year}{2010}).

\bibitem{grabert1979quantum}
\bibinfo{author}{Grabert, H.}, \bibinfo{author}{H\"anggi, P.} \&
  \bibinfo{author}{Talkner, P.}
\newblock \bibinfo{journal}{\bibinfo{title}{Is quantum mechanics equivalent to
  a classical stochastic process?}}
\newblock {\emph{\JournalTitle{Phys. Rev. A}}} \textbf{\bibinfo{volume}{19}},
  \bibinfo{pages}{2440--2445} (\bibinfo{year}{1979}).

\bibitem{van1992stochastic}
\bibinfo{author}{Van~Kampen, N.~G.}
\newblock \emph{\bibinfo{title}{Stochastic processes in physics and chemistry}}
  (\bibinfo{publisher}{Elsevier}, \bibinfo{year}{1992}).

\bibitem{risken1996fokker}
\bibinfo{author}{Risken, H.}
\newblock \emph{\bibinfo{title}{The {Fokker}-{Planck} {Equation}}}
  (\bibinfo{publisher}{Springer}, \bibinfo{year}{1996}).

\bibitem{born1928adiabatic}
\bibinfo{author}{Born, M.} \& \bibinfo{author}{Fock, V.}
\newblock \bibinfo{journal}{\bibinfo{title}{Beweis des adiabatensatzes}}.
\newblock {\emph{\JournalTitle{Z. Phys.}}} \textbf{\bibinfo{volume}{51}},
  \bibinfo{pages}{165--180} (\bibinfo{year}{1928}).

\bibitem{nichol2019antibiotic}
\bibinfo{author}{Nichol, D.} \emph{et~al.}
\newblock \bibinfo{journal}{\bibinfo{title}{Antibiotic collateral sensitivity
  is contingent on the repeatability of evolution}}.
\newblock {\emph{\JournalTitle{Nat. Commun.}}} \textbf{\bibinfo{volume}{10}},
  \bibinfo{pages}{334} (\bibinfo{year}{2019}).

\bibitem{li2019single}
\bibinfo{author}{Li, Y.}, \bibinfo{author}{Petrov, D.~A.} \&
  \bibinfo{author}{Sherlock, G.}
\newblock \bibinfo{journal}{\bibinfo{title}{Single nucleotide mapping of trait
  space reveals pareto fronts that constrain adaptation}}.
\newblock {\emph{\JournalTitle{Nature Ecol. Evol.}}} \bibinfo{pages}{1--13}
  (\bibinfo{year}{2019}).

\bibitem{vaikuntanathan2009lag}
\bibinfo{author}{Vaikuntanathan, S.} \& \bibinfo{author}{Jarzynski, C.}
\newblock \bibinfo{journal}{\bibinfo{title}{Dissipation and lag in irreversible
  processes}}.
\newblock {\emph{\JournalTitle{{EPL} (Europhysics Letters)}}}
  \textbf{\bibinfo{volume}{87}}, \bibinfo{pages}{60005} (\bibinfo{year}{2009}).

\bibitem{gillespie1983simple}
\bibinfo{author}{Gillespie, J.~H.}
\newblock \bibinfo{journal}{\bibinfo{title}{A simple stochastic gene
  substitution model}}.
\newblock {\emph{\JournalTitle{Theor. Popul. Biol.}}}
  \textbf{\bibinfo{volume}{23}}, \bibinfo{pages}{202--215}
  (\bibinfo{year}{1983}).

\bibitem{gerrish1998fate}
\bibinfo{author}{Gerrish, P.~J.} \& \bibinfo{author}{Lenski, R.~E.}
\newblock \bibinfo{journal}{\bibinfo{title}{The fate of competing beneficial
  mutations in an asexual population}}.
\newblock {\emph{\JournalTitle{Genetica}}} \textbf{\bibinfo{volume}{102}},
  \bibinfo{pages}{127} (\bibinfo{year}{1998}).

\bibitem{desai2007beneficial}
\bibinfo{author}{Desai, M.~M.} \& \bibinfo{author}{Fisher, D.~S.}
\newblock \bibinfo{journal}{\bibinfo{title}{Beneficial mutation--selection
  balance and the effect of linkage on positive selection}}.
\newblock {\emph{\JournalTitle{Genetics}}} \textbf{\bibinfo{volume}{176}},
  \bibinfo{pages}{1759--1798} (\bibinfo{year}{2007}).

\bibitem{sniegowski2010beneficial}
\bibinfo{author}{Sniegowski, P.~D.} \& \bibinfo{author}{Gerrish, P.~J.}
\newblock \bibinfo{journal}{\bibinfo{title}{Beneficial mutations and the
  dynamics of adaptation in asexual populations}}.
\newblock {\emph{\JournalTitle{Philosophical Transactions of the Royal Society
  B: Biological Sciences}}} \textbf{\bibinfo{volume}{365}},
  \bibinfo{pages}{1255--1263} (\bibinfo{year}{2010}).

\bibitem{martens2011interfering}
\bibinfo{author}{Martens, E.~A.} \& \bibinfo{author}{Hallatschek, O.}
\newblock \bibinfo{journal}{\bibinfo{title}{Interfering waves of adaptation
  promote spatial mixing}}.
\newblock {\emph{\JournalTitle{Genetics}}} \textbf{\bibinfo{volume}{189}},
  \bibinfo{pages}{1045--1060} (\bibinfo{year}{2011}).

\bibitem{magdanova2013heterogeneity}
\bibinfo{author}{Magdanova, L.} \& \bibinfo{author}{Golyasnaya, N.}
\newblock \bibinfo{journal}{\bibinfo{title}{Heterogeneity as an adaptive trait
  of microbial populations}}.
\newblock {\emph{\JournalTitle{Microbiology}}} \textbf{\bibinfo{volume}{82}},
  \bibinfo{pages}{1--10} (\bibinfo{year}{2013}).

\bibitem{krishnan2019range}
\bibinfo{author}{Krishnan, N.} \& \bibinfo{author}{Scott, J.~G.}
\newblock \bibinfo{journal}{\bibinfo{title}{Range expansion shifts clonal
  interference patterns in evolving populations}}.
\newblock {\emph{\JournalTitle{bioRxiv}}} \bibinfo{pages}{794867}
  (\bibinfo{year}{2019}).

\bibitem{Sella2005}
\bibinfo{author}{Sella, G.} \& \bibinfo{author}{Hirsh, A.~E.}
\newblock \bibinfo{journal}{\bibinfo{title}{The application of statistical
  physics to evolutionary biology}}.
\newblock {\emph{\JournalTitle{Proc. Natl. Acad. Sci.}}}
  \textbf{\bibinfo{volume}{102}}, \bibinfo{pages}{9541--9546}
  (\bibinfo{year}{2005}).

\bibitem{discreteCDpreprint}
\bibinfo{author}{Chiel, J.} \emph{et~al.} (\bibinfo{year}{2020}).
\newblock \bibinfo{note}{In preparation}.

\bibitem{kullback_information_1951}
\bibinfo{author}{Kullback, S.} \& \bibinfo{author}{Leibler, R.~A.}
\newblock \bibinfo{journal}{\bibinfo{title}{On {Information} and
  {Sufficiency}}}.
\newblock {\emph{\JournalTitle{Ann. Math. Stat.}}}
  \textbf{\bibinfo{volume}{22}}, \bibinfo{pages}{79--86}
  (\bibinfo{year}{1951}).

\bibitem{kaznatcheev2019computational}
\bibinfo{author}{Kaznatcheev, A.}
\newblock \bibinfo{journal}{\bibinfo{title}{Computational complexity as an
  ultimate constraint on evolution}}.
\newblock {\emph{\JournalTitle{Genetics}}} \textbf{\bibinfo{volume}{212}},
  \bibinfo{pages}{245--265} (\bibinfo{year}{2019}).

\bibitem{dobzhansky1973nothing}
\bibinfo{author}{Dobzhansky, T.}
\newblock \bibinfo{journal}{\bibinfo{title}{Nothing in biology makes sense
  except in the light of evolution}}.
\newblock {\emph{\JournalTitle{Am. Biol. Teach.}}}
  \textbf{\bibinfo{volume}{35}}, \bibinfo{pages}{125--129}
  (\bibinfo{year}{1973}).

\bibitem{romero2009exploring}
\bibinfo{author}{Romero, P.~A.} \& \bibinfo{author}{Arnold, F.~H.}
\newblock \bibinfo{journal}{\bibinfo{title}{Exploring protein fitness
  landscapes by directed evolution}}.
\newblock {\emph{\JournalTitle{Nat. Rev. Mol. Cell Biol}}}
  \textbf{\bibinfo{volume}{10}}, \bibinfo{pages}{866} (\bibinfo{year}{2009}).

\bibitem{tuerk1990systematic}
\bibinfo{author}{Tuerk, C.} \& \bibinfo{author}{Gold, L.}
\newblock \bibinfo{journal}{\bibinfo{title}{Systematic evolution of ligands by
  exponential enrichment: Rna ligands to bacteriophage t4 dna polymerase}}.
\newblock {\emph{\JournalTitle{Science}}} \textbf{\bibinfo{volume}{249}},
  \bibinfo{pages}{505--510} (\bibinfo{year}{1990}).

\bibitem{ellington1990vitro}
\bibinfo{author}{Ellington, A.~D.} \& \bibinfo{author}{Szostak, J.~W.}
\newblock \bibinfo{journal}{\bibinfo{title}{In vitro selection of rna molecules
  that bind specific ligands}}.
\newblock {\emph{\JournalTitle{Nature}}} \textbf{\bibinfo{volume}{346}},
  \bibinfo{pages}{818} (\bibinfo{year}{1990}).

\bibitem{bita2013plant}
\bibinfo{author}{Bita, C.} \& \bibinfo{author}{Gerats, T.}
\newblock \bibinfo{journal}{\bibinfo{title}{Plant tolerance to high temperature
  in a changing environment: scientific fundamentals and production of heat
  stress-tolerant crops}}.
\newblock {\emph{\JournalTitle{Front. Plant Sci.}}}
  \textbf{\bibinfo{volume}{4}}, \bibinfo{pages}{273} (\bibinfo{year}{2013}).

\bibitem{Baxter2007}
\bibinfo{author}{Baxter, G.~J.}, \bibinfo{author}{Blythe, R.~A.} \&
  \bibinfo{author}{McKane, A.~J.}
\newblock \bibinfo{journal}{\bibinfo{title}{Exact solution of the multi-allelic
  diffusion model}}.
\newblock {\emph{\JournalTitle{Math. Biosci.}}} \textbf{\bibinfo{volume}{209}},
  \bibinfo{pages}{124--170} (\bibinfo{year}{2007}).

\bibitem{Kimura1955}
\bibinfo{author}{Kimura, M.}
\newblock \bibinfo{journal}{\bibinfo{title}{Stochastic processes and
  distribution of gene frequencies under natural selection}}.
\newblock {\emph{\JournalTitle{Cold Spring Harb. Symp. Quant. Biol.}}}
  \textbf{\bibinfo{volume}{20}}, \bibinfo{pages}{33--53}
  (\bibinfo{year}{1955}).

\bibitem{Gillespie1996}
\bibinfo{author}{Gillespie, D.~T.}
\newblock \bibinfo{journal}{\bibinfo{title}{The multivariate {Langevin} and
  {Fokker}-{Planck} equations}}.
\newblock {\emph{\JournalTitle{Am. J. Phys.}}} \textbf{\bibinfo{volume}{64}},
  \bibinfo{pages}{1246--1257} (\bibinfo{year}{1996}).

\bibitem{sahoo2008}
\bibinfo{author}{Sahoo, S.}
\newblock \bibinfo{journal}{\bibinfo{title}{Inverse vector operators}}.
\newblock {\emph{\JournalTitle{arXiv}}} \bibinfo{pages}{0804.2239}
  (\bibinfo{year}{2008}).

\bibitem{gillespie2000chemical}
\bibinfo{author}{Gillespie, D.~T.}
\newblock \bibinfo{journal}{\bibinfo{title}{The chemical {Langevin} equation}}.
\newblock {\emph{\JournalTitle{J. Chem. Phys.}}}
  \textbf{\bibinfo{volume}{113}}, \bibinfo{pages}{297--306}
  (\bibinfo{year}{2000}).

\bibitem{griffiths2018introduction}
\bibinfo{author}{Griffiths, D.~J.} \& \bibinfo{author}{Schroeter, D.~F.}
\newblock \emph{\bibinfo{title}{Introduction to quantum mechanics}}
  (\bibinfo{publisher}{Cambridge University Press}, \bibinfo{year}{2018}).

\bibitem{ryter1987eigenfunctions}
\bibinfo{author}{Ryter, D.}
\newblock \bibinfo{journal}{\bibinfo{title}{On the eigenfunctions of the
  {Fokker}-{Planck} operator and of its adjoint}}.
\newblock {\emph{Physica A}} \textbf{\bibinfo{volume}{142}},
  \bibinfo{pages}{103--121} (\bibinfo{year}{1987}).


\bibitem{basanta2012exploiting}
\bibinfo{author}{Basanta, D.}, \bibinfo{author}{Gatenby, R.~A.} \&
  \bibinfo{author}{Anderson, A.~R.}
\newblock \bibinfo{journal}{\bibinfo{title}{Exploiting evolution to treat drug
  resistance: combination therapy and the double bind}}.
\newblock {\emph{\JournalTitle{Molecular pharmaceutics}}}
  \textbf{\bibinfo{volume}{9}}, \bibinfo{pages}{914--921}
  (\bibinfo{year}{2012}).

\bibitem{gerlee2017extinction}
\bibinfo{author}{Gerlee, P.} \& \bibinfo{author}{Altrock, P.~M.}
\newblock \bibinfo{journal}{\bibinfo{title}{Extinction rates in tumour public
  goods games}}.
\newblock {\emph{\JournalTitle{Journal of The Royal Society Interface}}}
  \textbf{\bibinfo{volume}{14}}, \bibinfo{pages}{20170342}
  (\bibinfo{year}{2017}).

\bibitem{imamovic2013use}
\bibinfo{author}{Imamovic, L.} \& \bibinfo{author}{Sommer, M.~O.}
\newblock \bibinfo{journal}{\bibinfo{title}{Use of collateral sensitivity
  networks to design drug cycling protocols that avoid resistance
  development}}.
\newblock {\emph{\JournalTitle{Science translational medicine}}}
  \textbf{\bibinfo{volume}{5}}, \bibinfo{pages}{204ra132--204ra132}
  (\bibinfo{year}{2013}).

\bibitem{dhawan2017collateral}
\bibinfo{author}{Dhawan, A.} \emph{et~al.}
\newblock \bibinfo{journal}{\bibinfo{title}{Collateral sensitivity networks
  reveal evolutionary instability and novel treatment strategies in alk mutated
  non-small cell lung cancer}}.
\newblock {\emph{\JournalTitle{Scientific Reports}}}
  \textbf{\bibinfo{volume}{7}}, \bibinfo{pages}{1--9} (\bibinfo{year}{2017}).

\bibitem{zhao2016exploiting}
\bibinfo{author}{Zhao, B.} \emph{et~al.}
\newblock \bibinfo{journal}{\bibinfo{title}{Exploiting temporal collateral
  sensitivity in tumor clonal evolution}}.
\newblock {\emph{\JournalTitle{Cell}}} \textbf{\bibinfo{volume}{165}},
  \bibinfo{pages}{234--246} (\bibinfo{year}{2016}).

\bibitem{barbosa2018antibiotic}
\bibinfo{author}{Barbosa, C.}, \bibinfo{author}{Beardmore, R.},
  \bibinfo{author}{Schulenburg, H.} \& \bibinfo{author}{Jansen, G.}
\newblock \bibinfo{journal}{\bibinfo{title}{Antibiotic combination efficacy
  (ace) networks for a pseudomonas aeruginosa model}}.
\newblock {\emph{\JournalTitle{PLoS biology}}} \textbf{\bibinfo{volume}{16}},
  \bibinfo{pages}{e2004356} (\bibinfo{year}{2018}).

\bibitem{maltas2019pervasive}
\bibinfo{author}{Maltas, J.} \& \bibinfo{author}{Wood, K.~B.}
\newblock \bibinfo{journal}{\bibinfo{title}{Pervasive and diverse collateral
  sensitivity profiles inform optimal strategies to limit antibiotic
  resistance}}.
\newblock {\emph{\JournalTitle{PLoS biology}}} \textbf{\bibinfo{volume}{17}}
  (\bibinfo{year}{2019}).

\bibitem{acar2020exploiting}
\bibinfo{author}{Acar, A.} \emph{et~al.}
\newblock \bibinfo{journal}{\bibinfo{title}{Exploiting evolutionary steering to
  induce collateral drug sensitivity in cancer}}.
\newblock {\emph{\JournalTitle{Nature Communications}}}
  \textbf{\bibinfo{volume}{11}}, \bibinfo{pages}{1--14} (\bibinfo{year}{2020}).

\bibitem{kaznatcheev2019computational}
\bibinfo{author}{Kaznatcheev, A.}
\newblock \bibinfo{journal}{\bibinfo{title}{Computational complexity as an
  ultimate constraint on evolution}}.
\newblock {\emph{\JournalTitle{Genetics}}} \textbf{\bibinfo{volume}{212}},
  \bibinfo{pages}{245--265} (\bibinfo{year}{2019}).

\bibitem{barbosa2019evolutionary}
\bibinfo{author}{Barbosa, C.}, \bibinfo{author}{Roemhild, R.},
  \bibinfo{author}{Rosenstiel, P.} \& \bibinfo{author}{Schulenburg, H.}
\newblock \bibinfo{journal}{\bibinfo{title}{Evolutionary stability of
  collateral sensitivity to antibiotics in the model pathogen pseudomonas
  aeruginosa}}.
\newblock {\emph{\JournalTitle{eLife}}} \textbf{\bibinfo{volume}{8}},
  \bibinfo{pages}{e51481} (\bibinfo{year}{2019}).

\bibitem{chenu_simulating_2009}
\bibinfo{author}{Chenu, K.} \emph{et~al.}
\newblock \bibinfo{journal}{\bibinfo{title}{Simulating the {Yield} {Impacts} of
  {Organ}-{Level} {Quantitative} {Trait} {Loci} {Associated} {With} {Drought}
  {Response} in {Maize}: {A} “{Gene}-to-{Phenotype}” {Modeling}
  {Approach}}}.
\newblock {\emph{\JournalTitle{Genetics}}} \textbf{\bibinfo{volume}{183}},
  \bibinfo{pages}{1507--1523}, \doiprefix\url{10.1534/genetics.109.105429}
  (\bibinfo{year}{2009}).

\bibitem{messina_yieldtrait_2011}
\bibinfo{author}{Messina, C.~D.}, \bibinfo{author}{Podlich, D.},
  \bibinfo{author}{Dong, Z.}, \bibinfo{author}{Samples, M.} \&
  \bibinfo{author}{Cooper, M.}
\newblock \bibinfo{journal}{\bibinfo{title}{Yield–trait performance
  landscapes: from theory to application in breeding maize for drought
  tolerance}}.
\newblock {\emph{\JournalTitle{Journal of Experimental Botany}}}
  \textbf{\bibinfo{volume}{62}}, \bibinfo{pages}{855--868},
  \doiprefix\url{10.1093/jxb/erq329} (\bibinfo{year}{2011}).


\end{thebibliography}
\end{document}